\pdfoutput=1

\documentclass[11pt]{article}
\usepackage[margin=1in]{geometry}
\usepackage{booktabs,graphicx,amsmath,xcolor,hyperref,caption}
\IfFileExists{xurl.sty}{\usepackage{xurl}}{}
\usepackage[numbers,sort&compress]{natbib}
\newcommand{\keeppub}{\ensuremath{-}193.68}
\newcommand{\keeprepro}{\ensuremath{-}191.34}
\newcommand{\keepreprolo}{\ensuremath{-}202.61}
\newcommand{\keepreprohi}{\ensuremath{-}180.07}
\newcommand{\keepfirst}{\ensuremath{-}196.06}
\newcommand{\hmarlpub}{\ensuremath{-}132.63}
\newcommand{\hmarlpubsd}{0.87}
\newcommand{\hmarlrepro}{\ensuremath{-}131.34}
\newcommand{\hmarlreprolo}{\ensuremath{-}140.61}
\newcommand{\hmarlreprohi}{\ensuremath{-}122.08}
\newcommand{\crossmachinesd}{76.37}
\newcommand{\pristinemean}{\ensuremath{-}154}
\newcommand{\pristinesd}{32.53}
\newcommand{\keepbusyattempt}{0.385}
\newcommand{\hmarlbusyattempt}{0.529}
\newcommand{\keepexecshare}{0.605}
\newcommand{\hmarlexecshare}{0.358}
\newcommand{\keepgainfired}{452.3}
\newcommand{\keepgainheld}{624.9}
\newcommand{\keepheldvsfired}{172.7}
\newcommand{\keepheldvsfiredwins}{79}
\newcommand{\hmarlgainfired}{98.5}
\newcommand{\hmarlgainheld}{414.8}
\newcommand{\hmarlheldvsfired}{316.3}
\newcommand{\heldclickkeep}{\ensuremath{-}1.2}
\newcommand{\heldclickkeepwins}{40}
\newcommand{\heldclickhmarl}{\ensuremath{-}29.9}
\newcommand{\heldclickhmarlp}{0.023}
\newcommand{\dirkeepfired}{0.487}
\newcommand{\dirhmarlfired}{0.140}
\newcommand{\dirkeepheld}{0.743}
\newcommand{\dirhmarlheld}{0.730}
\newcommand{\burstfired}{\ensuremath{-}874.27}
\newcommand{\burstheld}{\ensuremath{-}327.54}
\newcommand{\burstheldshare}{0.937}
\newcommand{\burstvsspread}{162.9}
\newcommand{\burstvsspreadwins}{96}
\newcommand{\decoyfiredgap}{243.6}
\newcommand{\decoyheldrfourtn}{\ensuremath{-}413.37}
\newcommand{\decoyheldrtwo}{\ensuremath{-}372.64}
\newcommand{\samplingsd}{59.39}
\newcommand{\samplingvarshare}{0.889}
\newcommand{\samplingsdwrecker}{320.96}
\newcommand{\difficultysdwrecker}{41.12}
\newcommand{\nullbandstock}{14.9}
\newcommand{\nullbandwins}{66}
\newcommand{\nullbandwrecker}{89.3}
\newcommand{\nullbandhmarl}{73.2}
\newcommand{\unseededa}{\ensuremath{-}196.5}
\newcommand{\unseededb}{\ensuremath{-}181}
\newcommand{\paralleldistinct}{12}
\newcommand{\hmarlbestofsix}{40.3}
\newcommand{\hmarlnullcentre}{\ensuremath{-}883.09}
\newcommand{\contsignone}{\ensuremath{-}65.7}
\newcommand{\contsigntwo}{92.9}
\newcommand{\contsignthree}{\ensuremath{-}119.3}
\newcommand{\trainenvs}{25}
\newcommand{\sleepscore}{\ensuremath{-}6592.68}
\newcommand{\bestepisode}{\ensuremath{-}22}
\newcommand{\greedyscore}{\ensuremath{-}552.22}
\newcommand{\greedyfactor}{2.95}
\newcommand{\restoreprecision}{45.59}
\newcommand{\removeprecision}{24.12}
\newcommand{\oonescore}{\ensuremath{-}520.29}
\newcommand{\oonedelta}{\ensuremath{-}332.97}
\newcommand{\oonewins}{1}
\newcommand{\evidenceprecision}{0.279}
\newcommand{\evidencerecall}{0.480}
\newcommand{\evidenceflagged}{0.345}
\newcommand{\evidencelift}{1.39}
\newcommand{\hmarlownmetric}{75.4}
\newcommand{\tridentkeepscore}{\ensuremath{-}202.34}
\newcommand{\publishedkeepbytes}{3536882}
\newcommand{\pyarrowpin}{15.0.2}
\newcommand{\episodes}{100}
\newcommand{\shaikhepisodes}{200}
\newcommand{\hmarlheldvsfiredwins}{83}
\newcommand{\samplingvarsharelo}{0.716}
\newcommand{\samplingvarsharewrecker}{0.984}
\newcommand{\samplingvarsharewreckerlo}{0.927}
\newcommand{\difficultysdstock}{21.03}
\newcommand{\widestepisodestock}{335.4}
\newcommand{\widestepisodewrecker}{625.5}
\newcommand{\burstperattemptfired}{3.1}
\newcommand{\spreadperattemptfired}{8.3}
\newcommand{\burstratiofired}{2.6}
\newcommand{\burstperexecheld}{11.2}
\newcommand{\spreadperexecheld}{13.3}
\newcommand{\spreadfired}{\ensuremath{-}680.42}
\newcommand{\chanceredshare}{0.240}
\newcommand{\hmarlnullsd}{28.23}
\newcommand{\hmarlgainheldnull}{455.1}
\newcommand{\hmarlgainfirednull}{138.9}
\newcommand{\keepreprose}{3.54}
\newcommand{\hmarlreprose}{2.91}
\newcommand{\heldclickkeepp}{0.826}
\newcommand{\heldclickhmarlwins}{32}
\newcommand{\decoyheldrtwod}{\ensuremath{-}380.56}
\newcommand{\evidenceprecisionstock}{0.285}
\newcommand{\evidencerecallstock}{0.521}
\newcommand{\evidenceflaggedstock}{0.380}
\newcommand{\evidenceliftstock}{1.37}
\newcommand{\greedysd}{343.75}
\newcommand{\dirkeepfiredceiling}{0.538}
\newcommand{\dirhmarlfiredceiling}{0.174}
\newcommand{\burstbudget}{77}
\newcommand{\decoyrestoresrfourtn}{26.4}
\newcommand{\decoyrestoresrtwo}{39.3}
\newcommand{\contsignonewins}{38}
\newcommand{\contsigntwowins}{63}
\newcommand{\contsignthreewins}{40}
\newcommand{\keepevidenceshare}{71.62}
\newcommand{\ooneevidenceshare}{70.12}
\newcommand{\ruleperiod}{32}
\newcommand{\restoresperepisode}{77.1}
\newcommand{\removesperepisode}{102.6}
\newcommand{\leaderboardkeepsd}{21.07}
\newcommand{\acdzeroepisodes}{100}
\newcommand{\acdzerorecpreclo}{61}
\newcommand{\acdzerorecprechi}{71}
\newcommand{\hcmarlepisodes}{10}
\newcommand{\hcmarlsinghscore}{\ensuremath{-}181.62}
\newcommand{\hcmarliocdelta}{242}
\newcommand{\hcmarlnoiocscore}{\ensuremath{-}388.5}
\newcommand{\acdzeroscore}{\ensuremath{-}150.03}
\newcommand{\acdzerosd}{19.85}
\newcommand{\leaderboardkeepsdhc}{84}
\newcommand{\hcmarlthreelayer}{\ensuremath{-}146.2}
\newcommand{\hcmarltwolayer}{\ensuremath{-}131.6}
\newcommand{\hmarlcompatseedzero}{\ensuremath{-}124.11}
\newcommand{\hmarlunseeded}{\ensuremath{-}131.99}
\newcommand{\smokescore}{\ensuremath{-}194}

\newcommand{\wreckerimpactp}{0.55}
\newcommand{\wreckerdegradep}{0.40}
\newcommand{\episodesteps}{500}
\newcommand{\samplingseeds}{4}
\newcommand{\tridentkeepsd}{1.42}
\newcommand{\stockimpactp}{0.25}
\newcommand{\stockdegradep}{0.25}
\newcommand{\stockdiscoverp}{0.50}
\newcommand{\cagerev}{8c3c50ca54b1}
\newcommand{\keeprev}{2afd652d80ce}
\newcommand{\hmarlrev}{6fe960f5931f}
\newcommand{\lbfirst}{\ensuremath{-}113.81}
\newcommand{\lbsecond}{\ensuremath{-}118.46}
\newcommand{\lbkeeprank}{5}
\newcommand{\lbentries}{14}

\newcommand{\hmarlauthorscore}{\ensuremath{-}129.53}
\newcommand{\hmarlauthorsd}{44.60}
\newcommand{\hmarlexpertprec}{73}
\newcommand{\decoybusyrfourtn}{0.317}
\newcommand{\decoybusyrtwo}{0.185}
\newcommand{\decoyproposedrfourtn}{58.9}
\newcommand{\decoyproposedrtwo}{55.9}
\newcommand{\oonetargets}{6466}
\newcommand{\keepremoveevidenceshare}{53.94}
\newcommand{\armcount}{123}
\newcommand{\cagetwoepisodes}{1000}
\newcommand{\cagerevdate}{2025-10-27}
\newcommand{\keeprevdate}{2025-11-18}
\newcommand{\hmarlrevdate}{2025-04-03}
\newcommand{\tridentruns}{3}
\newcommand{\tridentrunepisodes}{500}
\newcommand{\lbgapfirstsecond}{4.65}
\newcommand{\keepwreckercentre}{\ensuremath{-}1143.49}
\newcommand{\keepbestofthree}{27.84}
\newcommand{\keepcorpusmean}{\ensuremath{-}187.32}
\newcommand{\seedlo}{0}
\newcommand{\seedhi}{99}
\newcommand{\fsredmapline}{547}
\newcommand{\wreckerdiscoverp}{0.05}
\newcommand{\hmarlauthorepisodes}{100}
\newcommand{\decoyheldgap}{40.7}
\newcommand{\decoyheldgapwins}{41}
\newcommand{\decoyheldgapp}{0.089}
\newcommand{\decoyheldrtwodgap}{7.9}
\newcommand{\decoyheldrtwodwins}{49}
\newcommand{\spreadasserted}{52.7}
\newcommand{\spreadexecuted}{30.1}
\newcommand{\burstheldasserted}{71.0}
\newcommand{\burstheldexecuted}{70.6}
\newcommand{\wreckerrootpriority}{64}
\newcommand{\wreckerqrootpriority}{32}

\newcommand{\castrosteps}{499}
\newcommand{\castroclusterbusy}{101}
\newcommand{\tridentrev}{a4c7e34d9bf2}
\newcommand{\tridentrevdate}{2026-09-23}
\newcommand{\hmarlstockcontrol}{\ensuremath{-}136.43}
\newcommand{\leaderboardepisodes}{100}
\newcommand{\samplingvarsharehi}{0.998}
\newcommand{\samplingvarsharewreckerhi}{1.000}
\newcommand{\chanceredsharelatch}{0.201}
\newcommand{\nullbandwidestwins}{52}
\newcommand{\keepexecsharewb}{0.990}
\newcommand{\hmarlexecsharewb}{0.989}
\newcommand{\contwreckerone}{95.4}
\newcommand{\contwreckeronewins}{60}
\newcommand{\contwreckertwo}{167.2}
\newcommand{\contwreckertwowins}{70}
\newcommand{\contwreckerthree}{28.9}
\newcommand{\contwreckerthreewins}{55}

\newcommand{\burstfiredasserted}{77.0}
\newcommand{\burstfiredexecuted}{14.6}
\newcommand{\rfourtndecoysasserted}{772.5}
\newcommand{\rfourtndecoysexecuted}{203.7}
\newcommand{\oonedeficit}{332.97}

\newcommand{\hmarlwreckercontrol}{\ensuremath{-}842.76}
\newcommand{\chanceredsharelatchstock}{0.208}
\newcommand{\tridenthidden}{320}
\newcommand{\keephidden}{256}
\newcommand{\tridentconv}{280}
\newcommand{\keepconv}{128}
\newcommand{\tridentclip}{0.3}
\newcommand{\keepclip}{0.2}
\newcommand{\tridentlr}{3e-5}
\newcommand{\keeplr}{3e-4}
\newcommand{\oonerulerestoreshare}{0.650}
\newcommand{\oonepassthrough}{0.350}
\newcommand{\wreckerrdimpactp}{0.55}
\newcommand{\wreckerrddegradep}{0.45}

\newcommand{\tridentcriticlr}{1e-4}
\newcommand{\keepcriticlr}{1e-3}
\newcommand{\ooneinterventions}{401.8}
\newcommand{\keepheldvsfiredlo}{133.8}
\newcommand{\keepheldvsfiredhi}{212.2}
\newcommand{\hmarlheldvsfiredlo}{241.5}
\newcommand{\hmarlheldvsfiredhi}{393.9}
\newcommand{\heldclickkeeplo}{\ensuremath{-}23.2}
\newcommand{\heldclickkeephi}{20.4}
\newcommand{\heldclickkeepties}{17}
\newcommand{\heldclickhmarllo}{\ensuremath{-}53.4}
\newcommand{\heldclickhmarlhi}{\ensuremath{-}7.2}
\newcommand{\heldclickhmarlties}{14}
\newcommand{\burstvsspreadlo}{137.9}
\newcommand{\burstvsspreadhi}{189.4}
\newcommand{\decoyfiredgaplo}{138.7}
\newcommand{\decoyfiredgaphi}{345.0}
\newcommand{\decoyheldgaplo}{13.2}
\newcommand{\decoyheldgaphi}{68.6}
\newcommand{\decoyheldrtwodgaplo}{\ensuremath{-}13.2}
\newcommand{\decoyheldrtwodgaphi}{28.9}
\newcommand{\oonedeltalo}{\ensuremath{-}354.6}
\newcommand{\oonedeltahi}{\ensuremath{-}311.2}
\newcommand{\contsignonelo}{\ensuremath{-}141.4}
\newcommand{\contsignonehi}{9.7}
\newcommand{\contsigntwolo}{32.1}
\newcommand{\contsigntwohi}{153.6}
\newcommand{\contsignthreelo}{\ensuremath{-}203.2}
\newcommand{\contsignthreehi}{\ensuremath{-}41.4}
\newcommand{\contwreckeronelo}{15.1}
\newcommand{\contwreckeronehi}{173.8}
\newcommand{\contwreckertwolo}{87.9}
\newcommand{\contwreckertwohi}{245.5}
\newcommand{\contwreckerthreelo}{\ensuremath{-}61.8}
\newcommand{\contwreckerthreehi}{113.7}
\newcommand{\nulldeltasdlo}{62.1}
\newcommand{\nulldeltasdhi}{101.1}
\newcommand{\sheldvsfired}{25.3}
\newcommand{\sheldvsfiredlo}{15.6}
\newcommand{\sheldvsfiredhi}{35.2}
\newcommand{\sheldvsfiredwins}{77}
\newcommand{\hmsheldvsfired}{27.6}
\newcommand{\hmsheldvsfiredlo}{21.3}
\newcommand{\hmsheldvsfiredhi}{33.8}
\newcommand{\hmsheldvsfiredwins}{77}

\newcommand{\soonefired}{\ensuremath{-}12.8}
\newcommand{\soonefiredlo}{\ensuremath{-}27.1}
\newcommand{\soonefiredhi}{1.3}
\newcommand{\soonefiredwins}{35}

\newcommand{\sooneheld}{12.5}
\newcommand{\sooneheldlo}{0.1}
\newcommand{\sooneheldhi}{25.7}
\newcommand{\sooneheldwins}{57}
\newcommand{\sooneheldp}{0.129}
\newcommand{\hmsoonefired}{\ensuremath{-}24.3}
\newcommand{\hmsoonefiredlo}{\ensuremath{-}33.5}
\newcommand{\hmsoonefiredhi}{\ensuremath{-}14.3}
\newcommand{\hmsoonefiredwins}{22}
\newcommand{\hmsooneheld}{3.3}
\newcommand{\hmsooneheldlo}{\ensuremath{-}6.2}
\newcommand{\hmsooneheldhi}{13.8}
\newcommand{\hmsooneheldwins}{43}
\newcommand{\hmsooneheldp}{0.228}
\newcommand{\sooneunboundedheld}{\ensuremath{-}5.5}
\newcommand{\sooneunboundedheldlo}{\ensuremath{-}21.3}
\newcommand{\sooneunboundedheldhi}{10.1}
\newcommand{\sooneunboundedheldwins}{43}
\newcommand{\sooneunboundedheldp}{0.266}
\newcommand{\sooneunboundedheldinterventions}{74.3}
\newcommand{\ooneexecutedown}{73.8}
\newcommand{\curveqfour}{49.8}
\newcommand{\curveqfourlo}{36.0}
\newcommand{\curveqfourhi}{63.5}
\newcommand{\curveqfourwins}{80}
\newcommand{\curveqsixteen}{95.4}
\newcommand{\curveqsixteenlo}{76.9}
\newcommand{\curveqsixteenhi}{115.6}
\newcommand{\curveqsixteenwins}{85}
\newcommand{\curveqthirtytwo}{131.6}
\newcommand{\curveqthirtytwolo}{90.4}
\newcommand{\curveqthirtytwohi}{170.1}
\newcommand{\curveqthirtytwowins}{83}
\newcommand{\curveqfortyeight}{161.2}
\newcommand{\curveqfortyeightlo}{130.1}
\newcommand{\curveqfortyeighthi}{192.3}
\newcommand{\curveqfortyeightwins}{82}
\newcommand{\curveconvonly}{51.7}
\newcommand{\curveconvonlylo}{36.7}
\newcommand{\curveconvonlyhi}{67.6}
\newcommand{\curveconvonlywins}{84}
\newcommand{\hmarlsamplingsharewrecker}{0.907}
\newcommand{\hmarlsamplingsharewreckerlo}{0.847}
\newcommand{\hmarlsamplingsharewreckerhi}{0.972}
\newcommand{\samplingvarsharetrim}{0.720}
\newcommand{\samplingvarsharewreckertrim}{0.967}
\newcommand{\samplingsdmedian}{36.5}
\newcommand{\samplingsdmedianwrecker}{316.7}
\newcommand{\oneseedse}{6.3}
\newcommand{\oneseedsewrecker}{32.4}
\newcommand{\stockdrawsdlo}{49.40}
\newcommand{\stockdrawsdhi}{83.33}
\newcommand{\hmarlsamplingsharestock}{0.889}

\newcommand{\dirkeepfiredlo}{0.437}
\newcommand{\dirkeepfiredhi}{0.534}
\newcommand{\dirkeepheldloss}{0.673}
\newcommand{\dirkeepheldlosslo}{0.637}
\newcommand{\dirkeepheldlosshi}{0.706}
\newcommand{\dirhmarlfiredlo}{0.016}
\newcommand{\dirhmarlfiredhi}{0.249}
\newcommand{\dirhmarlheldloss}{0.587}
\newcommand{\dirhmarlheldlosslo}{0.516}
\newcommand{\dirhmarlheldlosshi}{0.645}
\newcommand{\dirkeepfiredceilinglo}{0.485}
\newcommand{\dirkeepfiredceilinghi}{0.587}
\newcommand{\dirkeepheldlo}{0.707}
\newcommand{\dirkeepheldhi}{0.777}
\newcommand{\dirhmarlfiredceilinglo}{0.021}
\newcommand{\dirhmarlfiredceilinghi}{0.304}
\newcommand{\dirhmarlheldlo}{0.651}
\newcommand{\dirhmarlheldhi}{0.793}
\newcommand{\keeprandomexec}{0.615}
\newcommand{\hmarlrandomexec}{0.471}
\newcommand{\contqseedonemean}{\ensuremath{-}52.1}
\newcommand{\contqseedtwomean}{69.8}
\newcommand{\contqseedthreemean}{\ensuremath{-}124.2}
\newcommand{\contqbetweensd}{98.1}
\newcommand{\contqwithinsd}{31.5}
\newcommand{\contqbetweenshare}{0.904}
\newcommand{\contqpermp}{0.004}
\newcommand{\contwseedonemean}{72.4}
\newcommand{\contwseedtwomean}{109.3}
\newcommand{\contwseedthreemean}{25.5}
\newcommand{\contwbetweensd}{42.0}
\newcommand{\contwwithinsd}{45.3}

\newcommand{\contwpermp}{0.127}
\newcommand{\wreckerexploitstock}{0.50}
\newcommand{\wreckerexploit}{0.80}
\newcommand{\wreckerescstock}{0.50}
\newcommand{\wreckeresc}{0.90}
\newcommand{\hcmarlsleepcount}{2265}
\newcommand{\hcmarlbusysteps}{2257}
\newcommand{\keepckptdigest}{55fb3e7f46540eeea63230b149e1744e}
\newcommand{\hmarlckptdigest}{fd9e9172fd56b723fedca72ce0e78a16}
\newcommand{\keepwreckercontrol}{\ensuremath{-}1115.65}
\newcommand{\stockkdrs}{0.50}
\newcommand{\stockkasd}{0.25}
\newcommand{\stockkssd}{0.25}
\newcommand{\stockkdasd}{0.50}
\newcommand{\stockkdssd}{0.50}
\newcommand{\stocksdrs}{0.25}
\newcommand{\stocksdd}{0.25}
\newcommand{\stocksers}{0.50}
\newcommand{\stocksddd}{0.25}
\newcommand{\stocksders}{0.75}
\newcommand{\stockudrs}{0.50}
\newcommand{\stockupe}{0.50}
\newcommand{\stockudpe}{1.00}
\newcommand{\stockrdrs}{0.50}
\newcommand{\stockrimp}{0.25}
\newcommand{\stockrdeg}{0.25}
\newcommand{\stockrdimp}{0.50}
\newcommand{\stockrddeg}{0.50}
\newcommand{\wreckkdrs}{0.30}
\newcommand{\wreckkasd}{0.50}
\newcommand{\wreckkssd}{0.20}

\newcommand{\wrecksdrs}{0.10}
\newcommand{\wrecksdd}{0.10}
\newcommand{\wrecksers}{0.80}
\newcommand{\wrecksddd}{0.20}
\newcommand{\wrecksders}{0.80}
\newcommand{\wreckudrs}{0.10}
\newcommand{\wreckupe}{0.90}

\newcommand{\wreckrdrs}{0.05}
\newcommand{\wreckrimp}{0.55}
\newcommand{\wreckrdeg}{0.40}
\newcommand{\wreckrdimp}{0.55}
\newcommand{\wreckrddeg}{0.45}
\newcommand{\sooneheldties}{2}
\newcommand{\hmsooneheldties}{1}
\newcommand{\sooneunboundedheldties}{2}
\newcommand{\curvectlconvonly}{\ensuremath{-}316.62}
\newcommand{\curvectlqfour}{\ensuremath{-}325.11}
\newcommand{\curvectlqsixteen}{\ensuremath{-}539.17}
\newcommand{\curvectlqthirtytwo}{\ensuremath{-}773.08}
\newcommand{\curvectlqfortyeight}{\ensuremath{-}928.16}

\newcommand{\lagcorrmin}{\ensuremath{-}0.150}
\newcommand{\lagcorrmax}{0.117}
\newcommand{\decoyfiredgapwins}{26}
\newcommand{\sheldvsfiredcharge}{11.9}
\newcommand{\sheldvsfiredrest}{13.4}
\newcommand{\sheldvsfiredrestlo}{4.0}
\newcommand{\sheldvsfiredresthi}{23.0}
\newcommand{\soonefiredcharge}{36.9}
\newcommand{\soonefiredrest}{24.0}
\newcommand{\soonefiredrestlo}{10.7}
\newcommand{\soonefiredresthi}{37.4}
\newcommand{\sooneheldcharge}{25.0}
\newcommand{\sooneheldrest}{37.5}
\newcommand{\sooneheldrestlo}{25.5}
\newcommand{\sooneheldresthi}{50.1}
\newcommand{\hmsheldvsfiredcharge}{20.6}
\newcommand{\hmsheldvsfiredrest}{7.0}
\newcommand{\hmsheldvsfiredrestlo}{1.1}
\newcommand{\hmsheldvsfiredresthi}{12.9}
\newcommand{\hmsoonefiredcharge}{46.0}
\newcommand{\hmsoonefiredrest}{21.7}
\newcommand{\hmsoonefiredrestlo}{12.5}
\newcommand{\hmsoonefiredresthi}{31.6}
\newcommand{\hmsooneheldcharge}{25.4}
\newcommand{\hmsooneheldrest}{28.7}
\newcommand{\hmsooneheldrestlo}{19.3}
\newcommand{\hmsooneheldresthi}{39.2}
\newcommand{\curveconvonlycharge}{13.6}
\newcommand{\curveconvonlyrest}{38.1}
\newcommand{\curveconvonlyrestlo}{23.2}
\newcommand{\curveconvonlyresthi}{53.9}
\newcommand{\curveqfourcharge}{14.7}
\newcommand{\curveqfourrest}{35.1}
\newcommand{\curveqfourrestlo}{21.9}
\newcommand{\curveqfourresthi}{48.7}
\newcommand{\curveqsixteencharge}{16.9}
\newcommand{\curveqsixteenrest}{78.5}
\newcommand{\curveqsixteenrestlo}{59.8}
\newcommand{\curveqsixteenresthi}{98.9}
\newcommand{\curveqthirtytwocharge}{12.0}
\newcommand{\curveqthirtytworest}{119.6}
\newcommand{\curveqthirtytworestlo}{79.2}
\newcommand{\curveqthirtytworesthi}{157.5}
\newcommand{\curveqfortyeightcharge}{13.6}
\newcommand{\curveqfortyeightrest}{147.7}
\newcommand{\curveqfortyeightrestlo}{117.0}
\newcommand{\curveqfortyeightresthi}{178.7}
\newcommand{\keepheldvsfiredcharge}{11.8}
\newcommand{\keepheldvsfiredrest}{160.9}
\newcommand{\keepheldvsfiredrestlo}{122.3}
\newcommand{\keepheldvsfiredresthi}{199.9}
\newcommand{\hmarlheldvsfiredcharge}{20.9}
\newcommand{\hmarlheldvsfiredrest}{295.3}
\newcommand{\hmarlheldvsfiredrestlo}{220.5}
\newcommand{\hmarlheldvsfiredresthi}{372.9}
\newcommand{\oonecharge}{364.4}
\newcommand{\oonerest}{31.4}
\newcommand{\oonerestlo}{16.4}
\newcommand{\ooneresthi}{46.2}
\newcommand{\oonefiredheldcharge}{327.4}

\newcommand{\sooneunboundedheldcharge}{36.9}

\newcommand{\oonesameexec}{100}
\newcommand{\keepchargeshare}{0.414}

\newcommand{\hmarlchargeshare}{0.273}
\newcommand{\keepchargesharewrecker}{0.066}
\newcommand{\hmarlchargesharewrecker}{0.049}
\newcommand{\refgapstock}{50.89}
\newcommand{\refgapcharge}{40.3}
\newcommand{\curvechargemin}{11.8}
\newcommand{\curvechargemax}{16.9}
\newcommand{\curverestsharestock}{0.072}
\newcommand{\curverestsharemin}{0.108}
\newcommand{\curverestsharemax}{0.159}
\newcommand{\hmarlstockdrawsd}{0.97}
\newcommand{\hmarlstockdrawsdexp}{4.71}
\newcommand{\hmarlstockdrawp}{0.012}
\newcommand{\billingline}{310}
\newcommand{\keeprestoreswrecker}{74.0}
\newcommand{\hmarlrestoreswrecker}{40.9}
\newcommand{\hcmarlrestores}{232.5}
\newcommand{\hcmarlrestoresmin}{227.4}
\newcommand{\hcmarlrestoresmax}{231.1}

\newcommand{\burstvsspreadrest}{173.0}
\newcommand{\burstvsspreadrestlo}{148.1}
\newcommand{\burstvsspreadresthi}{199.7}
\newcommand{\burstfirednever}{62.4}
\newcommand{\spreadfirednever}{22.6}
\newcommand{\rfourtnownrestores}{149.3}
\newcommand{\rtwoownrestores}{75.5}
\newcommand{\decoyownchargeshare}{0.303}
\newcommand{\decoyheldcharge}{1.6}
\newcommand{\decoyheldrest}{39.2}
\newcommand{\decoyheldrestlo}{11.6}
\newcommand{\decoyheldresthi}{67.3}
\newcommand{\rfourtnheldownrestores}{71.4}
\newcommand{\rtwoheldownrestores}{69.2}
\newcommand{\rtrestoreswrecker}{73.8}
\newcommand{\rthmarlchargegap}{32.9}
\newcommand{\hmarlsamplingsharestocksix}{0.831}
\newcommand{\hmarlsamplingsharestocksixlo}{0.743}
\newcommand{\hmarlsamplingsharestocksixhi}{0.917}
\newcommand{\hmarlstocksixdrawsd}{2.76}
\newcommand{\hmarlstocksixdrawsdexp}{4.63}
\newcommand{\hmarlstocksixdrawp}{0.121}
\newcommand{\relbusyscore}{\ensuremath{-}286.90}

\newcommand{\relbusyrest}{\ensuremath{-}3.87}
\newcommand{\relbusyrestlo}{\ensuremath{-}19.49}
\newcommand{\relbusyresthi}{10.55}
\newcommand{\relbusyrestoresbusy}{102.58}
\newcommand{\relbusyrestorespassedoff}{180.91}
\newcommand{\relbusyrestorespassedon}{77.46}

\newcommand{\relbusywinson}{92}
\newcommand{\relbusycost}{99.58}
\newcommand{\relbusycostlo}{82.74}
\newcommand{\relbusycosthi}{115.41}
\newcommand{\relbusychargecost}{103.45}
\newcommand{\relbusychargecostlo}{98.75}
\newcommand{\relbusychargecosthi}{108.22}
\newcommand{\relbusyshare}{0.532}
\newcommand{\relbusycostmin}{81.79}
\newcommand{\relbusycostmax}{155.05}
\newcommand{\relbusychargesharemin}{0.604}
\newcommand{\relbusychargesharemax}{1.069}

\newcommand{\relbusywreckerchargecost}{93.61}
\newcommand{\relbusywreckerrest}{61.44}
\newcommand{\relbusywreckerrestlo}{\ensuremath{-}16.15}
\newcommand{\relbusywreckerresthi}{138.78}
\newcommand{\keeppooledprecision}{33.33}
\newcommand{\keepactorparamsreleased}{476,167}
\newcommand{\keepactorparamstrainer}{459,783}
\newcommand{\seedruledate}{October 19, 2026}
\newcommand{\pilotseedlo}{4101}
\newcommand{\pilotseedhi}{4104}

\newcommand{\pilotdeltatwoh}{145.4}

\newcommand{\pilotdeltathreeh}{164.8}

\newcommand{\pilotffourh}{\ensuremath{-}457.8}
\newcommand{\pilotnownfourh}{\ensuremath{-}858.9}
\newcommand{\pilotnonfourh}{\ensuremath{-}707.6}
\newcommand{\pilotbusyfourh}{152.3}
\newcommand{\pilotdeltafourh}{249.8}

\newcommand{\pilotpairs}{12}

\newcommand{\pilotpairsexcl}{12}
\newcommand{\pilotleadmin}{62.6}
\newcommand{\pilotleadmax}{460.0}
\newcommand{\pilotnleadonehmin}{454.4}
\newcommand{\pilotnleadonehmax}{1654.0}
\newcommand{\pilotgapfourh}{266.5}
\newcommand{\chanceredshareowned}{0.189}
\newcommand{\oonerestoreprecision}{65.24}

\newcommand{\keeppaperepisodelen}{100}
\newcommand{\keeppaperepisodesperupdate}{100}
\newcommand{\keeppapertrainepisodes}{1,000,000}
\newcommand{\keeptraindefault}{500,000}
\newcommand{\keeptrainplateau}{50,000}

\newcommand{\hmrelbusyscore}{\ensuremath{-}282.41}
\newcommand{\hmrelbusycost}{145.98}
\newcommand{\hmrelbusycostlo}{142.03}
\newcommand{\hmrelbusycosthi}{149.97}

\newcommand{\hmrelbusyrest}{0.00}

\newcommand{\hmrelbusyrestorespassedon}{37.20}
\newcommand{\hmrelbusyrestoresbusy}{145.98}
\newcommand{\hmrelbusybusysteps}{1307.73}
\newcommand{\hmrelbusysentbusy}{1289.00}
\newcommand{\hmrelbusysamesteps}{100}

\newcommand{\pilotdeltasdtwoh}{67.1}
\newcommand{\pilotdeltatlotwoh}{38.6}
\newcommand{\pilotdeltathitwoh}{252.2}
\newcommand{\pilotdeltasdthreeh}{99.7}
\newcommand{\pilotdeltatlothreeh}{6.1}
\newcommand{\pilotdeltathithreeh}{323.5}
\newcommand{\pilotdeltasdfourh}{157.3}
\newcommand{\pilotdeltatlofourh}{\ensuremath{-}0.4}
\newcommand{\pilotdeltathifourh}{500.0}
\newcommand{\pilotsignp}{0.125}
\newcommand{\pilotdecisionsfourh}{351.5}
\newcommand{\longfaeighth}{\ensuremath{-}336.11}
\newcommand{\longfatwelveh}{\ensuremath{-}371.37}
\newcommand{\longfasixteenh}{\ensuremath{-}360.36}
\newcommand{\longfatwentyh}{\ensuremath{-}324.35}
\newcommand{\longfadraws}{\ensuremath{-}316.72}
\newcommand{\longfadrawslo}{\ensuremath{-}339.08}
\newcommand{\longfadrawshi}{\ensuremath{-}294.35}

\newcommand{\longfagap}{125.4}
\newcommand{\longseedsphrase}{three seeds}
\newcommand{\longseedlist}{4101, 4102 and 4103}

\newcommand{\longgapmax}{125.4}
\newcommand{\relbusytaopcost}{96.67}
\newcommand{\relbusytaoprest}{15.23}
\newcommand{\relbusytaoprestlo}{4.40}
\newcommand{\relbusytaopresthi}{26.59}
\newcommand{\longfafourh}{\ensuremath{-}482.58}

\newcommand{\longfbeighth}{\ensuremath{-}341.08}
\newcommand{\longfbtwelveh}{\ensuremath{-}286.33}
\newcommand{\longfbsixteenh}{\ensuremath{-}223.98}

\newcommand{\longfbdraws}{\ensuremath{-}218.71}
\newcommand{\longfbdrawslo}{\ensuremath{-}234.04}
\newcommand{\longfbdrawshi}{\ensuremath{-}203.38}

\newcommand{\longfbgap}{27.4}
\newcommand{\seedb}{4102}

\newcommand{\longfceighth}{\ensuremath{-}399.40}
\newcommand{\longfctwelveh}{\ensuremath{-}360.14}
\newcommand{\longfcsixteenh}{\ensuremath{-}221.64}

\newcommand{\longfcdraws}{\ensuremath{-}189.97}
\newcommand{\longfcdrawslo}{\ensuremath{-}192.39}
\newcommand{\longfcdrawshi}{\ensuremath{-}187.56}

\newcommand{\longfcabove}{1.4}
\newcommand{\seedc}{4103}
\newcommand{\longdrawsworst}{\ensuremath{-}316.72}
\newcommand{\longdrawsbest}{\ensuremath{-}189.97}
\newcommand{\longdrawsspread}{126.7}
\newcommand{\keepreprowidth}{22.5}

\newcommand{\longwithincount}{one}
\newcommand{\longbelowcount}{two}

\newcommand{\longsfdiffmin}{\ensuremath{-}52.1}
\newcommand{\longsfdiffmax}{48.1}
\newcommand{\longsfmatched}{five}
\newcommand{\relbusybothoffgap}{4.49}
\newcommand{\relbusycheckworthgap}{46.40}
\newcommand{\longsafourh}{\ensuremath{-}458.45}
\newcommand{\longsaeighth}{\ensuremath{-}305.72}
\newcommand{\longsatwelveh}{\ensuremath{-}352.95}
\newcommand{\longsasixteenh}{\ensuremath{-}312.22}
\newcommand{\longraoneh}{\ensuremath{-}349.73}
\newcommand{\longratwoh}{\ensuremath{-}223.06}
\newcommand{\longrathreeh}{\ensuremath{-}215.40}
\newcommand{\longrafourh}{\ensuremath{-}247.96}
\newcommand{\longrboneh}{\ensuremath{-}493.37}
\newcommand{\longrbtwoh}{\ensuremath{-}248.33}
\newcommand{\longrafourhdraws}{\ensuremath{-}254.45}
\newcommand{\longrafourhdrawslo}{\ensuremath{-}266.02}
\newcommand{\longrafourhdrawshi}{\ensuremath{-}242.88}

\newcommand{\longrafourhgap}{63.1}
\newcommand{\longsatwentyh}{\ensuremath{-}376.48}
\newcommand{\longsatwentyhdraws}{\ensuremath{-}387.12}
\newcommand{\longsatwentyhdrawslo}{\ensuremath{-}413.20}
\newcommand{\longsatwentyhdrawshi}{\ensuremath{-}361.04}

\newcommand{\longrbthreeh}{\ensuremath{-}206.70}
\newcommand{\longrbfourh}{\ensuremath{-}229.81}
\newcommand{\longrbfourhdraws}{\ensuremath{-}224.07}
\newcommand{\longrbfourhdrawslo}{\ensuremath{-}236.61}
\newcommand{\longrbfourhdrawshi}{\ensuremath{-}211.53}

\newcommand{\longrbfourhgap}{32.7}
\newcommand{\longsfdrawsgap}{70.4}
\newcommand{\longsfearlier}{four}
\newcommand{\longsfearlierother}{four}
\newcommand{\longfalaststep}{+43.6}
\newcommand{\longfaintgap}{91.7}
\newcommand{\longfblaststep}{+5.3}
\newcommand{\longfbintgap}{0.8}
\newcommand{\longfclaststep}{+31.7}
\newcommand{\longgapbelowmin}{27.4}
\newcommand{\longgapbelowmax}{125.4}
\newcommand{\longrafpair}{+76.4}
\newcommand{\longrafpairlo}{46.4}
\newcommand{\longrafpairhi}{106.6}
\newcommand{\longrafpairwins}{73}
\newcommand{\longrbfpair}{\ensuremath{-}8.0}
\newcommand{\longrbfpairlo}{\ensuremath{-}34.1}
\newcommand{\longrbfpairhi}{18.6}
\newcommand{\longrbfpairwins}{44}
\newcommand{\longsafpair}{\ensuremath{-}52.1}
\newcommand{\longsafpairlo}{\ensuremath{-}101.9}
\newcommand{\longsafpairhi}{\ensuremath{-}6.2}
\newcommand{\longsafpairwins}{43}

\newcommand{\arm}[1]{\texttt{#1}}

\title{What a CAGE 4 Score Measures: An Audit of a Learned Cyber-Defense Benchmark, Its Reference Defenders, and the Tables Built on Them}
\author{James Rice\thanks{Rochester Institute of Technology. jmrics@rit.edu.} \and Ji Hwan Park\thanks{Rochester Institute of Technology. jhpigm@rit.edu} \and Andres Kwasinski\thanks{Rochester Institute of Technology. axkeec@rit.edu}}
\date{}

\newcommand{\secref}[1]{Section~\ref{#1}}
\hypersetup{hidelinks}
\usepackage[T1]{fontenc}
\usepackage{lmodern}
\usepackage[htt]{hyphenat}
\usepackage{xfp}
\usepackage[section]{placeins}
\usepackage{enumitem}
\setlist{itemsep=0pt,parsep=0pt}
\usepackage{pgf}
\DeclareRobustCommand{\pct}[1]{\pgfmathparse{#1*100}\pgfmathprintnumber[fixed,precision=1,fixed zerofill]{\pgfmathresult}\%}
\newcommand{\pcu}[1]{#1\%}
\DeclareRobustCommand{\pco}[1]{\pgfmathparse{#1}\pgfmathprintnumber[fixed,precision=1,fixed zerofill]{\pgfmathresult}\%}
\usepackage{longtable}
\usepackage{array}
\newcolumntype{L}[1]{>{\raggedright\arraybackslash}p{#1}}
\makeatletter
\def\hash@split#1#2\hash@end{#1\ifx\relax#2\relax\else\allowbreak\hash@split#2\hash@end\fi}
\newcommand{\hashtt}[1]{\texttt{\expandafter\hash@split#1\hash@end}}
\makeatother
\begin{document}
\maketitle

\begin{abstract}
CAGE Challenge 4 is the benchmark on which learned multi-agent cyber defenders are compared, and its scores and the counts reported beside them are read as measures of what a defender does. We reproduce its two reference defenders in a pinned container, KEEP at \keeprepro\ against a published \keeppub\ and H-MARL Expert at \hmarlrepro\ against \hmarlauthorscore, and show that three things read from this benchmark measure less than they are taken to measure. First, the simulator discards, by design and without an error, any action asserted for an agent while its previous action is still running, charges the team one point for a discarded Restore as for one that runs, and leaves the busy flag out of the observation it documents. Both reference defenders read the flag in their own wrappers, and switching that check off at evaluation costs KEEP's released policy \relbusycost\ points per episode at the published setting and H-MARL's \hmrelbusycost, in both cases the charge for Restores sent to busy agents, with H-MARL's executed actions unchanged step for step. On a supervisory harness, holding interventions until the agent is idle is worth \sheldvsfired\ and \hmsheldvsfired\ points per episode on the two defenders at the published setting and \keepheldvsfired\ and \hmarlheldvsfired\ under an impact-focused adversary, and three readings of supervision drawn from fired interventions shrink or vanish once directives are held. In training the check decides how a busy step enters the trajectory, a choice the released recipe makes without stating it. Trained without it, KEEP leads at a hundred updates in all four paired seeds and trails from two hundred on in all four, and continued to two thousand updates on \longseedsphrase\ the repaired recipe ends inside the released interval on one and short of it on \longbelowcount, by \longgapbelowmin\ to \longgapbelowmax\ points, a spread across training seeds far wider than the released sampling interval. Second, the evaluation seeds the environment but not the policy, so a reported mean is one draw, and \pct{\samplingvarshare} of per-episode outcome variance at the stock adversary is the closed-loop response to KEEP's own sampling draw. The Restore charge is \pct{\keepchargeshare} of KEEP's loss at the published setting, and the README's reward tables do not list it. Third, recovery precision scored against the true state or against the evidence a defender's observations keep gives different numbers for the same Restores, and a rule arm correct \pco{\oonerestoreprecision} of the time on KEEP's own criterion against KEEP's \pco{\restoreprecision} recovers more than KEEP and trails it by \oonedeficit\ points when its directives are fired, almost all of it the charge. Across five published comparison tables, none states the artifact, sampling, criterion and adversary that would make its numbers comparable. We release the instrument and our patches to both defenders' code, and we give a five-item disclosure checklist that would make the next table comparable.
\end{abstract}

\section{Introduction}\label{sec:intro}
Network defense is a sequential decision problem run at machine speed. An intrusion moves across hosts in minutes, a security operations center sees it through alerts that are mostly noise, and the response actions available, isolating a host, removing a process, restoring from image, each cost something whether or not the host was compromised. Response is already automated in production by playbooks that fire on a signature and isolate or reimage without a person in the loop \citep{stevens2022playbooks, bridges2023soar}. Autonomous cyber defense proposes to learn the policy instead, and over the last five years reinforcement learning has become its main method \citep{nguyen2021drlcyber, vyas2023acdreview}. We know of no defender whose response policy was learned by reinforcement learning in production, and the surveys of the field describe it as at the research stage \citep{nguyen2021drlcyber, vyas2023acdreview}. The evidence that such a defender could work comes almost entirely from simulators and small emulated testbeds, because an agent cannot be trained against live intrusions on a production network, and a small number of simulators have become the shared ground on which the proposals are compared. What those simulators measure is therefore the case for learned defense, and this paper asks what one of them measures.

The CAGE challenge series \citep{standen2021cyborg, kiely2025cage4}, built on the CybORG simulator \citep{baillie2020cyborg}, is the most used of these, and its fourth challenge is the first with a multi-agent blue team, an enterprise topology and a scripted adversary with a kill chain. Its scores are how the field currently decides which defender is better. That makes the benchmark's validity a question for every result built on it, and the record in neighboring fields is that such questions are rarely asked early. Reinforcement learning results on standard benchmarks were shown to turn on seeds, implementation details and reporting choices before the field adopted the protocols that control for them \citep{henderson2018drlmatters, agarwal2021precipice}, and security machine learning has its own catalog of evaluation pitfalls that recur across a decade of papers \citep{arp2022dosdonts}. A defense benchmark sits at the intersection of both.

CAGE Challenge 4 gives a team of blue agents an enterprise network, a scripted red agent, and a reward that charges for the work its simulated users lose, the services they cannot reach and the impact actions red completes. The simulator also charges the team one point for every Restore a caller supplies for a blue agent, which the README's reward tables do not list, and the challenge ranks submissions by mean episode reward \citep{kiely2025cage4}. Two of its defenders released code and checkpoints, KEEP \citep{king2025keep} and H-MARL Expert \citep{singh2025hmarl}, and they have become the reference defenders of the field. One or both are the baselines in a red-teaming framework \citep{masukawa2026tridentbreakdeep}, a tree-search agent \citep{li2026acdzero}, a hierarchical architecture \citep{qu2026hcmarl}, whose H-MARL baseline is the Meta variant, and an LLM red-team evaluation \citep{shaikh2026noone}. The tree-search agent and the hierarchical architecture report a CAGE 4 score beside a published one and read the difference as a property of their method, and the red-teaming framework retrains a baseline and reports its own measurement of it.

The question this paper asks is what a CAGE 4 score measures, and which conclusions drawn from one survive a corrected evaluation. Answering it needs an instrument the benchmark does not ship, a record at every step of what was asked of each agent, what the simulator did, and what was true on the network, and that instrument is the first thing we built. We built it to measure what a human supervisor adds to a learned defender, and the question above had to be answered first.

We reproduce both reference defenders in a pinned container, with both published numbers inside our reproduction intervals, and run them on \episodes\ paired episodes under the stock adversary and an impact-focused one we specify, with KEEP also under five points between the two, and with several sampling draws of each defender, recording the executed actions at every step and, on KEEP's policy arms, the true state. We read the simulator's controller, the challenge's wrappers and the evaluation script against the README's contract, and we read the final leaderboard and every table from 2026 that compares a new defender or adversary against one of the two reference defenders.

Three things read from this benchmark measure less than they are taken to measure. The simulator discards, by design and without an error, any action asserted for an agent while its previous action is still running, and it charges the team one point for a discarded Restore as for one that runs. The observation the challenge documents does not expose the flag, so an agent built on it can comply only by tracking its own action durations. The two released defenders read the flag in their own wrappers and are not subject to it. Switching KEEP's check off at evaluation costs its released policy \relbusycost\ points per episode at the published setting and \relbusycostmin\ to \relbusycostmax\ at nine other adversaries, almost all of it the charge for Restores sent to busy agents, and training KEEP without the check puts it ahead at a hundred updates and behind from two hundred on, in all four of our paired training seeds. Continued to two thousand updates, the repaired recipe reaches the released score on one of \longseedsphrase\ and falls \longgapmax\ points short on another, so one released checkpoint does not characterize the variability of retraining under its procedure. A harness that injects directives is subject to the drop, and so is any agent built on the documented observation, and a count of such an agent's actions taken from its own outputs counts directives asserted. On our own supervisory harness, holding interventions until the agent is idle is worth \sheldvsfired\ points per episode on KEEP and \hmsheldvsfired\ on H-MARL at the published setting, of which \sheldvsfiredcharge\ and \hmsheldvsfiredcharge\ are the charge, and \keepheldvsfired\ and \hmarlheldvsfired\ under an impact-focused adversary we specify, where the charge is a small part. Three readings of supervision we had drawn from fired interventions shrink or vanish once directives are held until the agent is idle. The evaluation seeds the environment but not the policy, so \pct{\samplingvarshare} of per-episode outcome variance in a KEEP score at the stock adversary (interval \pct{\samplingvarsharelo} to \pct{\samplingvarsharehi}) is the closed-loop response to the policy's own sampling draw, a one-seed mean over \episodes\ episodes has a standard error of about \oneseedse\ points, and two identical policies differ by up to \nullbandstock\ points on the same seeds. The Restore charge is \pct{\keepchargeshare} of KEEP's loss at the published setting. And a recovery precision is not comparable without its criterion and its volume, since a rule arm that scores \pco{\oonerestoreprecision} on KEEP's own precision criterion against KEEP's \pco{\restoreprecision} recovers more than KEEP, yet is unresolved against it when its directives are held and trails it by hundreds of points when they are fired, because every Restore it sends is charged. None of the five published comparison tables states the four things that would make its numbers comparable.

\paragraph{Contributions.}
\begin{enumerate}
  \item \textbf{The silent drop and its charge.} The simulator drops a directive asserted at a busy agent by design and without an error, the challenge's documented observation hides the busy state, and the simulator still charges one point for a dropped Restore. We give the mechanism and its cost, measured three ways. On a supervisory harness, held delivery is worth \sheldvsfired\ and \hmsheldvsfired\ points per episode on the two reference defenders at the published setting, of which \sheldvsfiredcharge\ and \hmsheldvsfiredcharge\ are the charge. On KEEP's own released policy, switching its busy check off costs \relbusycost\ points at the published setting and \relbusycostmin\ to \relbusycostmax\ at nine other adversaries, almost all of it the charge, and on H-MARL's, switching its mask off costs \hmrelbusycost, the charge to the point, with the executed actions unchanged step for step. And three readings of supervision drawn from fired interventions shrink or vanish once directives are held (\secref{sec:drop}, \secref{sec:pilot}).
  \item \textbf{What a score is made of.} The evaluation seeds the environment and not the policy, so a reported mean is one draw. A variance decomposition puts \pct{\samplingvarshare} of KEEP's per-episode outcome variance on the policy's own sampling at the stock adversary, a null band describes the noise two identical policies show, and as an accounting split the Restore charge is \pct{\keepchargeshare} of KEEP's loss and \refgapcharge\ of the \refgapstock\ points between the two reference defenders (\secref{sec:tiers}, \secref{sec:noise}).
  \item \textbf{A trainer audited and its unstated choices measured.} KEEP's released trainer never seeds its workers, replays the same scenario draws every update, and learns from five of the twenty-five episodes each update generates. With the first two repaired and the third held as shipped, generating only the five episodes the learner reads, we train KEEP from scratch with and without its busy check on paired seeds, score every hundredth checkpoint both ways, and continue the check-on arm to two thousand updates, where the trainer's own comment places the plateau, under a seed rule fixed before the four-hundred-update scores of three of the four seeds had been read, against the released policy's reproduction interval. Trained without the check, KEEP is ahead at a hundred updates in all four seeds and behind from two hundred on in all four, by \pilotleadmin\ to \pilotleadmax\ points when both are scored with the check on. Continued to two thousand updates on \longseedsphrase, the check-on arm's four-draw means run from \longdrawsworst\ to \longdrawsbest\ against the released reproduction mean of \keeprepro, \longwithincount\ of them inside the released interval and \longbelowcount\ below it, so the released score is reachable at the budget the trainer's comment names and the spread across training seeds, \longdrawsspread\ points, exceeds the width of the released sampling interval.
 Two controls measure two more choices the shipped trainer makes, the single layout it trains on, which on the one seed it was run on leads fresh layouts at every earlier checkpoint and ends \longsfdrawsgap\ points below them, and the four fifths of each update's episodes its learner discards, which read in full, in a fifth as many updates on batches five times the size, close about half the shortfall against the released score at the endpoint on one seed and nothing the draws resolve on a second. The check is a choice about how a busy step enters training that the released recipe makes without stating it, its effect on learning is large early in training, and the audit's reading of the simulator is what makes it a choice that can be measured (\secref{sec:pilot}).
  \item \textbf{Precision without its criterion.} Two recovery-precision bases score the same Restores differently, and a rule arm scoring \pco{\oonerestoreprecision} on the policy's own precision criterion against its \pco{\restoreprecision} recovers more than the policy it supervises while unresolved against it held and far below it fired, because each Restore it sends is charged (\secref{sec:nonmeasures}).
  \item \textbf{An audit, a checklist and the instrument.} Five published tables read against four disclosure axes, the five disclosures a comparable number needs, and the released instrument, the pinned reproduction, the step-record capture, the charge accounting, the paired-seed protocol and our patches to both defenders' code, with a one-command reproduction of each defect and the change it calls for (\secref{sec:audit}, \secref{sec:checklist}, \secref{sec:method}, Appendix~\ref{app:issues}).
\end{enumerate}

Five scope statements hold throughout. All results are on CAGE 4, and we claim nothing about simulators that do not share CybORG's controller. The learned defenders are the two reference defenders. Our hand-written rule defenders are controls, not proposals. Every number is \episodes\ paired episodes on environment seeds \seedlo\ to \seedhi\ unless stated. And the supervision results concern a scripted supervisory rule, not a human, which is the work this instrument was built for.

\section{Environment and policy reproduction}\label{sec:method}
Everything in this paper is measured in one pinned environment, against two published defenders, under the stock adversary and an impact-focused one, with the other adversaries named where they are used, on paired seeds, with an instrument that records what the simulator did rather than what the agent asked for. This section fixes each of those. One script recurs throughout. The challenge distributes an evaluation script, \texttt{evaluation.py}, and each defender's repository carries its own copy, so below we distinguish the challenge's script, KEEP's copy and H-MARL's copy.

\paragraph{Container.} Two Docker images pin the two defenders' dependency sets, which are mutually incompatible and so cannot share one Python environment. Each image records two pins the defender's repository does not declare, \texttt{pyarrow} at \pyarrowpin, without which every RLlib import fails, and the CPU wheel index for \texttt{torch}, without which the pinned version installs as a CUDA build. After KEEP's copy of the evaluation script is patched to seed policy sampling (\secref{sec:tiers}, Tier B), a two-episode run at seed zero returns \smokescore\ with sd \crossmachinesd\ on two machines of different microarchitecture, identical to the last printed digit. Unpatched, the same command on one machine returns \pristinemean\ with sd \pristinesd, because the script never seeds \texttt{torch}. Determinism given a seed is therefore established before any number is reported.

\paragraph{Defenders.} KEEP \citep{king2025keep}, a graph-network policy, and H-MARL Expert \citep{singh2025hmarl}, a hierarchical one, are the two CAGE 4 defenders with released code and checkpoints, and so the two the field compares against. KEEP's published score is \keeppub, from the challenge leaderboard \citep{cage4leaderboard}. H-MARL's authors report \hmarlauthorscore\ with sd \hmarlauthorsd\ over \hmarlauthorepisodes\ episodes \citep{singh2025hmarl}, and an independent run of the released checkpoint reports \hmarlpub\ \citep{masukawa2026tridentbreakdeep}. We run the released checkpoints from the public repositories at the revisions in Appendix~\ref{app:instrument} \citep{keepcode, hmarlcode}, in the sampling mode their shipped evaluation scripts use. Before our instrument replaces H-MARL's copy of the evaluation script, we check that the instrument, run in a compatibility mode that follows that script's path, returns the same score as the script itself at seed zero, \hmarlcompatseedzero.

\begin{table}[tbp]
\centering
\caption{Red's transition rows, stock / Wrecker where they differ. A row is red's state for the host it has chosen to act on, known (K), scanned (S), user (U) or root (R), and a D marks a host whose subnet red has already scanned with DiscoverRemoteSystems. Columns are the actions, DiscoverRemoteSystems, AggressiveServiceDiscovery, StealthServiceDiscovery, DiscoverDeception, ExploitRemoteService, PrivilegeEscalate, Impact and DegradeServices, and Withdraw, zero in both, is omitted. Six rows differ. The KD and UD rows do not and print one value.}
\label{tab:wrecker}
\footnotesize\setlength{\tabcolsep}{5pt}\begin{tabular}{lcccccccc}
\toprule
State & DRS & ASD & SSD & DD & ERS & PE & Imp & Deg \\
\midrule
K & \stockkdrs\ / \wreckkdrs & \stockkasd\ / \wreckkasd & \stockkssd\ / \wreckkssd &  &  &  &  &  \\
KD &  & \stockkdasd & \stockkdssd &  &  &  &  &  \\
S & \stocksdrs\ / \wrecksdrs &  &  & \stocksdd\ / \wrecksdd & \stocksers\ / \wrecksers &  &  &  \\
SD &  &  &  & \stocksddd\ / \wrecksddd & \stocksders\ / \wrecksders &  &  &  \\
U & \stockudrs\ / \wreckudrs &  &  &  &  & \stockupe\ / \wreckupe &  &  \\
UD &  &  &  &  &  & \stockudpe &  &  \\
R & \stockrdrs\ / \wreckrdrs &  &  &  &  &  & \stockrimp\ / \wreckrimp & \stockrdeg\ / \wreckrdeg \\
RD &  &  &  &  &  &  & \stockrdimp\ / \wreckrdimp & \stockrddeg\ / \wreckrddeg \\
\bottomrule
\end{tabular}

\end{table}

\paragraph{Adversaries.} The stock adversary is the challenge's \texttt{FiniteStateRedAgent}, which reproduces the published setting exactly. The impact-focused adversary, which the rest of the paper calls Wrecker, is a point we hand-specified in the planner schema of Trident \citep{masukawa2026tridentbreakdeep}, with no search and no language model (\texttt{red\_agents/FSRedAgentWrecker.py}, against the transition map at \texttt{FiniteStateRedAgent.py} line \fsredmapline\ and the rows around it). It keeps the stock agent's states and actions and changes two things. First, six of the eight transition rows (Table~\ref{tab:wrecker}). Four rows move red to root faster, exploit from \wreckerexploitstock\ to \wreckerexploit\ and privilege escalation from \wreckerescstock\ to \wreckeresc. On the root row with discovery still to do, the stock agent spends \stockdiscoverp\ of its mass on further discovery, \stockimpactp\ on Impact and \stockdegradep\ on DegradeServices, and Wrecker spends \wreckerdiscoverp\ on discovery, \wreckerimpactp\ on Impact and \wreckerdegradep\ on DegradeServices. On the root row with discovery done, Wrecker moves the stock agent's even split to \wreckerrdimpactp\ on Impact and \wreckerrddegradep\ on DegradeServices. Second, it adds a host-state priority list, which the stock agent does not use, placing \pcu{\wreckerrootpriority} of the weight that picks red's next host on hosts where it already holds root. Impact and DegradeServices are the two red actions that move the reward directly, so Wrecker changes what a compromise costs, returns red to the hosts it already holds, and reaches root faster and spends more of its time there. Under it KEEP scores \keepwreckercontrol\ against \keepcorpusmean\ at stock and H-MARL \hmarlwreckercontrol\ against \hmarlstockcontrol, about six times the stock loss, and it is the adversary under which supervision has room to matter.

\paragraph{Design and instrument.} Every arm is \episodes\ episodes of the scenario's \episodesteps\ steps on the same \episodes\ environment seeds (the KEEP-side record holds one step fewer than the H-MARL side, a wrapper difference), so any two harness arms are paired episode by episode. A paired difference in score is reported with a 95\% percentile bootstrap interval over episodes (ten thousand resamples) and its wins of \episodes\ wherever a claim rests on it, with an interval on its rest wherever the body of the paper splits off the Restore charge and prints the rest, and with an exact sign test over the untied episodes and its tie count wherever the text prints a $p$ value. An arm run without its own policy seed seeds its policy's sampling from the environment seed, so a supervised arm shares its control's sampling stream until its first intervention changes the trajectory. A redraw sets a policy seed $P$ and seeds episode $i$ with $P+i$. KEEP is evaluated under four sampling seeds on the paired environment seeds at the stock adversary and six under Wrecker, which gives its reproduction interval below and the variance decompositions of \secref{sec:noise}. H-MARL's reproduction interval is over four seeds of the authors' evaluation path, which reseeds the environment and the policy together, so it checks the published number and is not a paired arm. Its paired arms run through our harness. Our instrument records, at every step, the action the simulator executed for every agent and the reward, on most arms the red agents' actions, and in one of two schemas (Appendix~\ref{app:instrument}) either a count of the compromised hosts the agents own and, on arms captured from mid-September 2026 onward, the intervention asserted when there is one, on every arm run through the supervisory harness (every supervised arm, every H-MARL arm run through it, the rule-defender arms and the continuation arms), or the agents' raw observations and the simulator's true state (which hosts hold a red session), on the KEEP policy arms captured by \texttt{scripts/capture.py}. The H-MARL reference arms run through the authors' evaluator and carry only its summary file. On the harness arms a learned policy's own directives are not in the record, which holds what executed, so the drop of \secref{sec:drop} is measured on interventions, whose assertion the harness controls, and on the rule defenders' own proposals, which the record keeps in a field of their own. The evaluation script writes neither the intervention, the reward nor the true state (\secref{sec:tiers}, Tier C). Executed actions are read from the record by the rule of \secref{sec:drop}.

\paragraph{Result.} KEEP reproduces at \keeprepro\ with standard error \keepreprose\ over the four sampling seeds, interval [\keepreprolo, \keepreprohi], which contains the published \keeppub. A first single draw through KEEP's copy of the evaluation script gave \keepfirst. H-MARL Expert reproduces at \hmarlrepro\ with standard error \hmarlreprose\ over four seeds of the authors' path, interval [\hmarlreprolo, \hmarlreprohi], which contains both the authors' \hmarlauthorscore\ and Trident's \hmarlpub. An unseeded \episodes-episode run lands at \hmarlunseeded, inside that interval. Both published numbers are inside their reproduction intervals, and both intervals are about twenty points wide, four times the gap between the leaderboard's first and second entries (\lbgapfirstsecond\ points).

\begin{table}[tbp]
\centering
\caption{Both published numbers lie inside our reproduction intervals. KEEP \keeprepro\ [\keepreprolo, \keepreprohi] against the published \keeppub. H-MARL Expert \hmarlrepro\ [\hmarlreprolo, \hmarlreprohi] against the authors' \hmarlauthorscore\ and Trident's \hmarlpub. KEEP over four sampling seeds on the \episodes\ paired environment seeds. H-MARL over four seeds of the authors' evaluation path, each of which reseeds environment and sampling together, so its interval is a check on the published number and not a paired arm. The arms are \arm{corpus\_v1}, \arm{keep\_null}, \arm{keep\_null\_9001} and \arm{keep\_null\_9002} for KEEP and \arm{hmarl\_s0} to \arm{hmarl\_s3} for H-MARL.}
\label{tab:repro}
\footnotesize\setlength{\tabcolsep}{4pt}\begin{tabular}{L{0.17\linewidth}L{0.19\linewidth}rlL{0.27\linewidth}}
\toprule
Defender & Published & Reproduced & 95\% interval & Protocol \\
\midrule
KEEP (graph PPO) & \keeppub\ (leaderboard) & \keeprepro & [\keepreprolo, \keepreprohi] & \samplingseeds\ sampling seeds $\times$ \episodes\ episodes, sampled actions \\
H-MARL Expert (hierarchical PPO) & \hmarlauthorscore\ $\pm$ \hmarlauthorsd\ (authors), \hmarlpub\ (Trident) & \hmarlrepro & [\hmarlreprolo, \hmarlreprohi] & \samplingseeds\ seeds $\times$ \episodes\ episodes, the authors' evaluator \\
\bottomrule
\end{tabular}

\end{table}

\section{Defects in the stack, tiers of consequence}\label{sec:tiers}
The stack a CAGE 4 result is computed through has three layers. CybORG is the simulator. The CAGE 4 template is the scenario, the wrappers and the evaluation script the challenge distributes. A defender is built on both, and we reproduce two. We read all three with the question of what a published number computed through them means, and we report the defects by consequence rather than by severity of code (Table~\ref{tab:defects}). Tier A drops directives and charges for some of them. Tier B would change numbers. Tier C is why the cost of the first and the defects of the second went unmeasured. Items that are disclosure choices rather than defects, the sampling mode of \secref{sec:noise} and the undeclared pins of \secref{sec:method}, are reported where they arise, and the two defects of our own harness that changed a number we report, the capture point of the true state and the execution-matching window, are reported where they arise (\secref{sec:nonmeasures} and \secref{sec:drop}).

\begin{table}[tbp]
\centering
\caption{Six defects in the CAGE 4 stack, in three tiers by consequence. Tier A drops directives without an error and still charges for a dropped Restore, which costs a harness that injects them and leaves an agent on the documented observation to track its own action durations or be dropped. Tier B would change a published number's value or its reproducibility. Every leaderboard mean went through the script that does not seed the policy, and we know of no published number that went through the parallel path. Tier C is why neither had been measured. Two of the six are in KEEP's repository rather than in what the challenge ships. The drop's mechanism is in the simulator's tracker as intended behavior \citep{cage4issue32}, and two other issues report that Restore is the one blue action with a cost and that an invalid action is charged at the cost of the action requested \citep{cage4issue40, cage4issue50}. The charge for a dropped directive, the drop's cost and the other five defects are unreported (Appendix~\ref{app:issues}).}
\label{tab:defects}
\scriptsize\setlength{\tabcolsep}{4pt}\begin{tabular}{lL{0.33\linewidth}L{0.20\linewidth}L{0.28\linewidth}}
\toprule
Tier & Defect & Lives in & What it changes \\
\midrule
A & A directive asserted while the agent's previous action is in progress is dropped by design and without an error, a dropped Restore is still charged one point, and the observation the challenge documents omits the busy flag (\secref{sec:drop}) & CybORG controller and challenge vector wrappers & Injected interventions are dropped unseen when the agent is busy and charged if they are Restores, costing our supervisory harness \sheldvsfired\ and \hmsheldvsfired\ points per episode at the stock adversary and \keepheldvsfired\ and \hmarlheldvsfired\ under Wrecker, the rule's silence after each drop included. An agent on the documented observation cannot see the busy state, and a count of its outputs counts directives. Both released defenders read the flag \\
\midrule
B & Evaluation seeds the environment but not policy sampling & CAGE 4 template & A reported mean is one draw that no one can reproduce. For KEEP, a stochastic policy evaluated in sampling mode, the within-episode sampling sd is \samplingsd\ at stock and \samplingsdwrecker\ under Wrecker, \pct{\samplingvarshare} (interval \pct{\samplingvarsharelo} to \pct{\samplingvarsharehi}) and \pct{\samplingvarsharewrecker} of per-episode outcome variance, and a one-seed \episodes-episode mean has a standard error of \oneseedse\ and \oneseedsewrecker\ points \\ \addlinespace
B & Parallel evaluation builds one environment per worker from one seed argument and rotates episodes over them & KEEP's copy of the evaluation script & A run with \paralleldistinct\ workers visits at most \paralleldistinct\ distinct episodes \\ \addlinespace
B & All \trainenvs\ training environments share one seed, the training workers' generators are never seeded, every update replays the parent's scenario draws, and the learner reads five of the twenty-five episodes an update generates & KEEP trainer & A training run learns from one scenario draw as shipped and is not reproducible from its seed (\secref{sec:pilot}) \\
\midrule
C & The evaluation record holds per-step actions and observations and per-episode totals, and no per-step reward, phase or true state & CAGE 4 template (KEEP's copy writes nothing from the command line) & No decision can be scored from what the script writes \\ \addlinespace
C & \texttt{get\_reward\_breakdown} raises, and \texttt{AgentInterface} has no reward calculator & CybORG & Per-host reward attribution, which the benchmark advertises, is unavailable \\
\bottomrule
\end{tabular}

\end{table}

\paragraph{Tier A.} The simulator discards a directive asserted while the agent's previous action is in progress, without an error, for every agent in the scenario, and it charges the team one point for a discarded Restore as for one that runs. The README tells agents to wait, says nothing of what happens when they do not, and the maintainers describe the drop as intended \citep{cage4issue32}. Every intervention a harness injects, and every action of an agent built on the challenge's documented observation, is dropped unseen when the agent is busy. The two released defenders read the flag in their own wrappers and are not subject to it, and with its check switched off KEEP's released policy loses \relbusycost\ points per episode at the published setting (\secref{sec:pilot}). We measured the cost on our own supervisory harness, which for its first month asserted directives without reading the flag, and three readings of supervision we had drawn from fired interventions shrink or vanish once directives are held until the agent is idle. \secref{sec:drop} gives the mechanism, the cost and the three readings.

\paragraph{Tier B.} Three defects would change the value of a published number, or its reproducibility, without changing its meaning. The challenge's evaluation script seeds the environment and not the policy's sampling, so a score is one draw from a distribution whose per-episode sd runs from tens to hundreds of points, and for KEEP, a policy evaluated in sampling mode, \pct{\samplingvarshare} to \pct{\samplingvarsharewrecker} of the per-episode variance in an outcome is that draw (\secref{sec:noise}). KEEP's copy of the script evaluates in parallel by building one environment per worker from one seed argument and rotating its episodes over them, so a run with \paralleldistinct\ workers visits at most \paralleldistinct\ distinct episodes. KEEP's \trainenvs\ parallel training environments share one seed, so a training run starts from one scenario draw, its workers sample from generators the trainer never seeds, so two runs with every setting equal diverge from their first updates, every update replays the parent's scenario draws, and the learner reads five of the twenty-five episodes an update generates (\secref{sec:noise}, \secref{sec:pilot}).\footnote{Three smaller items in the two defenders' code change no number in this paper. KEEP's \texttt{load()} clears the training flag, so a loaded checkpoint cannot be resumed as the trainer expects, which the shipped pipeline never does. The H-MARL wrapper's tick handling is off by one, which stops its trainer at random iteration boundaries. And the H-MARL wrapper's time-to-recover metric, which feeds nothing the paper reports, raises a \texttt{KeyError} when a red session on a contractor-network host ends. The wrapper as shipped leaves the metric switched off, the patch in our artifact switches it on, and our harness guards the call.}

\paragraph{Tier C.} Two defects prevent anyone from checking the others. The record the challenge's evaluation script writes holds per-step executed actions and observations and per-episode totals, and no per-step reward, phase or true state, so no decision can be scored from it. KEEP's copy writes nothing at all from the command line, since nothing in it sets the flag that enables the record. \texttt{get\_reward\_breakdown} raises, and \texttt{AgentInterface} carries no reward calculator, so the per-host attribution the benchmark advertises is unavailable. Tier C is our explanation for why the drop's cost, its charge and the Tier B defects went unreported for the lifetime of the benchmark. A defect that reverses a conclusion is visible only to someone who counts both the directives asserted and the actions executed, and the challenge's record holds only the second, with each dropped directive shown as a Sleep. The charge is hidden the same way, since the record holds no per-step reward and the challenge's wrappers fold the action-cost term into the one reward they return.

Every item in the table is reproduced in the pinned container of \secref{sec:method}, each with a one-command check in the artifact.

\section{The silent drop}\label{sec:drop}
\subsection{Action mechanism}

CAGE 4 gives every action a duration. Restore takes five steps, Remove three, Analyse and DeployDecoy two, and on the red side ExploitRemoteService (the README's Exploit Network Services) takes four and StealthServiceDiscovery three. The benchmark's README states the contract that follows from this. Agents ``must wait until their action is completed before they are prompted to launch another action.'' The simulator does not enforce that contract. It drops the offending directive, and its maintainers describe that as intended \citep{cage4issue32}. On every step, \texttt{SimulationController.step} takes an action for every agent interface, from the caller when one was supplied and from the agent object otherwise. We call the action asserted for an agent at a step a \emph{directive}. The controller stores the directive, and so \emph{executes} it, only if that agent has no action in progress.
\begin{quote}\small\texttt{if self.actions\_in\_progress.get(agent\_name, None) is None:\\ \hspace*{2em}self.actions\_in\_progress[agent\_name] = set\_item}\end{quote}
There is no else branch. A directive asserted while the previous action is still running is neither queued nor rejected. It is \emph{dropped}. The validity check runs before this point, so no \texttt{InvalidAction} results, as the same thread shows for an invalid action sent while busy. The busy agent receives an observation whose success flag reads \texttt{IN\_PROGRESS}, and a \texttt{Sleep} is recorded as its action for the step, so the flag describes the earlier action and the record shows nothing of the dropped directive.

The same step charges the team for what it was handed. After the reward calculators run, \texttt{step} adds an action-cost term summed over the directives the caller supplied (line \billingline), and \texttt{Restore}, the only action in a blue agent's action space with a cost, costs one point. Its docstring calls this a flat penalty for the host's downtime, and the README's reward tables do not list it. A Restore dropped because the agent is busy is charged exactly as one that runs. On the pinned code, four Restores asserted while the agent was still running an earlier one cost one point each, through the controller and through the documented \texttt{BlueFlatWrapper} alike, with the executed actions unchanged (\texttt{tools\_billing.py}). A defender that waits for its agents is charged once for each Restore it runs, and two unsupervised KEEP episodes were charged exactly their Restore counts. The issue tracker holds the pieces of this without the whole. One issue reports Restore as the only blue action with a cost \citep{cage4issue40}, and another reports that an invalid action is charged at the cost of the action requested \citep{cage4issue50}. The H-MARL paper also states that blue agents are penalized ``when they use a costly action like Restore machine'' \citep{singh2025hmarl}.

The stock \texttt{FiniteStateRedAgent} is written around the drop. It checks the flag and returns \texttt{Sleep} while busy, as do the variants of it we run. A submission built on the observation the challenge documents cannot do the same, because the flag is not part of it. The README's observation table has no success index, and none of the seven wrappers under \texttt{Agents/Wrappers} reads the flag or masks a busy agent's action space. The three vector wrappers, \texttt{BlueFlatWrapper}, \texttt{BlueEnterpriseWrapper} and \texttt{EnterpriseMAE}, build their vectors without it, so it does not survive them. \texttt{BlueFixedActionWrapper} forwards the simulator's raw observation dictionary, flag included, and masks by host only. A submission that reads the raw dictionary can see the flag, and both reference defenders do, H-MARL through a subclass of \texttt{BlueFixedActionWrapper} and KEEP by reading the controller's last observation from behind \texttt{EnterpriseMAE}. H-MARL's wrapper masks every action but Sleep while busy \citep{singh2025hmarl}, and KEEP's policy returns no action while busy, which its translator turns into \texttt{Monitor}, the default action \citep{keepcode}. The drop therefore falls on what their policies do not control. An intervention a harness injects bypasses the policy's check and is dropped whenever the agent is busy, and an agent built on the challenge's vector wrappers has no check to make. For every such agent, and for every intervention delivered to any agent while it is busy, a count taken from what was asserted counts directives rather than actions executed, and the challenge's record, which holds executed actions, shows each dropped directive as a Sleep and nothing of what was asked.

\subsection{What it costs}

\begin{figure}[t]
\centering
\includegraphics[width=\linewidth]{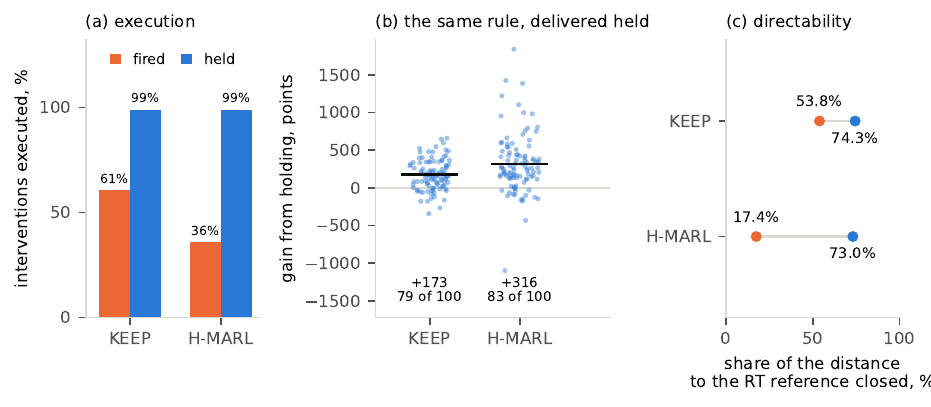}
\caption{(a) Fired at whatever agent owns the host, \pct{\keepexecshare} of interventions execute on KEEP and \pct{\hmarlexecshare} on H-MARL, and the rest are unmatched, almost all dropped at a busy agent and a few censored by the episode's end. (b) Delivered held, the same rule's interventions are worth \keepheldvsfired\ more per episode on KEEP (\keepheldvsfiredwins\ of \episodes\ paired episodes) and \hmarlheldvsfired\ more on H-MARL (\hmarlheldvsfiredwins\ of \episodes), of which \keepheldvsfiredcharge\ and \hmarlheldvsfiredcharge\ are the Restore charge. (c) Held, the two defenders close about the same share of the distance to the RT reference (\pct{\dirkeepheld} and \pct{\dirhmarlheld}, from \pct{\dirkeepfiredceiling} and \pct{\dirhmarlfiredceiling} when fired), on scores that carry the Restore charge. The arms are \arm{wo\_O1\_g32}, \arm{wo\_O1\_g32\_wb}, \arm{hm\_wo\_O1\_g32}, \arm{hm\_wo\_O1\_g32\_wb}.}
\label{fig:drop}
\end{figure}

We measured the drop with a supervisory rule, O1, which reads the true state and issues a Restore to the agent owning a host with a red session, with at least \ruleperiod\ steps between its interventions to the same agent, on \episodes\ paired environment seeds, under Wrecker, the impact-focused adversary of \secref{sec:method}, which costs both defenders about six times their stock loss. We chose it because under the stock adversary a Restore rule has little to recover. A directive that originates with the supervisor rather than the policy is an \emph{intervention}. The rule delivers it either \emph{fired}, asserted at the step the rule issues it whether or not the agent is busy, or \emph{held}, where the rule is consulted only at steps on which the agent is idle, so every intervention is asserted at an idle agent. An intervention asserted at step $t$ for an action of duration $d$ counts as executed if the step records show the same agent executing the same action on the same host at step $t+d-1$, the step at which an accepted directive completes. Held arms, in which every intervention is asserted at an idle agent, match at that step and nowhere else, which is the check on the rule. A wider window we used at first also matched the agent's own later actions on the same host and overstated the fired shares by four points on KEEP and five on H-MARL, as the artifact's record notes. Fired at whatever agent owns the host, \pct{\keepexecshare} of interventions execute on KEEP and \pct{\hmarlexecshare} on H-MARL (Figure~\ref{fig:drop}a). The gap is mostly a busy-rate gap. In the fired arms KEEP's agents are busy on \pct{\keepbusyattempt} of agent-steps and H-MARL's on \pct{\hmarlbusyattempt}, counting the steps inside executed multi-step actions. Held, \pct{\keepexecsharewb} of interventions execute on KEEP and \pct{\hmarlexecsharewb} on H-MARL. Held interventions are asserted at idle agents, so the one percent the match misses is consistent with directives asserted within an action's duration of the last step, whose completion step lies outside the record, a censoring at the episode's end and no drop.

The held interventions are worth \keepheldvsfired\ points more per episode on KEEP (bootstrap interval \keepheldvsfiredlo\ to \keepheldvsfiredhi, \keepheldvsfiredwins\ of \episodes\ paired episodes with no ties, sign test $p<0.001$) and \hmarlheldvsfired\ more on H-MARL (\hmarlheldvsfiredlo\ to \hmarlheldvsfiredhi, \hmarlheldvsfiredwins\ of \episodes, no ties), Figure~\ref{fig:drop}b. The rule, its budget and the adversary are the same. Only the delivery changed. In the fired arms a dropped directive also starts the rule's \ruleperiod-step gap, as it would for a supervisor who believes the directive landed, so the fired cost is the drop plus that silence, and a harness that retried on the next idle step would measure less. Every Restore the harness passes to the controller is charged a point whether it runs or not, so the charge enters the difference as the gap between the numbers of Restores the two arms send, the defender's own included, which the step records give (\texttt{tools\_charge.py}). Under Wrecker it is \keepheldvsfiredcharge\ points on KEEP and \hmarlheldvsfiredcharge\ on H-MARL, and the rest of the difference, \keepheldvsfiredrest\ (interval \keepheldvsfiredrestlo\ to \keepheldvsfiredresthi) and \hmarlheldvsfiredrest\ (\hmarlheldvsfiredrestlo\ to \hmarlheldvsfiredresthi), is the difference in every component of the return other than the charge. The delivery changed the work the dropped directives would have done and the rule's silence after them together, and the rest does not apportion it between them. Table~\ref{tab:curve} gives the same held-minus-fired difference at the stock adversary and along a ladder of adversaries between it and Wrecker. The difference grows with the loss the adversary inflicts, with one small reversal between adjacent points, from \sheldvsfired\ points at stock (interval \sheldvsfiredlo\ to \sheldvsfiredhi, \sheldvsfiredwins\ of \episodes) to the headline under Wrecker, and on H-MARL from \hmsheldvsfired\ at stock (\hmsheldvsfiredlo\ to \hmsheldvsfiredhi, \hmsheldvsfiredwins\ of \episodes) to \hmarlheldvsfired, with every interval clear of zero. On KEEP the charge stays between \curvechargemin\ and \curvechargemax\ points along the whole ladder, so it is \sheldvsfiredcharge\ of the \sheldvsfired\ at stock, and on H-MARL it is \hmsheldvsfiredcharge\ of the \hmsheldvsfired. The rest is clear of zero at every point, \sheldvsfiredrest\ on KEEP at stock (interval \sheldvsfiredrestlo\ to \sheldvsfiredresthi) and \hmsheldvsfiredrest\ on H-MARL (\hmsheldvsfiredrestlo\ to \hmsheldvsfiredresthi), the H-MARL figure being within a few points of what the count's own omission can reach (Appendix~\ref{app:instrument}, at most one Restore per agent per episode in either arm), so it is not separated from zero once that omission is allowed for, and on KEEP it is \pct{\curverestsharestock} of the unsupervised loss at stock and \pct{\curverestsharemin} to \pct{\curverestsharemax} at every other point (the Rest column of Table~\ref{tab:curve} over its Unsupervised column). At stock the held rule gains \sooneheld\ over KEEP (interval \sooneheldlo\ to \sooneheldhi, \sooneheldwins\ of \episodes\ with \sooneheldties\ tied, $p=\sooneheldp$) and \hmsooneheld\ over H-MARL (\hmsooneheldlo\ to \hmsooneheldhi, \hmsooneheldwins\ of \episodes\ with \hmsooneheldties\ tied, $p=\hmsooneheldp$), while the fired rule scores \soonefired\ against KEEP (\soonefiredlo\ to \soonefiredhi, \soonefiredwins\ of \episodes) and \hmsoonefired\ against H-MARL (\hmsoonefiredlo\ to \hmsoonefiredhi, \hmsoonefiredwins\ of \episodes). The interval is on the mean difference and the sign test on the direction of the episode differences alone, so the two can disagree, as they do for the held rule on KEEP, when a few large wins carry the mean. At the published setting, delivery alone moves the same rule on H-MARL from a cost to none we can detect. The rule issues nothing but Restores, and each is charged a point, though its interventions take the place of some of the defender's own. Fired, the rule's arm pays \soonefiredcharge\ points more in charges than KEEP alone and \hmsoonefiredcharge\ more than H-MARL alone, and held \sooneheldcharge\ and \hmsooneheldcharge. Apart from the charge it adds \soonefiredrest\ points to KEEP fired (interval \soonefiredrestlo\ to \soonefiredresthi) and \sooneheldrest\ held (\sooneheldrestlo\ to \sooneheldresthi), and \hmsoonefiredrest\ (\hmsoonefiredrestlo\ to \hmsoonefiredresthi) and \hmsooneheldrest\ (\hmsooneheldrestlo\ to \hmsooneheldresthi) to H-MARL. At the published setting the rule's Restores help both defenders in both deliveries, and the fired arms score below the unsupervised defenders because the charges for their Restores, dropped ones included, exceed what the Restores recover. A third delivery, \emph{queue first}, also consults the rule on the busy steps of a busy spell, keeps the first directive it issues and holds it until the agent's next idle step, which is what a screen does when it accepts a click on a busy agent. Relative to held delivery under Wrecker it is worth \heldclickkeep\ on KEEP (interval \heldclickkeeplo\ to \heldclickkeephi, \heldclickkeepwins\ of \episodes\ with \heldclickkeepties\ ties, $p=\heldclickkeepp$ on the untied) and \heldclickhmarl\ on H-MARL (\heldclickhmarllo\ to \heldclickhmarlhi, \heldclickhmarlwins\ of \episodes\ with \heldclickhmarlties\ ties, $p=\heldclickhmarlp$). On KEEP a decision made on a busy step and kept costs nothing we can detect, and on H-MARL it costs a small amount that the interval resolves.

\begin{table}[tbp]
\centering
\caption{Held minus fired for the O1 rule at gap \ruleperiod, by adversary, on the same \episodes\ environment seeds, split into the one-point Restore charge and the rest, the difference in the return's other components, which the split does not attribute to a mechanism, beside the unsupervised defender's score under that adversary. The charge is the difference in the Restores the two arms send, executed or dropped, read from the step records. The difference and the rest each carry a 95\% bootstrap interval over episodes, and wins count the paired episodes on which held scored higher. The difference grows with the loss the adversary inflicts, apart from ConvOnly and Wrecker Q04, whose intervals overlap almost entirely. ConvOnly takes Wrecker's transition rows and the stock agent's random choice of host, and Wrecker Q$n$ adds a priority list placing $n$ percent of red's host-choice weight on hosts it already holds at root (Appendix~\ref{app:arms}). The fired arms include the rule's gap after a dropped directive (\secref{sec:drop}).}
\label{tab:curve}
\footnotesize\setlength{\tabcolsep}{2.5pt}\begin{tabular}{L{0.2\linewidth}rlrrrlr}
\toprule
Defender and adversary & Held minus fired & 95\% interval & Wins & Charge & Rest & 95\% interval & Unsupervised \\
\midrule
KEEP, stock & \sheldvsfired & [\sheldvsfiredlo, \sheldvsfiredhi] & \sheldvsfiredwins & \sheldvsfiredcharge & \sheldvsfiredrest & [\sheldvsfiredrestlo, \sheldvsfiredresthi] & \keepcorpusmean \\
KEEP, ConvOnly & \curveconvonly & [\curveconvonlylo, \curveconvonlyhi] & \curveconvonlywins & \curveconvonlycharge & \curveconvonlyrest & [\curveconvonlyrestlo, \curveconvonlyresthi] & \curvectlconvonly \\
KEEP, Wrecker Q04 & \curveqfour & [\curveqfourlo, \curveqfourhi] & \curveqfourwins & \curveqfourcharge & \curveqfourrest & [\curveqfourrestlo, \curveqfourresthi] & \curvectlqfour \\
KEEP, Wrecker Q16 & \curveqsixteen & [\curveqsixteenlo, \curveqsixteenhi] & \curveqsixteenwins & \curveqsixteencharge & \curveqsixteenrest & [\curveqsixteenrestlo, \curveqsixteenresthi] & \curvectlqsixteen \\
KEEP, Wrecker Q32 & \curveqthirtytwo & [\curveqthirtytwolo, \curveqthirtytwohi] & \curveqthirtytwowins & \curveqthirtytwocharge & \curveqthirtytworest & [\curveqthirtytworestlo, \curveqthirtytworesthi] & \curvectlqthirtytwo \\
KEEP, Wrecker Q48 & \curveqfortyeight & [\curveqfortyeightlo, \curveqfortyeighthi] & \curveqfortyeightwins & \curveqfortyeightcharge & \curveqfortyeightrest & [\curveqfortyeightrestlo, \curveqfortyeightresthi] & \curvectlqfortyeight \\
KEEP, Wrecker & \keepheldvsfired & [\keepheldvsfiredlo, \keepheldvsfiredhi] & \keepheldvsfiredwins & \keepheldvsfiredcharge & \keepheldvsfiredrest & [\keepheldvsfiredrestlo, \keepheldvsfiredresthi] & \keepwreckercontrol \\
\midrule
H-MARL, stock & \hmsheldvsfired & [\hmsheldvsfiredlo, \hmsheldvsfiredhi] & \hmsheldvsfiredwins & \hmsheldvsfiredcharge & \hmsheldvsfiredrest & [\hmsheldvsfiredrestlo, \hmsheldvsfiredresthi] & \hmarlstockcontrol \\
H-MARL, Wrecker & \hmarlheldvsfired & [\hmarlheldvsfiredlo, \hmarlheldvsfiredhi] & \hmarlheldvsfiredwins & \hmarlheldvsfiredcharge & \hmarlheldvsfiredrest & [\hmarlheldvsfiredrestlo, \hmarlheldvsfiredresthi] & \hmarlwreckercontrol \\
\bottomrule
\end{tabular}

\end{table}

\subsection{Three conclusions revisited}

The three readings below are ours. They were drawn from fired arms in the first month of this study, before the harness read the busy flag, and no published work draws them. We report them because any supervision or human-in-the-loop study on this benchmark that injects directives will draw their like, and because the cost of the drop is only visible in what it changes. Held delivery changes several things together. These are the dropped directives, what the rule spends on them (the gap after each under O1 and the spread, the budget under the burst), the charge for the dropped Restores and, in the rule-defender arms, the drop of the defenders' own directives, since the same setting makes a rule defender sleep while its agent is busy. The readings below do not separate them.

\paragraph{Directability.} Fired, the rule recovered \pct{\dirhmarlfired} of the loss Wrecker adds to H-MARL over the stock adversary (interval \pct{\dirhmarlfiredlo} to \pct{\dirhmarlfiredhi}) and \pct{\dirkeepfired} of the loss it adds to KEEP (\pct{\dirkeepfiredlo} to \pct{\dirkeepfiredhi}), and our reading was that the hierarchical architecture is less directable than the graph one. Table~\ref{tab:dir} gives both defenders on two measures, fired and held. Held, on the same loss measure, the shares are \pct{\dirkeepheldloss} and \pct{\dirhmarlheldloss}, about nine points apart. Their intervals overlap, which does not settle whether that difference is real, and we have not bootstrapped the difference itself. Measured instead as the share of the distance to a perfect-information rule reference that the rule closes, the reference being RT, a rule defender that reads the true state with held delivery (\arm{rd\_RT\_w\_none\_wb} in Appendix~\ref{app:arms}), the fired numbers are \pct{\dirhmarlfiredceiling} and \pct{\dirkeepfiredceiling} and the held ones \pct{\dirhmarlheld} and \pct{\dirkeepheld}, within two percentage points (Figure~\ref{fig:drop}c). Both measures are differences of scores, so both carry the Restore charge. RT is a reference and no bound, since a policy with the same information could combine Restores, Removes and traffic controls and score higher, so a share of the distance to it can exceed one. Under Wrecker the reference defender asserts \rtrestoreswrecker\ Restores per episode, about as many as KEEP's control runs (\keeprestoreswrecker) and \rthmarlchargegap\ more than H-MARL's (\hmarlrestoreswrecker), so the charge narrows H-MARL's reference denominator by about \rthmarlchargegap\ points and leaves KEEP's almost unchanged. The gains above each control carry the charge as well, and the held shares' agreement may rest partly on these. We have not separated them. The fired reading rested on a point estimate whose interval reaches \pct{\dirhmarlfiredlo}. What the delivery changed is the fired column. What the normalization changed is the row. Under this rule and this adversary, what separated the two fired numbers was mostly the busy share, \pct{\hmarlbusyattempt} against \pct{\keepbusyattempt}. At random times an intervention would execute on one minus the busy share, \pct{\keeprandomexec} on KEEP against the observed \pct{\keepexecshare} and \pct{\hmarlrandomexec} on H-MARL against the observed \pct{\hmarlexecshare}, so the rule also reaches H-MARL's agents when they are busier than average.

\begin{table}[tbp]
\centering
\caption{Share of the damage the O1 rule recovers on two measures, with paired bootstrap intervals over episodes. Fired, the two defenders differ on both measures. Held, they are within two points on the reference measure and about nine points apart on the loss measure. Both measures are differences of scores and carry the Restore charge (\secref{sec:drop}).}
\label{tab:dir}
\footnotesize\setlength{\tabcolsep}{4pt}\begin{tabular}{L{0.40\linewidth}L{0.25\linewidth}L{0.25\linewidth}}
\toprule
Share recovered & Fired & Held \\
\midrule
Of the loss Wrecker adds, KEEP & \pct{\dirkeepfired} [\pct{\dirkeepfiredlo}, \pct{\dirkeepfiredhi}] & \pct{\dirkeepheldloss} [\pct{\dirkeepheldlosslo}, \pct{\dirkeepheldlosshi}] \\
Of the loss Wrecker adds, H\nobreakdash-MARL & \pct{\dirhmarlfired} [\pct{\dirhmarlfiredlo}, \pct{\dirhmarlfiredhi}] & \pct{\dirhmarlheldloss} [\pct{\dirhmarlheldlosslo}, \pct{\dirhmarlheldlosshi}] \\
Of the distance to the RT reference, KEEP & \pct{\dirkeepfiredceiling} [\pct{\dirkeepfiredceilinglo}, \pct{\dirkeepfiredceilinghi}] & \pct{\dirkeepheld} [\pct{\dirkeepheldlo}, \pct{\dirkeepheldhi}] \\
Of the distance to the RT reference, H\nobreakdash-MARL & \pct{\dirhmarlfiredceiling} [\pct{\dirhmarlfiredceilinglo}, \pct{\dirhmarlfiredceilinghi}] & \pct{\dirhmarlheld} [\pct{\dirhmarlheldlo}, \pct{\dirhmarlheldhi}] \\
\bottomrule
\end{tabular}

\end{table}

\paragraph{Burst against spread.} A second rule, O2, which issues the same Restores as O1 but stands aside whenever the policy's own action is already a Restore or Remove on a compromised host, was run two ways, as a burst with no minimum gap and a budget of \burstbudget\ interventions, which it asserts as fast as it finds targets, and as a spread with the \ruleperiod-step gap, under which the budget never binds and which asserted \spreadasserted\ per episode. Fired, the burst scored \burstfired\ and the spread \spreadfired, which read as spread being worth \burstratiofired\ times as much per intervention (\spreadperattemptfired\ against \burstperattemptfired\ points of gain over the control per asserted intervention). Fired, the burst executed \burstfiredexecuted\ of its \burstfiredasserted\ interventions per episode and the spread \spreadexecuted\ of its \spreadasserted. Each intervention is a Restore and is charged whether it runs or not, so the fired burst also paid for the \burstfirednever\ per episode that never ran, against the spread's \spreadfirednever. Held, the burst executes \burstheldexecuted\ of \burstheldasserted\ asserted, scores \burstheld, closes \pct{\burstheldshare} of the distance to the RT reference, and beats the held spread by \burstvsspread\ points (interval \burstvsspreadlo\ to \burstvsspreadhi, \burstvsspreadwins\ of \episodes), and by \burstvsspreadrest\ apart from the Restore charge (interval \burstvsspreadrestlo\ to \burstvsspreadresthi). Per executed intervention the two are close, \burstperexecheld\ against \spreadperexecheld\ points. The fired burst had been a budget asserted faster than busy agents could take it, and charged for every directive they could not. Under held delivery a burst cannot be built at all, since the rule is consulted only when an agent is idle, so the held comparison is between two schedules that both follow the agents' idle steps. The earlier reading does not reverse. It dissolves.

\paragraph{A decoy reversal.} Under the O1 rule, a hand-written rule defender that deploys decoys (R4tn in Appendix~\ref{app:instrument}) trailed its decoy-free twin (R2) by \decoyfiredgap\ points (interval \decoyfiredgaplo\ to \decoyfiredgaphi, with R4tn ahead on \decoyfiredgapwins\ of \episodes) when the interventions were fired, which read as decoys hurting under supervision. The rule issued almost the same number of Restores to each, \decoyproposedrfourtn\ against \decoyproposedrtwo\ per episode, but the decoy defender's agents were busy on \pct{\decoybusyrfourtn} of agent-steps against \pct{\decoybusyrtwo}, so only \decoyrestoresrfourtn\ of its interventions executed per episode against \decoyrestoresrtwo. The defenders' own directives were dropped in the same arms and booked as done, R4tn executing \rfourtndecoysexecuted\ of the \rfourtndecoysasserted\ decoys it proposed per episode, so the fired gap was two drops at once. The defenders' own Restores were charged as well, whether they ran or not. Counting its Restore proposals less those the supervisor's interventions replaced, R4tn asserted \rfourtnownrestores\ Restores of its own per episode against R2's \rtwoownrestores, so as an accounting split the charge for R4tn's extra Restores is about \pct{\decoyownchargeshare} of the fired gap, an estimate and no bound, since the counts of both arms are open to the omission Appendix~\ref{app:instrument} describes and the difference of two undercounts can fall either way. Those Restores follow R4tn's extra alerts. Held, R4tn scores \decoyheldrfourtn\ and R2 \decoyheldrtwo, a gap of \decoyheldgap\ points (interval \decoyheldgaplo\ to \decoyheldgaphi) with R4tn ahead on \decoyheldgapwins\ of \episodes\ episodes ($p=\decoyheldgapp$, no ties), and a third variant with a decoy on every server (R2d) scores \decoyheldrtwod, \decoyheldrtwodgap\ below R2 (interval \decoyheldrtwodgaplo\ to \decoyheldrtwodgaphi, which includes zero) and ahead on \decoyheldrtwodwins\ of \episodes. Most of the fired gap went with the drop, on the supervisor's interventions and on the defenders' own actions. What remains held is a gap of about a sixth of the fired one, whose interval excludes zero and whose sign test does not reach the conventional level. Held, the two defenders assert about as many Restores of their own, \rfourtnheldownrestores\ against \rtwoheldownrestores\ per episode, net of the proposals the interventions replaced, and with the supervisor's held interventions counted as well the two arms differ by \decoyheldcharge\ Restores per episode. The charge part of that gap is therefore \decoyheldcharge\ points and the rest is \decoyheldrest\ (interval \decoyheldrestlo\ to \decoyheldresthi).

\subsection{Scope}

The drop is not specific to supervision or to blue. It is specific to CAGE 4, whose actions carry durations. CAGE 2's actions complete in one step, so the defect catalog of \citet{emerson2024cyborgpp} could not have contained it. The loop in \texttt{step} is over every agent in the scenario, and red actions carry durations. A learned or language-model red agent that asserts a directive on every step would have every directive during an exploit's trailing steps dropped in the same way, unless it keeps the stock red agent's busy check. The executors of a recent red-team evaluation act every step \citep{shaikh2026noone}, and the paper does not say whether they keep that check. Kill-chain reach is read from outcomes, so its measurement is not inflated by counting dropped directives, though the outcomes themselves depend on which directives executed. We have not measured a red-side loss, because no such agent is released, and the claim rests on the controller code and the action durations. The charge is blue's alone, since Restore is the only action whose cost the controller charges, so a red agent's dropped directives cost it nothing. On the blue side there is a published instance of the mechanism. The language-model defender of \citet{castro2025llmdefenders} sees the \texttt{IN\_PROGRESS} flag in its observation and is prompted without a Sleep action so that it stays active, and the authors' own clustering of \castrosteps\ action-reason pairs from one episode yields a cluster of \castroclusterbusy\ whose theme is that the agent's DeployDecoy is in progress and it issues an Analyse anyway (their Table IV, Cluster 3), and the controller drops any such directive asserted while the agent is busy. The per-episode action counts the authors compare with KEEP's (their Figure 7) credit KEEP with Sleep, which is not in KEEP's action space \citep{keepcode}, so they were most likely read from the simulator's record, where each dropped directive appears as a Sleep. The authors note that defining Sleep ``could improve the performance of the LLM agents when actions are in progress,'' which is the same conclusion reached from the agent's side.

HC-MARL reports per-episode action counts for its seven bottom-level agents \citep[Table~2]{qu2026hcmarl}. Its paper does not describe a busy flag in their observation, but the counts sum to within half a percent of the agent-steps, and the Sleep count is within eight of the trailing busy steps its other actions would occupy, \hcmarlsleepcount\ against \hcmarlbusysteps, so they are consistent with executed actions, whether from a wrapper that sleeps while busy or from the simulator's record. That is the arithmetic a reader can run on any action table, and the disclosure of the fifth item of \secref{sec:checklist} would make it unnecessary. The same table's Restore row is \hcmarlrestores\ per episode with no preference set and \hcmarlrestoresmin\ to \hcmarlrestoresmax\ under the three preferences. Those trials fix the top-level guidance by hand and report no reward, but under the pinned reward, with every agent's action passed to the controller, Restores at any of those rates would alone put the reward below the \hcmarlthreelayer\ the same three-level cluster is reported to score, so either its rewards exclude the Restore charge, as they would for agents the simulator steps itself \citep{cage4issue40}, or its scored runs restore far less often than any setting of its action table. We have not asked its authors. An agent trained on the documented vector observation also receives reward for directives the simulator did not run, and a charge of one point for each Restore among them, which is a credit-assignment cost during training that we name and have not measured.

The design is sound and the contract is stated. What is missing is any enforcement or error when an agent breaks it, and any exposure of the flag in the observation the challenge documents, so the drop is invisible to an agent built on it. A count of the directives such an agent asserts overstates what it did, and the challenge's record of executed actions shows each dropped directive as a Sleep without saying that anything was asked, so neither count alone shows the drop. What the stack needs, with the cost measured here as the case, is an \texttt{InvalidAction} on a directive asserted while the agent is busy, the flag in the vector observations the challenge documents, an action cost computed from the directives the controller accepts, and the Restore charge in the README's reward tables (Appendix~\ref{app:issues}).

\section{What a score is made of}\label{sec:noise}
\begin{figure}[t]
\centering
\includegraphics[width=\linewidth]{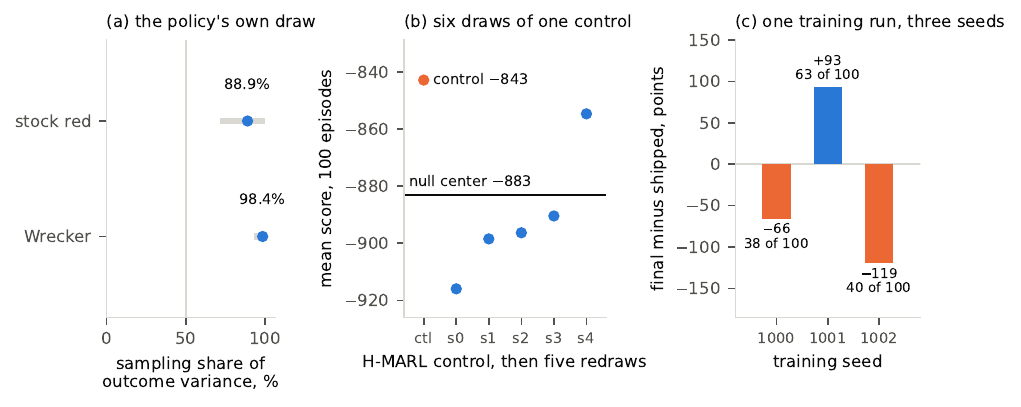}
\caption{An episode's outcome is mostly the policy's own sampling draw. (a) With episode as the factor and policy seeds as replicates, sampling accounts for \pct{\samplingvarshare} of KEEP's outcome variance at the stock adversary and \pct{\samplingvarsharewrecker} under Wrecker (bars are 95\% bootstrap intervals over episodes). (b) Six draws of one H-MARL control under Wrecker have an sd of \hmarlnullsd\ points, and the draw we had been using as the control is the best of them, \hmarlbestofsix\ above the center. (c) Three continuations of one KEEP training run from three seeds land at \contsignone, \contsigntwo\ and \contsignthree\ against the shipped weights, on the same \episodes\ paired episodes under Wrecker Q32, one evaluation draw each. Two further draws of each continuation agree in sign with the first (text).}
\label{fig:noise}
\end{figure}

A CAGE 4 score is read as a property of the defender. For KEEP at the stock adversary, most of the variation in a single episode's outcome is a property of the policy's draw. We separate the two with a one-way random-effects decomposition over the \episodes\ paired episodes of \secref{sec:method}, taking episode as the factor and policy sampling seeds as replicates, so that between-episode variance is scenario difficulty and within-episode variance is the policy's own sampling. Two runs with the same environment and policy seeds reproduce to the digit, so the within component is the closed-loop response to the policy's draw, red and green included, and not run-to-run noise (Appendix~\ref{app:arms}, \texttt{corpus\_v1} against \texttt{red\_FiniteStateRedAgent}). Every redraw in this paper uses an adjacent block of policy seeds, so two redraws whose $P$ differ by one share a stream on adjacent episodes, applied to different scenarios. The KEEP harness reseeds only \texttt{torch} with the policy seed, and the H-MARL harness reseeds \texttt{torch}, \texttt{numpy} and Python's \texttt{random}, so an H-MARL redraw also varies anything in the closed loop that draws from those two generators. Within each of the four sets of redraws, KEEP and H-MARL at the stock adversary and under Wrecker, the outcomes of episodes that share a stream correlate at \lagcorrmin\ to \lagcorrmax, so the shared streams leave no measurable dependence there. The two later H-MARL stock draws and the continuation redraws below share streams the same way and were not checked. At the stock adversary the pooled within-episode sd is \samplingsd\ points against a between-episode sd of \difficultysdstock, and sampling accounts for \pct{\samplingvarshare} of outcome variance (bootstrap interval over episodes \pct{\samplingvarsharelo} to \pct{\samplingvarsharehi}). Under Wrecker, over six draws, the two sds are \samplingsdwrecker\ and \difficultysdwrecker, and the sampling share is \pct{\samplingvarsharewrecker} (\pct{\samplingvarsharewreckerlo} to \pct{\samplingvarsharewreckerhi}), Figure~\ref{fig:noise}a. H-MARL Expert, whose redraws also vary those two generators, gives \pct{\hmarlsamplingsharewrecker} under Wrecker over six harness draws (\pct{\hmarlsamplingsharewreckerlo} to \pct{\hmarlsamplingsharewreckerhi}). At stock its six draws give \pct{\hmarlsamplingsharestocksix} (\pct{\hmarlsamplingsharestocksixlo} to \pct{\hmarlsamplingsharestocksixhi}). Its first four had given \pct{\hmarlsamplingsharestock}, the same as KEEP's to the printed digit, with draw means closer together than a share that high allows (an sd of \hmarlstockdrawsd\ points against the \hmarlstockdrawsdexp\ expected for independent draws, $p=\hmarlstockdrawp$), and over six draws the means are no closer than chance allows (\hmarlstocksixdrawsd\ against \hmarlstocksixdrawsdexp, $p=\hmarlstocksixdrawp$). At stock the within-episode spread is heavy-tailed and the share is carried by a few episodes. The widest single episode has a sampling sd of \widestepisodestock\ at stock and \widestepisodewrecker\ under Wrecker, the median episode \samplingsdmedian\ and \samplingsdmedianwrecker, and without the five widest episodes the share is \pct{\samplingvarsharetrim} at stock and \pct{\samplingvarsharewreckertrim} under Wrecker. A one-seed mean over \episodes\ episodes therefore has a standard error of about \oneseedse\ points at stock and \oneseedsewrecker\ under Wrecker, which is enough to separate KEEP from H-MARL and, if the leading entries' scores carry sampling noise like KEEP's (the leaderboard does not say), too coarse to separate its first entry from its second, \lbgapfirstsecond\ points apart, and no leaderboard carries it. The reproduction interval we report for KEEP, \keeprepro\ with standard error \keepreprose\ over four sampling seeds, is an interval over the sampling draw, which no leaderboard carries and which among the published tables only Trident's run-to-run $\pm$ approaches.

The same draws give the null band. The per-episode difference between two identical KEEP draws at stock has an sd of \nulldeltasdlo\ to \nulldeltasdhi\ points, so the standard error of a paired mean over \episodes\ episodes is about a tenth of that, and a 95\% interval on a null difference reaches about a fifth of it either side of zero. All but one of the six realized gaps sit inside that, and that one sits at its edge. Over the KEEP draws at stock the widest gap between two identical draws is \nullbandstock\ points, with the better draw ahead on \nullbandwidestwins\ of \episodes, and the most lopsided pair has the better draw ahead on \nullbandwins\ of \episodes, a margin a sign test at the nominal level would call significant. Under Wrecker the realized band is \nullbandwrecker\ points and for H-MARL \nullbandhmarl, the latter measured against a control that is the best of six draws. The band is the largest gap observed among four or six draws, not a quantile, so it has no fixed error rate. We use it to describe the noise, and we judge each comparison by its own paired interval. The paper prints some thirty paired differences, splits most of them into charge and rest, and splits part of the charge out of one more, the fired decoy gap. The comparisons its main claims rest on, held against fired on both defenders at the stock adversary and under Wrecker, the held burst against the held spread and the fired O1 arm against KEEP, have sign-test $p$ below one in a million and survive any correction for that number. The printed rests carry bootstrap intervals and no sign test, and the held-minus-fired rests at the stock adversary, \sheldvsfiredrest\ and \hmsheldvsfiredrest, and the held decoy rest, \decoyheldrest, would not survive a family-wise correction. The continuation result of this section rests on a permutation $p$ of \contqpermp, the smallest that test can return with nine draw means, which would not survive a correction for thirty, and the rest are reported with their intervals as what they are.

Two features of the shipped evaluation make a score harder to reproduce or to read. First, the evaluation seeds the environment but not policy sampling, so one command run twice on the same seeds returns different scores. We observed \unseededa\ and \unseededb\ from one two-episode invocation repeated. Second, KEEP's copy of the script evaluates in parallel by constructing one environment per worker in the parent process, all from the same seed argument, and handing each of the \episodes\ tasks a pickled copy of one of them in rotation. The parent never advances an environment and the task resets its copy without a seed, so every task on the same environment replays one scenario. Without a seed, our \paralleldistinct-worker run can contain at most \paralleldistinct\ distinct episodes, each sampled eight or nine times. With a seed given, the environments are constructed alike and the run replays a single scenario \episodes\ times. A published mean produced through that path averages over as many scenarios as there were workers. We know of no published number that passed through it. The leaderboard was scored by the organizers' script, and whether the KEEP paper's own figure used the parallel path is not stated.

Our own controls show how far a single control draw can sit from its center. Six evaluation draws of the H-MARL control under Wrecker, the arm the H-MARL gains and reference shares of \secref{sec:drop} are measured from, center at \hmarlnullcentre\ with sd \hmarlnullsd, and the draw we had been using as the control is the best of the six, \hmarlbestofsix\ points above the center (Figure~\ref{fig:noise}b). Against that control the held and fired gains of \secref{sec:drop} are \hmarlgainheld\ and \hmarlgainfired, and against the center they are \hmarlgainheldnull\ and \hmarlgainfirednull. KEEP's Wrecker control sits \keepbestofthree\ points above the center of its six draws, \keepwreckercentre, so measured against the center its held and fired gains of \keepgainheld\ and \keepgainfired\ would each be larger by that amount, a deviation within the noise of a single draw. Each gain is also net of the extra Restore charge the supervised arm pays, the rule's interventions less the defender's own Restores they replace, which these gains do not separate. The paired results in \secref{sec:drop} are reported against the controls as run, and the centers are given here so a reader can shift them.

Training adds a layer that evaluation cannot remove. KEEP trains on \trainenvs\ parallel environments that share one seed, so a training run starts from one scenario draw, and its workers sample actions from generators the trainer never seeds, so two runs with every setting equal diverge from their first updates and a continuation is a draw rather than a reproducible run. Two further defects of the same trainer, that every update replays the parent's scenario draws and that its learner reads five of the twenty-five episodes an update generates, are given in \secref{sec:pilot}, where KEEP is retrained with the first repaired and the second held as shipped and measured by a control that reads every episode. We ran three such draws with the environments reseeded. Three continuations of the shipped checkpoint from three training seeds, trained and scored under Wrecker Q32, a Wrecker variant with half its root-host priority (Appendix~\ref{app:arms}), on the same \episodes\ paired episodes against the shipped weights under the same adversary, land at \contsignone\ (interval \contsignonelo\ to \contsignonehi), \contsigntwo\ (\contsigntwolo\ to \contsigntwohi) and \contsignthree\ (\contsignthreelo\ to \contsignthreehi) (Figure~\ref{fig:noise}c), winning \contsignonewins, \contsigntwowins\ and \contsignthreewins\ of \episodes. Three continuations trained and scored under Wrecker itself land at \contwreckerone\ (\contwreckeronelo\ to \contwreckeronehi), \contwreckertwo\ (\contwreckertwolo\ to \contwreckertwohi) and \contwreckerthree\ (\contwreckerthreelo\ to \contwreckerthreehi), with \contwreckeronewins, \contwreckertwowins\ and \contwreckerthreewins\ of \episodes. Those are single evaluation draws, so we scored each continuation under two further policy seeds (Appendix~\ref{app:arms}) and took the training seed as the factor and the three draws as replicates. Under Wrecker Q32 the three draws of every continuation agree in sign, the seed means are \contqseedonemean, \contqseedtwomean\ and \contqseedthreemean, the sd of those means across training seeds is \contqbetweensd\ against a within-seed sd across draws of \contqwithinsd, the training seed accounts for \pct{\contqbetweenshare} of the variance of a draw's paired mean with the draw's own contribution removed, and a permutation test over the nine draw means puts the between-seed spread at $p=\contqpermp$. Under Wrecker itself the seed means are \contwseedonemean, \contwseedtwomean\ and \contwseedthreemean, the within-seed sd across draws (\contwwithinsd) exceeds the between-seed sd (\contwbetweensd), one continuation's draws disagree in sign, and the permutation test gives $p=\contwpermp$. Under one adversary the sign of ``more training'' is a property of the training run and reproduces across evaluation draws, though with three runs the share rests on two degrees of freedom and its interval is wide. Under the other, all three continuations gain on average and their differences in size cannot be separated from the evaluation draw. The continuations here were trained at the shipped learning rates. A single evaluation draw of a single training run says little about any of these, and whether a continuation beats its parent can turn on a run the trainer cannot reproduce. Across groups a single number says less. Trident retrained KEEP from the official repository and reports \tridentkeepscore\ \citep{masukawa2026tridentbreakdeep}, nine points below the released checkpoint's \keeppub\ and inside our reproduction interval for the release, so the number alone cannot tell a retrain from a redraw. Its text states the official recipe, and the checkpoint its repository carries records a different one (\secref{sec:audit}).

The scale on which all of this is read runs from \sleepscore, the score of doing nothing, to \bestepisode, the best single episode under any of the ten Trident adversaries we ran against KEEP. The shipped sampling mode matters within that range. The same KEEP checkpoint scores \greedyscore\ in greedy mode (sd \greedysd) against \keepcorpusmean\ sampled on the same seeds, a factor of \greedyfactor, and which mode a published number used is not disclosed.

A score also contains a term the README's reward tables do not list. Every Restore a caller asserts for a blue agent costs the team one point (\secref{sec:drop}), so a defender that waits for its agents pays one point for each Restore it runs. At the stock adversary KEEP passes \relbusyrestorespassedon\ Restores per episode to the controller, of which \restoresperepisode\ run and the rest are still running when the episode ends, and the charge is \pct{\keepchargeshare} of its loss. H-MARL passes \hmrelbusyrestorespassedon, \pct{\hmarlchargeshare} of its loss. The two reference defenders are \refgapstock\ points apart on our paired arms, and as an accounting split the difference in their Restore charges is \refgapcharge\ of those points, which says where the points are booked and nothing about why one defender restores more than the other. Under Wrecker the charge is \pct{\keepchargesharewrecker} of KEEP's loss and \pct{\hmarlchargesharewrecker} of H-MARL's. At the published setting, then, most of the gap between the two reference defenders is a difference in how much each is charged for restoring, at a price the README's reward tables do not list, and a table that prints Restores per episode beside the score lets a reader see it (\secref{sec:checklist}). Table~\ref{tab:busyreds} prints them for the two released policies, and the arms of Appendix~\ref{app:arms} carry them in the artifact.

\section{The check in training}\label{sec:pilot}
Both released defenders handle the drop in their own wrappers. H-MARL's masks every action but Sleep while an agent is busy, and KEEP's policy returns no action while busy, which its translator turns into \texttt{Monitor} (\secref{sec:drop}). That check is the only thing between each policy and the charge. This section measures what the check is worth at evaluation, by switching KEEP's check and H-MARL's mask off, and what KEEP's way of using it does in training, by training KEEP from scratch with and without it. The second needs the released trainer, whose defects we repair first.

\subsection{The released policy without its check}

A one-line switch in KEEP's wrapper (\texttt{KEEP\_BUSY} in the artifact) reports every agent as free, so the policy chooses an action every step and the controller drops what it sends to a busy agent. With the check switched off, the released policy scores \relbusyscore\ at the published setting on the \episodes\ paired episodes, against \keepcorpusmean\ with its check as shipped, a cost of \relbusycost\ points per episode (interval \relbusycostlo\ to \relbusycosthi, the check ahead on \relbusywinson\ of \episodes\ with no ties), which is \pct{\relbusyshare} of the policy's loss. The policy passes \relbusyrestorespassedoff\ Restores per episode to the controller against \relbusyrestorespassedon\ with the check, a difference of \relbusychargecost, and \relbusyrestoresbusy\ of the check-off arm's Restores per episode go to busy agents, every one dropped and charged. The charge, \relbusychargecost\ points (interval \relbusychargecostlo\ to \relbusychargecosthi), accounts for the whole of the \relbusycost, and the rest, the cost less the charge as in \secref{sec:drop}, is \relbusyrest\ (interval \relbusyrestlo\ to \relbusyresthi), indistinguishable from zero. Table~\ref{tab:busyreds} repeats the pair under Wrecker and the eight heuristic adversaries of the Trident framework \citep{masukawa2026tridentbreakdeep}, named in its rows. The cost runs from \relbusycostmin\ to \relbusycostmax\ points, the charge is \pct{\relbusychargesharemin} to \pct{\relbusychargesharemax} of it, and the rest's interval covers zero at eight of the nine. The one it does not is TargetA operational, where the rest is \relbusytaoprest\ points (interval \relbusytaoprestlo\ to \relbusytaopresthi) of a cost of \relbusytaopcost. Wrecker is the adversary at which the rest is largest, \relbusywreckerrest\ points (interval \relbusywreckerrestlo\ to \relbusywreckerresthi, which covers zero) beside a charge of \relbusywreckerchargecost, so under the adversary that punishes a missed recovery most the return's other components move with the check by an amount the interval does not settle. For a policy that reads the flag, then, the check is worth the charge, and at one adversary of nine the return's other components move with it by an amount the interval resolves, which the ablation does not attribute to a mechanism, since removing the check changes which sampling calls occur and the actions sampled after them. The actions it would have sent while busy would have been dropped either way. H-MARL's mask does the same work for its policy, and the record shows it exactly. With the mask switched off (\texttt{HMARL\_BUSY} in the artifact) its sub-policies send an action other than Sleep on \hmrelbusysentbusy\ of the \hmrelbusybusysteps\ busy agent-steps per episode, \hmrelbusyrestoresbusy\ of them Restores, every one dropped and charged, and the policy scores \hmrelbusyscore\ against \hmarlstockcontrol\ with the mask as shipped (last row of Table~\ref{tab:busyreds}). The cost, \hmrelbusycost\ points (interval \hmrelbusycostlo\ to \hmrelbusycosthi), is the charge to the point, the rest is \hmrelbusyrest\ on every episode, and the executed actions with the mask off are the executed actions with it on, step for step, on \hmrelbusysamesteps\ of the \episodes\ episodes. A dropped directive leaves nothing behind, and the mask is worth its charge and nothing else. The two checks are worth \relbusycost\ and \hmrelbusycost\ points to their policies, and the difference between those figures, \relbusycheckworthgap\ points, is most of the \refgapstock\ between the two released scores at the published setting. Without their checks the two policies would be \relbusybothoffgap\ points apart, H-MARL still ahead.

\begin{table}[tbp]
\centering
\caption{The released KEEP policy with its busy check as shipped and switched off, at the stock adversary and nine others, and the released H-MARL policy with its busy mask as shipped and switched off at the stock adversary (last row), \episodes\ paired episodes each. Cost is check on minus check off, split into the charge, the difference in Restores passed to the controller, and the rest, the cost less the charge, the difference in every component of the return other than the charge, with 95\% bootstrap intervals over episodes on the cost and the rest (the charge's are within a few points of it and are in the artifact). The last column is the Restores per episode the policy sends to busy agents with the check off, every one dropped and charged. The check-on arms reproduce the red-agent arms of Appendix~\ref{app:arms} to the digit. Recomputed from the traces by \texttt{fig/make\_busyreds.py}.}
\label{tab:busyreds}
\footnotesize\setlength{\tabcolsep}{2.5pt}\begin{tabular}{L{0.16\linewidth}rrrlrrlr}
\toprule
Adversary & Check on & Check off & Cost & 95\% interval & Charge & Rest & 95\% interval & To busy \\
\midrule
stock & \ensuremath{-}187.32 & \ensuremath{-}286.90 & 99.58 & [82.74, 115.41] & 103.45 & \ensuremath{-}3.87 & [\ensuremath{-}19.49, 10.55] & 102.58 \\
Wrecker & \ensuremath{-}1115.65 & \ensuremath{-}1270.70 & 155.05 & [77.91, 232.30] & 93.61 & 61.44 & [\ensuremath{-}16.15, 138.78] & 95.36 \\
TargetA restricted & \ensuremath{-}98.52 & \ensuremath{-}185.88 & 87.36 & [74.72, 100.48] & 79.95 & 7.41 & [\ensuremath{-}3.12, 18.58] & 78.35 \\
TargetA operational & \ensuremath{-}99.89 & \ensuremath{-}196.56 & 96.67 & [82.63, 111.21] & 81.44 & 15.23 & [4.40, 26.59] & 81.46 \\
TargetA & \ensuremath{-}104.18 & \ensuremath{-}185.97 & 81.79 & [68.44, 94.91] & 79.53 & 2.26 & [\ensuremath{-}8.47, 12.83] & 79.54 \\
TargetB & \ensuremath{-}116.25 & \ensuremath{-}206.86 & 90.61 & [77.47, 104.65] & 85.23 & 5.38 & [\ensuremath{-}5.46, 16.50] & 83.99 \\
TargetB restricted & \ensuremath{-}117.93 & \ensuremath{-}213.24 & 95.31 & [80.93, 110.47] & 87.33 & 7.98 & [\ensuremath{-}3.53, 20.07] & 86.39 \\
TargetB operational & \ensuremath{-}127.31 & \ensuremath{-}211.39 & 84.08 & [65.98, 100.63] & 83.66 & 0.42 & [\ensuremath{-}16.28, 15.13] & 83.71 \\
Combined & \ensuremath{-}165.87 & \ensuremath{-}267.09 & 101.22 & [79.74, 122.14] & 101.17 & 0.05 & [\ensuremath{-}20.67, 20.82] & 99.85 \\
FailureTracker & \ensuremath{-}205.66 & \ensuremath{-}299.18 & 93.52 & [72.07, 113.64] & 99.98 & \ensuremath{-}6.46 & [\ensuremath{-}28.39, 13.77] & 100.41 \\
\midrule
H-MARL, stock & \ensuremath{-}136.43 & \ensuremath{-}282.41 & 145.98 & [142.03, 149.97] & 145.98 & 0.00 & [0.00, 0.00] & 145.98 \\
\bottomrule
\end{tabular}

\end{table}

\subsection{Three defects in the released trainer}

KEEP's repository ships a trainer beside the released checkpoint, nothing in the repository ties the checkpoint to a run of that trainer, and we read it before retraining. It has three defects beyond the shared seed of \secref{sec:tiers}, listed with it in one row of Table~\ref{tab:defects}. First, the trainer seeds \texttt{torch} in its own process and never in the worker processes that run the episodes, so every worker starts from a seed of its own, and the policy's sampling during training is unseeded whatever seed is set. Second, the trainer builds its \trainenvs\ environments once, never steps them, and sends each worker a pickled copy at every update, so every update starts its episodes from the same scenario draws. As shipped, with one seed for every environment, every training episode starts from one network layout, and our harness's earlier repair, one seed per environment, gives \trainenvs\ fixed layouts. Third, the trainer's memory buffer is built with one slot per agent, five, and the trainer hands it one slot per episode, twenty-five, so the learner reads the first five episodes of each update and the other twenty are generated and discarded. As shipped, the released trainer therefore learns from a single layout. The trainer also builds a smaller network than the released checkpoint records, \keepactorparamstrainer\ actor parameters against \keepactorparamsreleased, by one flag (\texttt{concat\_edges}). Each is demonstrated by a one-command check in the artifact, and we repair the first two and build the released architecture for the runs below, generating only the five episodes the learner reads, so that an update is the same computation as the shipped trainer's at a fifth of the cost. The defects are of the same kind as the evaluation's. Each makes a number harder to reproduce or to read, and none was visible from the checkpoint.

\subsection{Training with and without the check}

\begin{figure}[tbp]
\centering
\includegraphics[width=\linewidth]{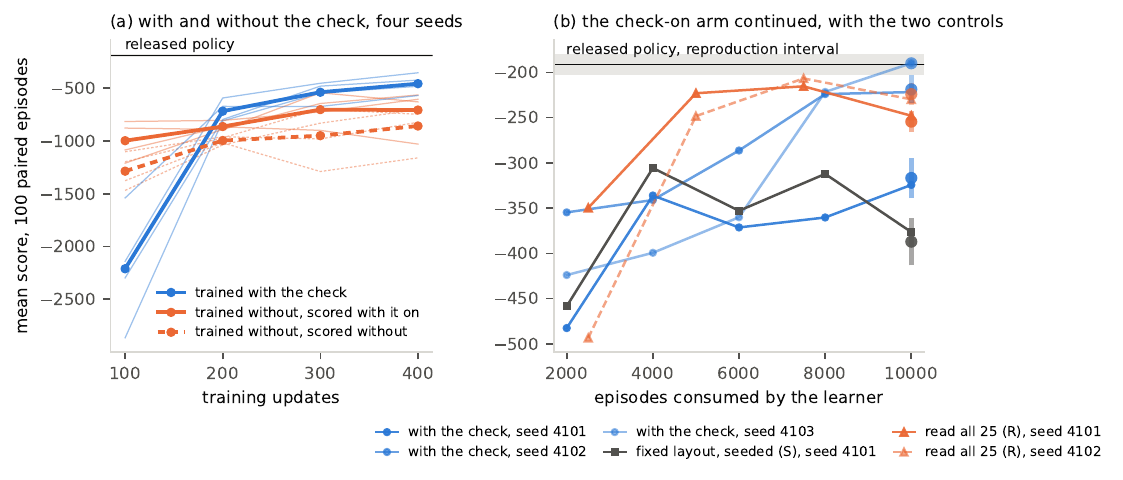}
\caption{The check in training. (a) Mean score on the \episodes\ paired episodes at every hundredth update, the four seeds as thin lines and their mean as the heavy one, for the arm trained with the check and for the arm trained without it scored both ways, against the released policy's reproduction mean. (b) The check-on arm continued past the paired range, on an axis of episodes consumed by the learner (five per update for the check-on arm and S, twenty-five for R), against the released policy's reproduction interval. Small markers are single-draw checkpoint scores, the heavier dots with bars are means and $t$ intervals over four sampling draws, squares are the fixed-layout, seeded control (S) and triangles the read-all control (R), each where scored. Drawn by \texttt{fig/fig\_curves.py}.}
\label{fig:curves}
\end{figure}

The trainer skips a busy agent. It computes no action for it, stores no transition, and adds the rewards earned while the agent is busy to the agent's next decision. That is a choice about credit assignment as much as about compute. An action's consequences are booked to the decision after it, a long action is discounted per decision and not per tick, and an agent busy at the last step receives no end-of-episode marker. With the check off, every tick is a transition with its own reward, and the policy is charged for every Restore it sends to a busy agent. We trained KEEP from scratch both ways on four paired training seeds, \pilotseedlo\ to \pilotseedhi. Within a seed the two arms start from identical weights and see identical scenario draws at every update, so they differ only in the treatment, and their trajectories diverge from the first decision the check changes. An update is five episodes of the scenario's \episodesteps\ steps, the five the shipped learner reads of the twenty-five it generates, so the two arms and the shipped trainer are matched on updates and on the episodes the learner consumes, and the shipped trainer generates five times as many. How long the released checkpoint trained is recorded nowhere. The paper's training description, which it applies to all its models, gives \keeppaperepisodelen-step episodes, \keeppaperepisodesperupdate\ episodes per update and \keeppapertrainepisodes\ episodes \citep[Table~V]{king2025keep}, and the released trainer uses \episodesteps-step episodes, \trainenvs\ per update, a default of \keeptraindefault\ episodes, and a comment that the policy stops improving around \keeptrainplateau, which is two thousand updates. Every hundredth update's checkpoint is scored on the \episodes\ paired episodes at the stock adversary, with the check as trained and crossed, so that the agent trained without the check is also scored with it on. The paired comparison stops at four hundred updates, a fifth of where the trainer's own comment places the plateau, and the check-on arm continues alone to two thousand, scored at every four hundredth update and at the last with four policy draws, against the released policy's reproduction interval of \secref{sec:method}, with two controls, a fixed-layout, seeded control on the first seed and a read-all control on the first two, described with their scores below. Which seeds continue was fixed on \seedruledate, before the four-hundred-update scores of three of the four seeds had been read, as the seeds in seed order, with their number set by the compute available before posting and never by a score. Table~\ref{tab:pilot} gives every pilot score, Table~\ref{tab:cont} the continuation's, and Figure~\ref{fig:curves} draws them. At a hundred updates the arm trained without the check is ahead in all four seeds, by \pilotnleadonehmin\ to \pilotnleadonehmax\ points when both are scored with the check on. At two hundred the order reverses in all four, and the arm trained with the check stays ahead at three and four hundred in all four, by \pilotleadmin\ to \pilotleadmax\ points, with the paired interval excluding zero in \pilotpairsexcl\ of the \pilotpairs\ comparisons from two hundred on. Those intervals describe the evaluation of each trained pair on its \episodes\ episodes. The treatment's independent replications are the four training seeds. Across them the lead is \pilotdeltatwoh\ points at two hundred updates (sd \pilotdeltasdtwoh\ across seeds, $t$ interval \pilotdeltatlotwoh\ to \pilotdeltathitwoh), \pilotdeltathreeh\ at three hundred (sd \pilotdeltasdthreeh, \pilotdeltatlothreeh\ to \pilotdeltathithreeh) and \pilotdeltafourh\ at four hundred (sd \pilotdeltasdfourh, \pilotdeltatlofourh\ to \pilotdeltathifourh), and four differences of one sign give an exact sign test of $p=\pilotsignp$, so the reversal is observed in every seed and the inference to other seeds rests on four replications. At four hundred updates the means are \pilotffourh\ with the check and \pilotnonfourh\ without it, scored with the check on. Scored its own way, with the check off, the arm trained without the check sends \pilotbusyfourh\ Restores per episode to busy agents, each dropped and charged, and scores \pilotnownfourh, so for this policy as for the released one the check is worth about its charge at evaluation. The arm trained with the check is still improving at four hundred updates and sits \pilotgapfourh\ points below the released policy's reproduction mean there, and the arm trained without it has leveled when scored with the check on.

On seed \pilotseedlo\ the check-on arm scores \longfadraws\ at two thousand updates over four sampling draws (interval \longfadrawslo\ to \longfadrawshi), \longfagap\ points below the released policy's reproduction mean of \keeprepro\ (interval \keepreprolo\ to \keepreprohi), and its interval lies below the released one, \longfaintgap\ points under it. Its checkpoints at eight hundred, twelve hundred and sixteen hundred updates score \longfaeighth, \longfatwelveh\ and \longfasixteenh, and its last step, from sixteen hundred to the four-draw mean, is \longfalaststep\ points, so it is still moving when the budget stops it. On seed \seedb\ the check-on arm scores \longfbdraws\ at two thousand updates over four sampling draws (interval \longfbdrawslo\ to \longfbdrawshi), \longfbgap\ points below the released mean, and its interval lies below the released one, \longfbintgap\ points under it, a separation four draws barely make. Its checkpoints at eight hundred, twelve hundred and sixteen hundred updates score \longfbeighth, \longfbtwelveh\ and \longfbsixteenh, and its last step, from sixteen hundred to the four-draw mean, is \longfblaststep\ points, so it has leveled. On seed \seedc\ the check-on arm scores \longfcdraws\ at two thousand updates over four sampling draws (interval \longfcdrawslo\ to \longfcdrawshi), \longfcabove\ points above the released mean, and the two intervals overlap. Its checkpoints at eight hundred, twelve hundred and sixteen hundred updates score \longfceighth, \longfctwelveh\ and \longfcsixteenh, and its last step, from sixteen hundred to the four-draw mean, is \longfclaststep\ points, so it is still moving when the budget stops it. Across the \longseedsphrase\ (\longseedlist) the four-draw means run from \longdrawsworst\ to \longdrawsbest, a spread of \longdrawsspread\ points across training seeds against a released reproduction interval \keepreprowidth\ points wide across sampling draws of one checkpoint, two different spreads, the comparison saying only that the seed-to-seed movement is larger than the sampling noise a reported score carries. Of the \longseedsphrase, \longwithincount\ overlaps the released interval and \longbelowcount\ lie below it. The fixed-layout, seeded control on seed \pilotseedlo\ (S, the shipped trainer's single layout kept and only the workers' sampling seeded) scores \longsafourh\ at four hundred, \longsaeighth\ at eight hundred, \longsatwelveh\ at twelve hundred, \longsasixteenh\ at sixteen hundred and \longsatwentyh\ at two thousand updates, and over four draws at two thousand \longsatwentyhdraws\ (interval \longsatwentyhdrawslo\ to \longsatwentyhdrawshi). The check-on arm on the same seed scores \longfafourh, \longfaeighth, \longfatwelveh, \longfasixteenh\ and \longfatwentyh\ at the same updates, so the control less the check-on arm runs from \longsfdiffmin\ to \longsfdiffmax\ points over the \longsfmatched\ matched updates, and at two thousand the control's four-draw interval lies below the check-on arm's, so at the endpoint on this seed the single layout scores lower than a fresh layout each update, by \longsfdrawsgap\ points, after sitting above it at \longsfearlierother\ of the \longsfearlier\ earlier matched checkpoints, a reversal of the size either arm moves between its own checkpoints. Paired over the same \episodes\ episodes on the tied draw, the control less the check-on arm at the endpoint is \longsafpair\ (interval \longsafpairlo\ to \longsafpairhi, the control ahead on \longsafpairwins\ of \episodes), one seed and one endpoint. The read-all control on seed \pilotseedlo\ (R, twenty-five episodes generated per update and all of them read, where the shipped learner reads five) scores \longraoneh\ at one hundred, \longratwoh\ at two hundred, \longrathreeh\ at three hundred and \longrafourh\ at four hundred updates. Each of its updates consumes five times the episodes of a check-on update, so its score at one hundred updates is on the experience of the check-on arm at five hundred, its two hundred on a thousand and its four hundred on two thousand, reached in a fifth as many gradient updates on batches five times the size, and the comparison does not separate the read from the batch size. At two hundred it scores \longratwoh, above both the check-on arm's \longfaeighth\ at eight hundred and \longfatwelveh\ at twelve hundred, the two checkpoints that bracket it in experience. At four hundred, matched to the check-on arm at two thousand on episodes consumed, its four-draw mean is \longrafourhdraws\ (interval \longrafourhdrawslo\ to \longrafourhdrawshi), \longrafourhgap\ points below the released policy's reproduction mean, against the check-on arm's \longfadraws\ (interval \longfadrawslo\ to \longfadrawshi) on the same seed and the same episodes consumed, and it sits between the check-on arm's interval and the released one, so on this seed the read per update accounts for part of the distance at matched experience and for less than all of it. Paired over the same \episodes\ episodes on the tied draw, the control less the check-on arm is \longrafpair\ (interval \longrafpairlo\ to \longrafpairhi, the control ahead on \longrafpairwins\ of \episodes). On seed \seedb\ the read-all control scores \longrboneh\ at one hundred, \longrbtwoh\ at two hundred, \longrbthreeh\ at three hundred and \longrbfourh\ at four hundred updates, the score at two hundred above both the check-on arm's \longfbeighth\ at eight hundred and \longfbtwelveh\ at twelve hundred, the two checkpoints that bracket it in experience. At four hundred, matched to the check-on arm at two thousand on episodes consumed, its four-draw mean is \longrbfourhdraws\ (interval \longrbfourhdrawslo\ to \longrbfourhdrawshi), \longrbfourhgap\ points below the released policy's reproduction mean, against the check-on arm's \longfbdraws\ (interval \longfbdrawslo\ to \longfbdrawshi) on the same seed and the same episodes consumed, and its interval overlaps the check-on arm's, so on this seed reading five times the episodes per update changes the score at matched experience by nothing the four draws resolve. Paired over the same \episodes\ episodes on the tied draw, the control less the check-on arm is \longrbfpair\ (interval \longrbfpairlo\ to \longrbfpairhi, the control ahead on \longrbfpairwins\ of \episodes). Where the arm ends is a test of reproducibility and convergence under the tested protocol, and the answer is seed by seed, as the sentences above give it. An interval that overlaps the released policy's is compatible evaluation performance at that checkpoint under this protocol and no equivalence test. An interval below it is a shortfall at this budget, resolved by the four draws where the two intervals are well apart and barely where they nearly touch. On seeds \pilotseedlo\ and \seedc\ the arm is still moving at two thousand updates, so the endpoint is where the budget stopped it and no plateau, and the spread across seeds may be partly a spread in how fast each seed climbs. The seeds were fixed before the runs, the released weights carry no seed, and the repository records neither the training length nor how the released checkpoint was chosen, so the comparison establishes that the released score is reachable at this budget from the shipped recipe on at least one seed, and it says nothing about how the released weights were produced.

\begin{table}[tbp]
\centering
\caption{KEEP trained from scratch with its busy check as shipped (F) and removed (N) on four paired training seeds, every hundredth checkpoint scored on the \episodes\ paired episodes at the stock adversary. N is scored its own way, with the check off, and with the check on. Restores at busy is the Restores per episode N sends to busy agents when scored its own way, each dropped and charged. The last columns are the paired difference between F and N scored with the check on, with a 95\% bootstrap interval over episodes, and the episodes of \episodes\ on which F scored higher. Recomputed from the traces by \texttt{fig/fig\_curves.py}.}
\label{tab:pilot}
\scriptsize\setlength{\tabcolsep}{1.5pt}\begin{tabular}{llrrrrrr}
\toprule
Updates & Seed & F & N, check off & N, check on & Restores at busy & F minus N, check on & Wins \\
\midrule
100 & 4101 & $-$2864.72 & $-$1469.00 & $-$1210.73 & 227.0 & $-$1654.0 [$-$1846, $-$1467] & 4 \\
100 & 4102 & $-$2140.24 & $-$1102.42 & $-$815.98 & 232.2 & $-$1324.3 [$-$1445, $-$1208] & 3 \\
100 & 4103 & $-$2297.17 & $-$1197.18 & $-$878.42 & 301.0 & $-$1418.8 [$-$1565, $-$1273] & 1 \\
100 & 4104 & $-$1539.99 & $-$1376.62 & $-$1085.62 & 282.8 & $-$454.4 [$-$542, $-$368] & 14 \\
100 & mean & $-$2210.5 & $-$1286.3 & $-$997.7 & 260.7 & $-$1212.8 (sd 524.3 across seeds) & F leads in 0 of 4 \\
\addlinespace
200 & 4101 & $-$809.76 & $-$1037.22 & $-$890.65 & 180.4 & 80.9 [2, 161] & 54 \\
200 & 4102 & $-$593.43 & $-$960.57 & $-$805.77 & 222.9 & 212.3 [138, 290] & 67 \\
200 & 4103 & $-$796.96 & $-$970.76 & $-$891.80 & 161.2 & 94.8 [6, 184] & 58 \\
200 & 4104 & $-$672.80 & $-$1017.77 & $-$866.40 & 186.1 & 193.6 [120, 268] & 75 \\
200 & mean & $-$718.2 & $-$996.6 & $-$863.7 & 187.7 & 145.4 (sd 67.1 across seeds) & F leads in 4 of 4 \\
\addlinespace
300 & 4101 & $-$546.44 & $-$831.65 & $-$645.60 & 170.5 & 99.2 [45, 156] & 59 \\
300 & 4102 & $-$454.01 & $-$981.40 & $-$725.28 & 201.2 & 271.3 [200, 351] & 85 \\
300 & 4103 & $-$481.34 & $-$697.76 & $-$543.95 & 171.3 & 62.6 [20, 106] & 61 \\
300 & 4104 & $-$673.97 & $-$1289.66 & $-$900.08 & 224.5 & 226.1 [138, 315] & 67 \\
300 & mean & $-$538.9 & $-$950.1 & $-$703.7 & 191.9 & 164.8 (sd 99.7 across seeds) & F leads in 4 of 4 \\
\addlinespace
400 & 4101 & $-$482.58 & $-$703.26 & $-$564.69 & 148.0 & 82.1 [30, 132] & 61 \\
400 & 4102 & $-$354.73 & $-$823.35 & $-$605.62 & 173.1 & 250.9 [191, 312] & 83 \\
400 & 4103 & $-$423.98 & $-$748.42 & $-$630.28 & 107.6 & 206.3 [136, 276] & 73 \\
400 & 4104 & $-$569.98 & $-$1160.40 & $-$1029.95 & 180.6 & 460.0 [375, 548] & 88 \\
400 & mean & $-$457.8 & $-$858.9 & $-$707.6 & 152.3 & 249.8 (sd 157.3 across seeds) & F leads in 4 of 4 \\
\bottomrule
\end{tabular}

\end{table}

\begin{table}[tbp]
\centering
\caption{The check-on arm continued past the paired range on \longseedsphrase, with the fixed-layout, seeded control (S), which matches the check-on arm update for update, and the read-all control (R), each of whose updates consumes the experience of five of the check-on arm's, every checkpoint scored on the \episodes\ paired episodes at the stock adversary. The difference is against the released policy's tied draw on the same episodes (\texttt{corpus\_v1}, mean \keepcorpusmean), with a 95\% bootstrap interval over episodes, and wins are the episodes on which the checkpoint scored higher, while the text's gaps are to the four-draw reproduction mean \keeprepro. The four-draw rows give the mean and $t$ interval over four sampling draws, the tied draw and three reseeded ones, as the released policy's reproduction interval is built. Recomputed from the traces by \texttt{fig/fig\_curves.py}.}
\label{tab:cont}
\footnotesize\begin{tabular}{llrlr}
\toprule
Updates & Arm & Mean & Minus the released policy & Wins \\
\midrule
800 & F 4101 & $-$336.11 & $-$148.8 [$-$173, $-$125] & 11 \\
1200 & F 4101 & $-$371.37 & $-$184.1 [$-$220, $-$149] & 8 \\
1600 & F 4101 & $-$360.36 & $-$173.0 [$-$205, $-$144] & 8 \\
2000 & F 4101 & $-$324.35 & $-$137.0 [$-$163, $-$111] & 12 \\
2000 & F 4101, four draws & $-$316.72 & $t$ interval [$-$339, $-$294] & \\
\addlinespace
800 & F 4102 & $-$341.08 & $-$153.8 [$-$179, $-$130] & 6 \\
1200 & F 4102 & $-$286.33 & $-$99.0 [$-$122, $-$77] & 14 \\
1600 & F 4102 & $-$223.98 & $-$36.7 [$-$56, $-$18] & 34 \\
2000 & F 4102 & $-$221.79 & $-$34.5 [$-$58, $-$13] & 41 \\
2000 & F 4102, four draws & $-$218.71 & $t$ interval [$-$234, $-$203] & \\
\addlinespace
800 & F 4103 & $-$399.40 & $-$212.1 [$-$245, $-$181] & 5 \\
1200 & F 4103 & $-$360.14 & $-$172.8 [$-$204, $-$142] & 7 \\
1600 & F 4103 & $-$221.64 & $-$34.3 [$-$54, $-$15] & 40 \\
2000 & F 4103 & $-$189.83 & $-$2.5 [$-$19, 14] & 50 \\
2000 & F 4103, four draws & $-$189.97 & $t$ interval [$-$192, $-$188] & \\
\addlinespace
400 & S 4101 & $-$458.45 & $-$271.1 [$-$307, $-$236] & 2 \\
800 & S 4101 & $-$305.72 & $-$118.4 [$-$148, $-$91] & 14 \\
1200 & S 4101 & $-$352.95 & $-$165.6 [$-$208, $-$126] & 15 \\
1600 & S 4101 & $-$312.22 & $-$124.9 [$-$174, $-$82] & 28 \\
2000 & S 4101 & $-$376.48 & $-$189.2 [$-$233, $-$150] & 11 \\
2000 & S 4101, four draws & $-$387.12 & $t$ interval [$-$413, $-$361] & \\
\addlinespace
100 & R 4101 & $-$349.73 & $-$162.4 [$-$187, $-$138] & 4 \\
200 & R 4101 & $-$223.06 & $-$35.7 [$-$54, $-$18] & 34 \\
300 & R 4101 & $-$215.40 & $-$28.1 [$-$48, $-$9] & 39 \\
400 & R 4101 & $-$247.96 & $-$60.6 [$-$81, $-$41] & 22 \\
400 & R 4101, four draws & $-$254.45 & $t$ interval [$-$266, $-$243] & \\
\addlinespace
100 & R 4102 & $-$493.37 & $-$306.1 [$-$343, $-$272] & 2 \\
200 & R 4102 & $-$248.33 & $-$61.0 [$-$85, $-$38] & 27 \\
300 & R 4102 & $-$206.70 & $-$19.4 [$-$41, 0] & 40 \\
400 & R 4102 & $-$229.81 & $-$42.5 [$-$62, $-$24] & 31 \\
400 & R 4102, four draws & $-$224.07 & $t$ interval [$-$237, $-$212] & \\
\bottomrule
\end{tabular}

\end{table}

\subsection{Reading}

The pilot shows a crossing. Without the check the policy's transitions are every tick, a Restore sent to a busy agent is charged on the spot, and the first hundred updates go faster. One possible mechanism is that the arm without the check meets the penalty for a Restore at a busy agent in every episode's reward, and the experiment does not separate that signal from the changes in trajectory structure, reward attribution and discounting that come with it. With the check an agent's transitions are its decisions, \pilotdecisionsfourh\ per episode at four hundred updates against \episodesteps\ without it, so the two arms are matched on episodes consumed and differ in transitions per episode, the reward earned during a long action is booked to the next decision, and the first hundred updates go slower. From two hundred updates on the order is reversed in every seed and the gap has not closed by four hundred, and the arm trained without the check trails even when it is scored with the check on, so what it learned without the check is not made good by giving it the check at evaluation. The claim is bounded in three ways. It concerns KEEP's handling of the flag as shipped, under the released architecture and recipe. The paired comparison covers the first four hundred updates, so it is a result about learning early in training, and the continuation covers the check-on arm alone, so it is a result about where that arm ends and says nothing about where the other would. Where it ends is a test of reproducibility and convergence. On its own it settles nothing about how the released checkpoint was produced, since the repairs change the process as well as the recipe, and the architecture flag of the previous subsection is the direct evidence of an undocumented choice where a final score is indirect. The fixed-layout, seeded control, which keeps the shipped trainer's single layout and seeds only the workers' sampling, separates the layout repair's effect from the rest, and the read-all control changes the read per update and the update schedule together. It rests on four paired seeds, reported with the paired difference and its spread across seeds, and on \longseedsphrase\ continued. And the mechanism is read from the code and not measured. A trainer that kept the check and credited the rewards earned while busy to the action that earned them, discounting per tick, would separate the credit assignment from the check itself, and that arm is future work. What the result does show is the use of the audit. The reading of the simulator that found the charge says the check is worth the charge at evaluation and at the published setting nothing else the interval resolves. The reading of the trainer says the check is a choice about how a busy step is handled, which steps enter the trajectory, how the reward earned during an action is attributed and discounted, and whether the policy samples and is charged while an action runs, and the released recipe makes it without stating it. The paired runs say the choice as a whole has a large effect on learning early in training, with the arm trained without the check ahead at a hundred updates and behind from two hundred on, in every seed, and they do not apportion that effect among the mechanisms. Any agent trained on this benchmark makes the same choice, and most make it by inheriting a wrapper. The audit turns that default into a choice with a measured cost, which is what an audit of a benchmark is for.
\section{Recovery precision without its criterion}\label{sec:nonmeasures}
Two recovery-precision figures are published on CAGE 4 beside a reward. The H-MARL authors score a recovery against the true state at the previous step, report the true and false positive counts beside it, and use it to rank defenders whose rewards are similar \citep[Section~5.5]{singh2025hmarl}. ACDZero prints the same formula beside its KEEP baseline without saying what counts as a true positive \citep{li2026acdzero}. We add a second basis, which no published table reports and which we construct from the latch the H-MARL wrapper keeps \citep[Section~4.2]{singh2025hmarl} and the first-placed team describe for their entry, whose flags ``remain until either Restore or Remove has been used on the host'' \citep{kiely2025cage4}, scoring a recovery against the evidence the defender's own observation had latched. We show with one instrument that the two bases give different numbers for the same Restores, that a rule arm scoring \pco{\oonerestoreprecision} on the same criterion recovers more than the policy it supervises, is unresolved against it held and far below it fired, because the reward charges a point for every Restore a caller asserts, dropped ones included, and that a precision is therefore not comparable across defenders unless its truth, its volume and its delivery are stated. The H-MARL authors propose their metrics as complements to the reward, and their own Table 3 already ranks one policy above another on precision and below it on reward. Our finding is the disclosure the complement needs, not a claim that precision should track the reward.

\paragraph{Ground-truth precision.} The criterion scores a Restore as correct if the host held a red session on the step before the Restore completed, the last state the record shows before the host is wiped, which is later than the step the defender decided by the action's duration less two. On \episodes\ episodes KEEP executes \restoresperepisode\ Restores and \removesperepisode\ Removes per episode, the step record holding executed actions only. The Restores are correct \pco{\restoreprecision} of the time and the Removes \pco{\removeprecision}, against a prevalence of \pct{\chanceredshare} over every host including the contractor network no blue agent owns, and \pct{\chanceredshareowned} over the hosts blue agents own, the only ones a Restore can target, each the share of host-steps on that arm on which a red session is present (the prevalence differs by arm and by host set). KEEP's Restores are more than twice the owned-host prevalence and its Removes are near it. The H-MARL authors also count a Remove on a host with a root session as a false positive, which our record cannot distinguish, so our Remove figure is an upper bound on theirs, and their published figure pools Restores and Removes in one precision, on which basis KEEP scores \pco{\keeppooledprecision} under our criterion. A supervisory rule that restores only hosts with a red session, so that each of its own directives is correct at the moment it is issued, scores \oonescore\ under the stock adversary, \oonedelta\ points against KEEP (interval \oonedeltalo\ to \oonedeltahi), and wins \oonewins\ of \episodes\ paired episodes. That arm is fired, with no gap and no budget. It asserts \ooneinterventions\ interventions per episode, of which about \ooneexecutedown\ execute, so it is a substitute policy subject to the drop of \secref{sec:drop} rather than a supervisor, and it is the extreme of the record. Every intervention is a Restore and is charged whether it runs or not, and net of the KEEP Restores its interventions take the place of, the arm pays \oonecharge\ points more in charges than KEEP, and apart from the charge it is \oonerest\ points ahead of KEEP (interval \oonerestlo\ to \ooneresthi). The same rule delivered held, so that every directive is asserted at an idle agent, asserts \sooneunboundedheldinterventions\ per episode and scores \sooneunboundedheld\ against KEEP (interval \sooneunboundedheldlo\ to \sooneunboundedheldhi, \sooneunboundedheldwins\ of \episodes\ with \sooneunboundedheldties\ tied, $p=\sooneunboundedheldp$), and at a \ruleperiod-step gap it gains \sooneheld\ (\sooneheldlo\ to \sooneheldhi, \sooneheldwins\ of \episodes\ with \sooneheldties\ tied, $p=\sooneheldp$). Held and fired, the unbounded rule's arms execute the same actions on \oonesameexec\ of \episodes\ episodes, so the \oonefiredheldcharge\ points between them are the charge for the directives the fired arm sent to busy agents, and held the rule still pays \sooneunboundedheldcharge\ points more than KEEP for the extra Restores it runs. The held arm's record carries a count of compromised hosts and not the host set, but since it runs the fired arm's actions, the fired arm's precision is its precision too. The fired arm is not the rule alone. On steps where an agent owns no compromised host the rule stands aside and KEEP acts, and those pass-through Restores, wrong at the moment the rule checked, are \pct{\oonepassthrough} of the arm's executed Restores, and on KEEP's own basis, the step before each Restore completed, the arm's Restores are correct \pco{\oonerestoreprecision} of the time, measured on a recapture of the arm that records the true state and reproduces the original on every one of its \episodes\ episodes, within a quarter of a point of the rule's own share. A pass-through Restore is wrong at the moment the rule checked and right on KEEP's basis only if its host was compromised in the four steps before it completed, a directive is right when issued and almost always still right when it completes, and the two corrections net to a quarter of a point. Its Restores recover more than KEEP's, and held its score against KEEP is \sooneunboundedheld\ points with an interval that covers zero, so that pair is unresolved, and fired it trails KEEP, because the rule runs more Restores than KEEP and, fired, sends hundreds that never run, and the reward charges a point for each. A precision says what the Restores hit, and the reward nets what they recover against what they are charged, so a precision says little about the reward without the volume behind it. The H-MARL authors' own implementation of this criterion, their metrics script run on the released checkpoint at seed zero for \episodes\ episodes, reports H-MARL Expert at \pcu{\hmarlownmetric}, beside the \pcu{\hmarlexpertprec} they publish.

\paragraph{A correction.} Our first scoring of this criterion read the true state from the same record as the action. The capture writes that state after the step has run, so every successful Restore scored as wrong and KEEP's precision came out at a fraction of a percent. The corrected basis is stated above and recorded in the artifact, and we report the error because it is of exactly the kind this paper is about.

\paragraph{Evidence precision.} The second criterion scores a recovery action as correct if the defender's own observation carried evidence for the host at the time. CAGE 4 emits an observation event on the step it occurs and forgets it. The H-MARL wrapper latches those events, adding each to the agent's history and removing it only when a recovery action handles it \citep[Section~4.2]{singh2025hmarl}. A flag therefore records that something was seen on the host at some point, not that red is present now, and some of what is seen is not red, since green agents' local work raises false alerts by design \citep{cage4code}. The flagged set grows through an episode, every recovery of a flagged host scores as correct, and a defender that clears hosts slowly is scored against a larger set of its own making.

The O2f rule of Appendix~\ref{app:instrument} keeps that latch from the raw per-agent observations as the arm runs, with the two exclusions given there, and scores it against the true state on the supervised KEEP arms under Wrecker, taking precision as the share of flagged host-steps on which a red session is present and recall as the share of host-steps with a red session that are flagged. The latched set is a weak detector. \pct{\evidenceflagged} of owned host-steps are flagged. Of those, \pct{\evidenceprecision} carry a red session, against \pct{\chanceredsharelatch} of owned host-steps on that arm (\pct{\chanceredsharelatchstock} on its stock counterpart), a lift of \evidencelift$\times$, and the flags reach \pct{\evidencerecall} of the host-steps that do. The stock adversary gives the same picture within a few points (Table~\ref{tab:criteria}). A set that covers a third of the owned host-steps and is right on about a quarter of them is not a measure of compromise, so a recovery scored as correct for landing in it is a count of recoveries on flagged hosts, not of recoveries.

On KEEP's own Restores the two criteria give \pco{\keepevidenceshare} under evidence and \pco{\restoreprecision} under ground truth. The evidence figures come from a reconstruction with no exclusions, which also counts the echo of the agent's own last action as evidence and may overstate them (Appendix~\ref{app:instrument}). At the \oonetargets\ KEEP Restore steps on which a compromised owned host was available, the target a ground-truth rule would have chosen scores \pco{\ooneevidenceshare} under evidence, close to the \pco{\keepevidenceshare} of KEEP's own Restores over all of its Restore steps (two shares over different sets of steps), while the rule itself, fired, scores \oonedelta\ points against KEEP, a deficit its Restore charge more than accounts for (Table~\ref{tab:criteria}). Neither criterion is wrong as arithmetic, and each answers a different question from the reward. A recovery measure needs the true state at decision time, which the simulator has and the challenge does not expose, and the benchmark's own evaluator writes neither the true state nor the evidence (\secref{sec:tiers}, Tier C). \secref{sec:audit} records which of the five published tables reports a precision, and under which criterion, when that is stated at all.

\begin{table}[tbp]
\centering
\caption{Two recovery-precision criteria, one published and one we construct, give different numbers for the same Restores. KEEP scores \pco{\restoreprecision} on the ground-truth criterion. In the O1 column, the ground-truth figure scores the executed Restores of the fired O1 arm of \secref{sec:nonmeasures}, with no gap and no budget, on the same criterion as KEEP's, from a recapture that records the true state and reproduces the arm on every episode, the share row is the part of those Restores that are the rule's own directives, correct at decision time, and the evidence figure scores the target the rule would choose at KEEP's own Restore steps, on the KEEP arm. The O1 arm's deficit against KEEP is more than accounted for by the Restore charge, most of it for the directives it sent to busy agents (\secref{sec:nonmeasures}). The O2f rule's latched set flags \pct{\evidenceflagged} of owned host-steps at \evidencelift$\times$ chance.}
\label{tab:criteria}
\footnotesize\setlength{\tabcolsep}{4pt}\begin{tabular}{L{0.30\linewidth}L{0.13\linewidth}L{0.13\linewidth}L{0.13\linewidth}L{0.14\linewidth}}
\toprule
Criterion & KEEP Restore & KEEP Remove & O1 & O1 vs.\ KEEP \\
\midrule
Ground truth (red session on the host before the step, for Remove an upper bound on the H-MARL authors' figure) & \pco{\restoreprecision} & \pco{\removeprecision} & \pco{\oonerestoreprecision} & \oonedelta\ points, \oonewins\ of \episodes \\
The rule's own directives, as a share of the arm's executed Restores, correct at decision time & & & \pct{\oonerulerestoreshare} & \\
Latched evidence (host named by an alert since its last Restore or Remove) & \pco{\keepevidenceshare} & \pco{\keepremoveevidenceshare} & \pco{\ooneevidenceshare}$^\dagger$ & \\
\multicolumn{5}{L{0.92\linewidth}}{\scriptsize $^\dagger$The target the rule would choose at each of KEEP's own Restore steps, scored on the KEEP arm, a different set of steps from the rest of the row.} \\
\midrule
\multicolumn{5}{l}{\emph{The latched set against the true state, supervised KEEP arms}} \\
Adversary & Flagged & Precision & Recall & Lift over chance \\
\midrule
Wrecker & \pct{\evidenceflagged} & \pct{\evidenceprecision} & \pct{\evidencerecall} & \evidencelift$\times$ \\
Stock & \pct{\evidenceflaggedstock} & \pct{\evidenceprecisionstock} & \pct{\evidencerecallstock} & \evidenceliftstock$\times$ \\
\bottomrule
\end{tabular}

\end{table}

\section{Five tables, four disclosure axes}\label{sec:audit}
A CAGE 4 score is comparable to another only if four things are known about each.
\begin{enumerate}
\item \textbf{Artifact identity.} Which checkpoint, which code and which version produced it.
\item \textbf{Seed and dispersion.} How it was sampled, over how many episodes, and how far it spreads.
\item \textbf{Criterion and scenario.} What was scored, on which scenario, by what rule.
\item \textbf{Adversary.} Which red agent it faced.
\end{enumerate}
We read the final leaderboard and every table from 2026 that compares a new defender or a new adversary against one of the two reference defenders, five tables, counting each paper's tables as one (Table~\ref{tab:audit}), and record for each which of the four it states. None states all four. The reference defenders' own papers, whose numbers these tables cite, are outside the audit, the organizers' paper is read with the leaderboard it reports, and two further CAGE 4 papers from 2025 are outside it too. \citet{castro2025llmdefenders} compare language-model defenders with a KEEP they retrained under their own communication protocol, in a figure, over two episodes against five adversaries, and cite no published score, so the comparison is internal to one harness. \citet{jiang2025agentic} train hierarchical agents on an LLM-generated reward and report that reward rather than the benchmark's, so their table carries no CAGE 4 score.

Part of the pattern is explained by \secref{sec:tiers}. The challenge script's record holds executed actions per step and nothing an intervention count needs, no per-step reward, true state or asserted directive, and KEEP's copy writes nothing at all, so no author could have reported the quantities of Sections~\ref{sec:drop} and \ref{sec:nonmeasures} without an instrument. Checkpoint hashes, seeds, episode counts, sampling modes and adversaries need none, and their absence is not forced by the stack.

\begin{table}[tbp]
\centering
\caption{Five published tables, 2024 to 2026, and the disclosure each omits. Without artifact identity, seed and dispersion, criterion and scenario, and adversary, two scores in one table cannot be shown to be comparable. The newest \citep{shaikh2026noone} is the only one that does all three of stating its criterion, excluding the reward from it and recomputing its numbers from its own traces, and it still omits dispersion and artifact identity.}
\label{tab:audit}
\footnotesize\setlength{\tabcolsep}{4pt}\begin{tabular}{L{0.14\linewidth}L{0.19\linewidth}L{0.19\linewidth}L{0.21\linewidth}L{0.16\linewidth}}
\toprule
Table & Artifact & Seed, dispersion & Criterion, scenario & Adversary \\
\midrule
CAGE 4 leaderboard (2024) & checkpoints not identified, no code or checkpoint released for entries above fifth & one mean per entry on the page, mean and unlabeled $\pm$ over \leaderboardepisodes\ episodes for the best agent of each of four teams in the organizers' paper, no seeds or sampling mode & scored by one script that sums the reward each submission's own wrapper returns (Appendix~\ref{app:issues}) & stock, stated in the challenge README and the organizers' paper, not on the page \\ \addlinespace
Trident, Table 2 (2026) & H-MARL weights md5-identical to the copy in Trident's repository (Table~\ref{tab:env}), whether its table used them unstated. The GNN is a disclosed retrain at \tridentkeepscore, from a checkpoint whose recorded recipe differs from the text's & $\pm$ stated as sd across \tridentruns\ runs of \tridentrunepisodes\ episodes & stated & varied, stated \\ \addlinespace
ACDZero, Tables I and II (2026) & KEEP baseline is the leaderboard number to the digit & mean $\pm$ sd over \acdzeroepisodes\ episodes stated. Baseline sd \leaderboardkeepsd\ is a third of the checkpoint's across-episode sd. Ablation row equals the baseline row & recovery precision reported with the criterion unstated, and no Restore counts & stock, stated as the official protocol \\ \addlinespace
HC-MARL, Tables 3, 4 and 5 (2026) & code available on request only & mean and sd over \hcmarlepisodes\ test runs, in a table titled best rewards & Singh row is H-MARL Meta, not Expert. Seven agents on a five-agent scenario. The action mask is worth \hcmarliocdelta\ points. Its own Restore count alone exceeds its reported loss under the pinned reward & stock, stated in its limitations \\ \addlinespace
Shaikh et al., Table 2 (2026) & H-MARL checkpoint not identified & \shaikhepisodes\ episodes, no dispersion, though a rate over that many episodes has a standard error under four points, no seeds, sampling mode unstated & predicate stated, recomputed independently, ``may end earlier'' unspecified & six configurations, stated \\
\bottomrule
\end{tabular}

\end{table}

\paragraph{Artifact identity.} The Trident paper reports KEEP at \tridentkeepscore\ beside H-MARL at \hmarlpub\ \citep{masukawa2026tridentbreakdeep}. It states that the H-MARL weights are the released ones, and we confirm that its sixteen checkpoint files are byte-identical to the published ones, and it states that the GNN defender was retrained from the official repository, with the official recipe stated in its appendix. The checkpoint in its repository records a different recipe, a larger network and different PPO settings,\footnote{Hidden width \tridenthidden\ against \keephidden, node embedding \tridentconv\ against \keepconv, PPO clip \tridentclip\ against \keepclip, actor learning rate \tridentlr\ against \keeplr, and critic learning rate \tridentcriticlr\ against \keepcriticlr, read from the agent tuple stored in each checkpoint (\texttt{tools\_ckpt\_recipe.py}).} but the repository does not say whether that checkpoint produced its Table 2, and we have not yet asked its authors, so the disagreement between checkpoint and text is a question for them rather than a finding. The retrain scores \tridentkeepscore\ against the released checkpoint's \keeppub\ on the leaderboard and \keeprepro\ in our reproduction, so the row compares a retraining to a release, which it discloses, and the retrained weights are not identified beyond that. The leaderboard shows how the field's baselines were chosen \citep{cage4leaderboard}. KEEP placed \lbkeeprank{}th of \lbentries\ at \keeppub, behind a leader at \lbfirst, and none of the four entries above it released code or a checkpoint. The released H-MARL Expert checkpoint is a post-challenge artifact whose \hmarlauthorscore\ would place third behind \lbfirst\ and \lbsecond. The two reference defenders are the two that can be run, not the two that scored best.

\paragraph{Seed and dispersion.} The previous challenge's leaderboard did better. CAGE 2 reported each entry as a total over three episode lengths and three red agents, \cagetwoepisodes\ episodes per condition, with a 95\% confidence interval and the seed stated \citep{cage2code, kiely2023autonomous}. The CAGE 4 leaderboard page reports one mean per entry \citep{cage4leaderboard}, and the organizers' paper adds a $\pm$ over \leaderboardepisodes\ episodes, unlabeled, for the best agent of each of the top four teams \citep{kiely2025cage4}, which is the \leaderboardkeepsdhc\ that HC-MARL prints beside KEEP's mean as a standard deviation \citep{qu2026hcmarl}. Neither states seeds or sampling mode. ACDZero states its protocol, the official evaluation against the stock red over \acdzeroepisodes\ episodes with mean and sd, and reports a GCN baseline at \keeppub\ with sd \leaderboardkeepsd\ and its own agent at \acdzeroscore\ with sd \acdzerosd\ \citep{li2026acdzero}. The baseline mean is the leaderboard's to the digit, and its sd is a third of the across-episode sd we measure for the released checkpoint over the same number of episodes at the same adversary (\stockdrawsdlo\ to \stockdrawsdhi\ across four draws, Appendix~\ref{app:arms}), which the paper does not explain and we have not asked its authors about. Its ablation row, the agent without tree search, carries the same mean and sd, which the paper does not discuss. Trident states its $\pm$ as the sd across three independent runs of \tridentrunepisodes\ episodes each, $\pm$\tridentkeepsd\ for KEEP and $\pm$\hmarlpubsd\ for H-MARL. That is a run-to-run spread, and it sits in the literature beside per-episode sds an order of magnitude or more larger on the tables cited here, with the same symbol, so a reader comparing across tables has to know which kind each is.

\paragraph{Criterion and scenario.} Two of the five tables place numbers side by side that were not scored the same way. HC-MARL publishes an ablation of its action mask, and the ablation is informative. It reports its two-level agent at \hcmarltwolayer\ points and its three-level agent at \hcmarlthreelayer, each a mean with sd over \hcmarlepisodes\ test runs of the model after its last training iteration, in a table titled best rewards, beside H-MARL at \hcmarlsinghscore\ and ACDZero at \acdzeroscore\ \citep{qu2026hcmarl}. Two things differ across that row. The H-MARL figure is H-MARL Meta, the learned-master variant, where the authors' own headline is H-MARL Expert at \hmarlauthorscore, which is above both HC-MARL agents, by a margin inside the noise of a hundred-episode mean for the two-level one. And the HC-MARL agent fields seven bottom-level defenders, one per subnet, where the benchmark's scenario defines five blue agents, so its row and its neighbors' were produced on different scenarios. Its own action table raises a third question, whether its reward includes the Restore charge (\secref{sec:drop}). The paper's own ablation shows that the indicator-of-compromise action mask carries all of its lead over the baselines. Without the mask the same architecture scores \hcmarlnoiocscore\ points, \hcmarliocdelta\ below itself with it, and what the hierarchy adds on top of the mask is not reported. ACDZero reports a recovery precision of \pcu{\acdzerorecpreclo} for its KEEP baseline and \pcu{\acdzerorecprechi} for itself, with the criterion unstated. Its column header, TP over TP plus FP, is the H-MARL authors' formula, and what counts as a true positive is not said. The baseline is the agent of \secref{sec:nonmeasures}, which scores \pco{\restoreprecision} under the ground-truth criterion and \pco{\keepevidenceshare} under the evidence criterion, so without the criterion the two precisions carry no information about which agent recovers better. Nor does the reward column beside them settle it, since KEEP pays \relbusyrestorespassedon\ points per episode in Restore charges at this adversary and ACDZero reports neither agent's Restore count.

\paragraph{Adversary.} The three tables that rank defenders, the leaderboard, ACDZero and HC-MARL, evaluate under the stock red alone. ACDZero names it as the official protocol, HC-MARL names it in its list of limitations, the leaderboard page does not name it though the challenge README does, and none carries it to the table where the ranking is read. The organizers' own paper does vary it, scoring the best agent of each of the top four teams under stealthy and aggressive reds among other changes to the environment \citep[Table~2]{kiely2025cage4}, which is the disclosure the leaderboard page lacks and the papers that cite the leaderboard do not carry forward. Trident and Shaikh vary the adversary, since the adversary is their subject, and state it. The same H-MARL checkpoint scores \hmarlstockcontrol\ against the stock red and \hmarlwreckercontrol\ against Wrecker (the center of six draws is \hmarlnullcentre), on the same seeds through our harness. A ranking made under one adversary is not evidence about another, and the benchmark's own red agent is one of several the simulator ships.

\paragraph{The fifth table.} A red-team evaluation posted in September 2026 \citep{shaikh2026noone} runs \shaikhepisodes\ episodes per configuration against an H-MARL defender and reports disruption and compromise rates and detection counts. It is the only table of the five that does all three of stating its criterion, excluding the reward from it and recomputing its numbers from its own traces, where the others cite them, and Table~\ref{tab:audit} credits it for that. It reports no dispersion and no seeds, and does not say which H-MARL artifact it ran or how it sampled. Its outcome rates are read from the realized state, so the drop of \secref{sec:drop} cannot inflate their count, though what its executors managed to execute shapes the outcomes themselves, and it reports no per-episode action counts. The pattern of the first four holds in the fifth.

\section{Recommendations and a disclosure checklist}\label{sec:checklist}
The defects of \secref{sec:tiers} have fixes, and the omissions of \secref{sec:audit} have a remedy that costs a paragraph. We separate the two, because the first falls to the benchmark's maintainers and the second to anyone who publishes a number on it, and the second does not wait on the first.

\paragraph{For the stack.} Four changes, each with a one-command reproduction of the defect behind it in the artifact (Appendix~\ref{app:issues}). The artifact carries our patches to the two defenders' code. They seed KEEP's evaluation script, H-MARL's two evaluation scripts and KEEP's training environments, make KEEP's script write its record on both paths and offer argmax actions, add a choice of adversary to both trainers and to H-MARL's two evaluation scripts, let KEEP's trainer restore a checkpoint with its training flag set, scale its learning rates and set its training seed, length, worker count and checkpoint interval, and in H-MARL also seed the trainer, let it resume, set its batch size, iteration count and log and model directories, load the agent's checkpoint from another directory and switch on the wrapper's metrics. The pilot patches of \secref{sec:pilot} seed each training episode's sampling and environment, step to a fresh layout per update, let the learner read every generated episode, and switch KEEP's busy check off in training and at evaluation and H-MARL's mask off at evaluation. The parallel path's defect remains, and the trainer as shipped still seeds no worker.
\begin{enumerate}
\item A directive asserted while an action is in progress should be rejected with \texttt{InvalidAction}, or the agent should not be asked, as the documentation says, and the busy flag should be part of the documented vector observation. Either makes the drop visible. The action cost should be computed from the directives the controller accepts, and the README's reward tables should list the Restore charge (\secref{sec:drop}).
\item The evaluation script should seed policy sampling, which is one line, the parallel path of KEEP's copy should give each task a fresh environment or its own seed rather than a copy of a parent environment, and KEEP's trainer should seed each training environment and each worker's generators (\secref{sec:noise}).
\item The evaluation script should write a per-step record, with the executed action, the reward and the true state, so that a published score can be decomposed and a decision scored by anyone who did not produce it (\secref{sec:tiers}, Tier C).
\item \texttt{get\_reward\_breakdown} should work, so that the per-host attribution the benchmark advertises exists.
\end{enumerate}

\paragraph{For a published number.} Five disclosures, one per axis of \secref{sec:audit} and one for the method. A table that states them can be judged for comparability against another that does. A table that omits one cannot, whatever its sd.
\begin{enumerate}
\item \textbf{Artifact.} The checkpoint's hash and size, the code revision, and whether the checkpoint is the released one or a retraining.
\item \textbf{Sampling and spread.} The sampling mode, the environment and policy seeds, the number of episodes, and what the reported spread is a spread over.
\item \textbf{Criterion and scenario.} What was scored, by what rule, on which scenario, and when the criterion is a precision, against what truth.
\item \textbf{Adversary.} Which red agent, as a scope limit on every claim in the table.
\item \textbf{Counting.} Whether action counts are directives asserted or actions executed, from what record executions were read, the Restores asserted and executed per episode, since the reward charges a point for each one asserted, and whether every agent's actions were passed to the controller's step, since only those are charged.
\end{enumerate}

\paragraph{For a supervisory harness.} Four practices that measured the defects above, and that we would now treat as defaults on this benchmark. Deliver interventions held, so that none is asserted at a busy agent, unless the experiment is about delivery itself. Run two or more null draws, the same policy on the same seeds, before reading any arm, report every control as a draw with its null center rather than as a point, and judge each comparison by its own paired interval, using the null draws to check the noise that interval assumes. Pair every comparison episode by episode on shared environment seeds and report the paired mean with its interval and win count. Pairing removes the scenario component, which here is a small share of the variance, so it narrows an interval only modestly, and it is a policy seed that makes a run reproducible. Count the Restores each arm sends, executed or dropped, and split every paired difference into the Restore charge and the rest, since an intervention that is a Restore costs a point whether or not it runs.

\paragraph{For training an agent.} Three practices from \secref{sec:pilot}, for anyone who retrains a defender on this benchmark. Seed the workers that sample actions and seed each training episode's environment, so that a run is reproducible and two runs differ by their seeds and nothing else. Step or reseed the environments an update is built from, so that every update sees a fresh scenario draw, and state how many distinct layouts a run trained on. And state how the trainer treats an agent whose action is in progress, whether it skips the step, masks the choice, or credits the rewards earned while busy, and to which decision, since that treatment is a modeling choice with a measured effect on learning and it is inherited from the wrapper unless it is chosen.

\paragraph{This paper against its checklist.} Each table here is meant to pass the five items, and a reader can check. Artifact. The two checkpoints are identified by digest and size and the code by revision (Table~\ref{tab:env}), and every retrained or continued policy is named as such with its training seed. Sampling and spread. Every arm is \episodes\ paired episodes on environment seeds \seedlo\ to \seedhi, each draw's policy seed is in Appendix~\ref{app:arms} or the artifact, every interval over episodes is a 95\% bootstrap and every interval over seeds or draws is a $t$ interval that says so, every comparison is paired with its win count, and the null draws of \secref{sec:noise} are the spread two identical policies show. Criterion and scenario. Every score is the episode reward the controller sums, on the challenge scenario as shipped, and every precision names the truth it is scored against (Table~\ref{tab:criteria}). Adversary. Every table names its red agent. Counting. Every action count is read from the step record, executed actions on the supervised arms and directives passed on the released-policy and pilot arms, and every agent's actions pass through the controller's step. Restores passed with and without the check are printed for the two released policies in \secref{sec:pilot}, those sent to busy agents in Table~\ref{tab:busyreds}, and the rule arms' asserted and executed interventions in \secref{sec:nonmeasures}, the charges in Tables~\ref{tab:curve} and \ref{tab:busyreds}, and the counts behind every other split, and the all-arms table of Appendix~\ref{app:arms}, leave the counts to the artifact. Where an item is met by the artifact and not by the page, as the per-arm policy seeds and counts are, the table says so.

\section{Related work and limitations}\label{sec:related}
\paragraph{Evaluation critiques.} The nearest work is in two literatures that have not met. In reinforcement learning, \citet{henderson2018drlmatters} showed that reported results on standard benchmarks turned on seeds, implementations and the choice of runs to report, and \citet{agarwal2021precipice} showed that point estimates from a handful of runs reverse under interval estimates, proposing the paired, interval-based reporting we adopt. In security, \citet{arp2022dosdonts} cataloged ten evaluation pitfalls across thirty papers from top venues, \citet{flood2024smells} traced the same class of flaw into seven benchmark intrusion datasets, and \citet{bilot2025simpler} reimplemented eight provenance-based detectors in one framework and found that reported rankings did not survive a common evaluation. Closest to this benchmark, \citet{emerson2024cyborgpp} reimplemented the CAGE 2 version of CybORG and documented three classes of defect in the original, invalid actions, an inconsistent reward and reduced observation visibility, two of which are of the same kind as ours, a reported quantity not measuring what it is read as and an observation that withholds what the agent needs in order to act on it. The security audits examine classifier benchmarks, where the unit is a labeled record. A defense benchmark is a sequential decision problem with a simulator in the loop, so the defects it admits are of a different kind. In this case, a controller that discards actions and charges for some of them, an evaluator that does not seed, and a criterion that scores against a set the agent built. Beyond that reimplementation, we know of no audit of how results on a sequential-decision security benchmark are reported, and the checklist of \secref{sec:checklist} is the sequential counterpart of the ones those papers propose. The first four checklist items adopt what \citet{henderson2018drlmatters}, \citet{agarwal2021precipice} and the cooperative MARL protocol of \citet{gorsane2022protocol} already ask for. The fifth is specific to a simulator whose actions carry durations.

\paragraph{Actions that take time.} An action with a duration turns the decision problem into a semi-Markov one, and the treatment is standard. The rewards earned during the action are summed, discounted by its length, and credited to the decision that started it \citep{bradtke1994smdp, sutton1999options}. Frame skipping in the Atari benchmarks is the everyday case, and it credits the summed reward to the repeated action. KEEP's trainer credits the rewards earned while an agent is busy to that agent's next decision and discounts per decision, which is the half of the treatment \secref{sec:pilot} examines. Masking invalid actions is the nearest practice on the other side, and it helps policy-gradient learning \citep{huang2022masking}. A busy check looks like a mask, and the difference matters. A mask leaves every step in the trajectory and restricts the choice at it, where KEEP's check removes the step.

\paragraph{CAGE and CybORG.} The CAGE challenges \citep{standen2021cyborg, kiely2025cage4} are the reference environments for learned cyber defense, and CAGE 4 is the first with a multi-agent blue team. The two defenders we reproduce are the two that released code and checkpoints \citep{king2025keep, singh2025hmarl, keepcode, hmarlcode}, which is why the work since has baselined against them. That work includes an LLM red-teaming framework \citep{masukawa2026tridentbreakdeep}, a tree-search agent \citep{li2026acdzero}, a hierarchical architecture \citep{qu2026hcmarl} and an LLM red-team evaluation \citep{shaikh2026noone}, and \secref{sec:audit} reads their tables. \citet{saika2026crossing} ask whether an agent trained in one simulator transfers to another, which presupposes that a score in each means the same thing, and this paper is about the presupposition. Cyberwheel \citep{oesch2024cyberwheel} is the other open simulator at this scale, and whether its controller shares the drop is an open question.

\paragraph{Limitations.} All results are on one environment and two learned defenders. The drop's cost on the supervised arms is measured on interventions, since both released policies read the busy flag, and its cost on the policies themselves is measured by switching that check off (\secref{sec:pilot}), an ablation of each wrapper and no measurement of an agent built without one. Agents built on the challenge's vector wrappers are subject to the drop and to its charge, and we did not run an agent on the documented observation itself. \secref{sec:pilot} also trains KEEP without the check. Every redraw uses an adjacent block of policy seeds, and the two harnesses reseed different generators with the policy seed, so the two defenders' sampling shares are not measured on the same footing. The H-MARL harness reseeds numpy's and Python's generators as well as torch's, and the pinned CybORG draws its simulator's and its agents' randomness from generators of its own (a read of its Simulator and Agents packages finds global draws in two test agents the scenario does not use), so the reseeding reaches the policy's sampling path and nothing in the environment. The evidence-criterion shares are reconstructed from the raw per-agent observations of the KEEP corpus, under one reading of a latch that each team implemented differently. The parallel-path defect in KEEP's copy of the evaluation script is reported, not repaired. Every arm in this paper was captured by our harness, which builds a fresh environment per episode, or by the H-MARL evaluator, which runs its episodes in sequence, and KEEP's first draw ran its copy of the script with one worker, so the defect touches none of them. The H-MARL authors' own metric is regenerated from one seeded run of their metrics script whose log is in the artifact. What changes when the supervisor is a person, and what changes on a network that is not simulated, are questions this instrument was built to ask and this paper does not answer.

\paragraph{AI Disclosure Statement.} The authors used a large language model assistant (Claude, Anthropic) throughout the preparation of this paper, for drafting and editing prose from a ledger of measured values built during the research, for writing the analysis, table and figure scripts and the patches to the released code, and for review passes against the tables. Every number in the paper regenerates from the released traces through the artifact’s checks, every claim was read against its evidence by the authors, and the authors take full responsibility for the content. The research question, the instrument, every decision about what to run and what to claim, and the intended applications of this work are the authors’ own.

\bibliographystyle{plainnat}
\bibliography{refs}

\appendix
\section{The instrument}\label{app:instrument}
\paragraph{Pinned environment.} Table~\ref{tab:env} lists the three repositories at the revisions every number in this paper was computed from, the two checkpoints, and the two dependency pins neither defender's repository declares. The copy of the H-MARL weights that \secref{sec:audit} compares by md5 is from the Trident repository, \texttt{BiasLabProjects/Trident} at \texttt{\tridentrev} (\tridentrevdate). The digests are in Table~\ref{tab:env}, each an md5 of the sorted per-file md5s, over KEEP's five checkpoint files and over H-MARL's fifteen policy files and \texttt{algorithm\_state.pkl}. The two defenders' dependency sets are mutually incompatible, so each runs in its own image, and a wrapper script (\texttt{bin/run}) selects the image by arm. On the arms of the supervisory harness the header record carries the rule and budget, the red agent and minimum gap from the arms of mid-September onward, and the two delivery flags from the late-September dates the experiment log gives for each, so the fired arms captured before then carry no delivery field and the log records what their headers lack. The \texttt{capture.py} arms carry the red agent, the policy seed and the scenario's host and reward maps, and no rule or budget, since they run none.

\begin{table}[h]
\centering
\caption{The pinned environment.}
\label{tab:env}
\footnotesize\setlength{\tabcolsep}{4pt}\begin{tabular}{L{0.26\linewidth}L{0.36\linewidth}L{0.14\linewidth}L{0.14\linewidth}}
\toprule
Component & Source & Revision & Date \\
\midrule
CybORG and CAGE 4 template & \texttt{cage-challenge/cage-challenge-4} & \texttt{\cagerev} & \cagerevdate \\ \addlinespace
KEEP & \texttt{cybermonic/cage-4-submission}, five checkpoints, one per agent, \texttt{gnn\_ppo-*.pt}, \publishedkeepbytes\ bytes each & \texttt{\keeprev} & \keeprevdate \\ \addlinespace
H-MARL Expert & \texttt{adityavs14/Hierarchical-MARL}, released checkpoint (digest below), byte-identical to Trident's copy at \texttt{\tridentrev} & \texttt{\hmarlrev} & \hmarlrevdate \\ \addlinespace
Undeclared pins & \texttt{pyarrow} \pyarrowpin\ and the CPU wheel index for \texttt{torch} & & \\ \addlinespace
Checkpoint digests & md5 of the sorted per-file md5s.\newline KEEP, its five checkpoints\newline \hashtt{\keepckptdigest}\newline H-MARL, its fifteen policy files and \texttt{algorithm\_state.pkl}\newline \hashtt{\hmarlckptdigest} & & \\ \addlinespace
Images & \texttt{cage-keep:dev} and \texttt{cage-hmarl:dev}, since the two dependency sets do not coexist & & \\
\bottomrule
\end{tabular}

\end{table}

\paragraph{Step record.} The instrument wraps the simulator and writes one JSON line per step per episode, in one of two schemas. On the KEEP policy arms captured by \texttt{scripts/capture.py}, a step record carries, for every blue agent, the raw observation the simulator returned, the action the simulator executed that step (\texttt{get\_last\_action}), and whether the agent was busy (\texttt{success} equal to \texttt{IN\_PROGRESS}), the per-agent and total step reward, and the true state, as the set of hosts holding a red session, and arms captured with later versions of the script also carry the red agents' actions and the action index each policy returned. On every arm run through \texttt{scripts/oracle.py} or \texttt{scripts/capture\_hmarl.py}, which is every supervised arm, every H-MARL arm of our harness, the rule-defender arms and the continuation arms, a step record carries every blue agent's executed action, the red agents' actions, the step and cumulative reward and a count of the compromised hosts the agents own, and no raw observation or host set. Arms captured from mid-September 2026 onward also carry the supervisor's interventions (agent, the policy's choice, the supervisor's), in a field the code names \texttt{overrides}, and the experiment log names the reruns that added it to earlier arms. A footer record carries the episode total, the intervention and agreement counts and the wall time, and on the rule-defender arms the rule's own decisions counted by type, the Restores it proposed among them. Because the action field holds executed actions, a learned policy's own directives are not in the harness record, while a rule defender's are, in the \texttt{defender} field, which is how R4tn's decoy proposals are counted in \secref{sec:drop}, where a defender's own Restores asserted are the footer's Restore proposals less the Restore proposals an intervention replaced, which the intervention record names. An intervention's execution is read by the rule of \secref{sec:drop}, an executed action of the same type on the same host by the same agent at step $t+d-1$ for an action of duration $d$ asserted at $t$. The charge accounting of \secref{sec:drop} (\texttt{tools\_charge.py}) counts executed Restores and the recorded interventions that never ran. It assumes a defender never asserts a Restore at a busy agent, which holds for KEEP and H-MARL. It would undercount rule defenders delivered fired, a defender's own Restores the controller replaces as invalid and arms captured before the intervention record, and it misses a defender's own Restore still running when an episode ends, at most one per agent per episode. Two accounting families follow. The first counts directives at submission. On the released-policy arms and the pilot checkpoints of \secref{sec:pilot}, captured with the policy field, every Restore the policy passed is in the record, busy or not, the count is exact and nothing is censored at the episode's end. The second reconstructs submissions from completions. On the supervised arms of \secref{sec:drop} and Table~\ref{tab:curve}, the count is executed Restores plus the recorded interventions that never ran, complete except for the terminal omission below for the policy's own Restores under the assumption above, for the rule's held directives, which never meet a busy agent, and for its fired directives on arms that carry the intervention record, which is every arm Table~\ref{tab:curve} prints. On the rule-defender arms delivered held, a defender's own Restores are the footer's proposals less those an intervention replaced, and on the fired rule-defender arms the count can miss a proposal the controller dropped at a busy agent, so their charges are lower bounds. On every arm of the second family the Restore still running at the end of an episode is missed, at most one per agent, and the charges and rests of \secref{sec:drop} carry that error, which is why the small rest of the H-MARL stock pair there is read beside it and the rests under Wrecker are not. In every split the charge is counted from what was passed to the controller, the rest is what remains of the paired difference after it, and a charged directive and an executed Restore are never the same accounting unit.

\paragraph{Latch reconstruction.} The evidence criterion of \secref{sec:nonmeasures} follows the H-MARL authors' own description of their observation memory. New events are added to the agent's history at each step and removed only when a recovery action on that host handles them \citep[Section~4.2]{singh2025hmarl}, and the first-placed team describe the same mechanism for theirs \citep{kiely2025cage4}. We reconstruct it from the raw per-agent observations in the step record. A host enters an agent's flagged set at the first step the agent's observation names it, and leaves it at that agent's next executed Restore or Remove on it. Two implementations of this latch appear in the paper. The O2f rule keeps its own, with the exclusions given below, and clears a host when it asserts or agrees a recovery on it, and the latched-set rows of Table~\ref{tab:criteria} sum that rule's footers. The evidence shares of KEEP and O1 in \secref{sec:nonmeasures} come from a reconstruction over the raw observations with no exclusions, cleared at the executed recovery. The two differ in the exclusions, in when a host leaves (assertion or agreement against the executed recovery) and in scope (the agent's owned hosts against every host named), and we have not measured the difference. The reconstruction runs on the KEEP corpus only and the O2f latch on supervised arms of both defenders, and our ground-truth scorer runs on KEEP's policy arms only and the H-MARL authors' metric on H-MARL only, so the two latches have never run on the same arm and the two ground-truth scorers have never run on the same defender. Some of what the observation names is not red. The challenge's green agents raise a false alert on \texttt{GreenAccessService} and \texttt{GreenLocalWork} with a configured probability (\texttt{fp\_detection\_rate}, README Appendix A) \citep{cage4code}.

\paragraph{Supervisory rules and rule defenders.} The arms of Appendix~\ref{app:arms} name their rules by code. O1 reads the true state and issues a Restore to the agent owning a host with a red session, with a minimum gap between interventions to the same agent. O2 issues the same Restores but stands aside when the policy's own action is already a Restore or Remove on a compromised host, within a budget. O2f is O2 reading, in place of the true state, the flagged set that an agent's own observations build. A host is flagged by any substantive observation entry naming it, apart from the first step's listing and the echo of the agent's own action the step before, whoever chose it, a file the simulator marks as confirmed malware always counts, and the rule targets the oldest flag first. The rule defenders replace the learned policy. R2 restores a host on its second alert since its last Restore or on a confirmed file, and analyzes once-alerted hosts. R2d is R2 with a decoy deployed, on steps where it has nothing else to do, to an owned server without one. R4tn deploys decoys on such steps, up to three per host, to the owned host with the fewest, analyzes on the first alert, restores on the second or on a confirmed file, and sets no firewall posture. RT restores any owned host the true state shows compromised, at most once in thirty steps per host, and is the perfect-information rule reference of \secref{sec:drop}, a reference and no bound, since a policy with the same information could do better. It evicts only by Restore, asserting \rtrestoreswrecker\ per episode under Wrecker, so its score carries a point for each.

\paragraph{Variance decomposition and intervals.} The sampling share of \secref{sec:noise} is a one-way random-effects decomposition with episode as the factor and policy sampling seeds as replicates. The within-episode variance is pooled across episodes, the between-episode variance is the variance of episode means less the sampling share of that mean, and the share is the within component over their sum, with a bootstrap over episodes for the interval. Reproduction intervals are $t$ intervals over four seed means with three degrees of freedom, KEEP's over sampling seeds on the paired environment seeds and H-MARL's over seeds of the authors' evaluation path. Every measured or cited number printed in digits in this paper is a macro generated from a ledger row that records its arm, its source and the command that regenerates it, apart from the all-arms table and Table~\ref{tab:busyreds}, which their scripts recompute from the traces, and the few counts written in words (action durations, the Restore charge, window lengths, caps, the window correction of \secref{sec:drop}, the episodes of the smoke runs and of Castro et al.\ and the four reproduction seeds) are ledger values or are read from the source or record named beside them. \texttt{make recompute} reruns every command except derivations, citations, parameters and container runs, and checks that the stored value, with or without its sign, appears in the command's output, a consistency check between the ledger and the tools and no proof that the right statistic was computed. \texttt{make check} tests the signed identities every split must satisfy, cost equals charge plus rest for each pair, the four-draw and four-seed intervals against their means and sds, and the H-MARL mask-off pair's zero rest and identical executions, so a sign or quantile slip in a tool fails the build.

\paragraph{What the artifact reproduces.} Three levels, which a reader should tell apart. The instrument, the patches, the ledger and the scripts behind every figure and table are at \url{https://github.com/jrice-ai/cage} at tag \texttt{arxiv-v1}, and the traces and checkpoints every number is computed from are deposited at \url{https://doi.org/10.5281/zenodo.23681828}.  Every printed number regenerates from the released traces and the ledger by \texttt{make recompute}, with no container. A fresh evaluation of any arm needs the two images, the pinned checkpoints and the harness scripts, and reproduces a released arm to the digit on the same seeds, as the H-MARL arm of \secref{sec:pilot}, captured on a second machine from a copy of the image, reproduces its control. Retraining needs the patched trainer and days of CPU per seed, and a pilot run is reproducible from its seed up to the learner's thread count, which moves the weights in their seventh decimal. The artifact's README names the commands for each level.

\section{All arms}\label{app:arms}
Table~\ref{tab:arms} lists every arm the paper cites, \episodes\ episodes each, with mean and sd recomputed by \texttt{fig/make\_arms.py} from its step-record footers, or for the five arms run through the H-MARL evaluator from the summary file that evaluator writes, including arms that appear only in a ledger row, as a control, a denominator or a draw the text does not report. Two single runs of \secref{sec:method} are not listed, KEEP's first draw through its own copy of the evaluation script and H-MARL's unseeded run through the authors' evaluator. Neither is a paired arm, and the experiment record gives both. Defender, adversary, supervisory rule (with its minimum gap in steps and its budget in interventions) and delivery mode are parsed from the arm name, \armcount\ arms in all. The released policy's busy-check arms of \secref{sec:pilot} are listed with the check's state as the defender, and the pilot's checkpoint arms are listed in Tables~\ref{tab:pilot} and \ref{tab:cont} and the record and are not repeated here. Arm names encode the defender (\texttt{hm} for H-MARL, \texttt{rd\_R*} for the rule defenders of Appendix~\ref{app:instrument}, otherwise KEEP), the adversary (\texttt{w} for Wrecker), the supervisory rule (\texttt{O1}, \texttt{O2}, \texttt{O2f}), the minimum gap (\texttt{g32}), the budget (\texttt{b77}), held delivery (\texttt{\_wb}) and queue-first delivery (\texttt{\_qf\_wb}). \texttt{arm\_keep}, \texttt{corpus\_v1} and \texttt{red\_FiniteStateRedAgent} are three runs of one configuration on three days, and their agreement to the digit is the determinism check of \secref{sec:noise}. Wrecker Q32 is Wrecker with the root-host priority halved, \pcu{\wreckerqrootpriority} of the weight in place of \pcu{\wreckerrootpriority}, on the same transition table, and it is the adversary of the Q32 continuation arms and of their baseline. The other points of Table~\ref{tab:curve} follow the same pattern. ConvOnly takes Wrecker's transition table with the stock agent's random choice of host, and Wrecker Q04, Q16 and Q48 place four, sixteen and forty-eight percent of the host-choice weight on hosts held at root. \texttt{arm\_monitor} and \texttt{arm\_analyse} agree to the digit, as expected when neither action changes the network and the seeds are shared.

\begingroup\scriptsize\setlength{\tabcolsep}{3pt}
\begin{longtable}{L{0.24\linewidth}L{0.15\linewidth}L{0.13\linewidth}L{0.14\linewidth}L{0.09\linewidth}rr}
\caption{Arms cited in this paper, \episodes\ episodes each. Reference arms are KEEP's four sampling seeds and H-MARL's four seeds of the authors' evaluation path. The null draws and redraws of \secref{sec:noise} are \texttt{keep\_null*}, \texttt{wrecker\_null*}, \texttt{hm\_s\_none\_p*} and \texttt{hm\_w\_none\_p*}, labeled by policy seed, and KEEP cont.\ $n$ is a continuation of the shipped checkpoint trained from seed $n$. Arms named \texttt{red\_*} run KEEP against the named red agent through the harness.}\label{tab:arms}\\
\toprule
Arm & Defender & Adversary & Supervision & Delivery & Mean & SD \\
\midrule
\endfirsthead
\multicolumn{7}{l}{\footnotesize Table~\ref{tab:arms}, continued}\\
\toprule
Arm & Defender & Adversary & Supervision & Delivery & Mean & SD \\
\midrule
\endhead
\bottomrule
\endfoot
\texttt{arm\textunderscore\allowbreak{}analyse} & Sleep and Analyse alternating & stock & none &  & $-$6592.68 & 1340.47 \\
\texttt{arm\textunderscore\allowbreak{}keep} & KEEP & stock & none &  & $-$187.32 & 61.24 \\
\texttt{arm\textunderscore\allowbreak{}keep\textunderscore\allowbreak{}greedy} & KEEP, argmax actions & stock & none &  & $-$552.22 & 343.75 \\
\texttt{arm\textunderscore\allowbreak{}monitor} & Monitor every step & stock & none &  & $-$6592.68 & 1340.47 \\
\texttt{arm\textunderscore\allowbreak{}sleep} & Sleep policy & stock & none &  & $-$6592.68 & 1340.47 \\
\texttt{corpus\textunderscore\allowbreak{}v1} & KEEP & stock & none &  & $-$187.32 & 61.24 \\
\texttt{hm\textunderscore\allowbreak{}released\textunderscore\allowbreak{}busy\textunderscore\allowbreak{}none} & H-MARL Expert, busy mask off & stock & none &  & $-$282.41 & 55.79 \\
\texttt{hm\textunderscore\allowbreak{}released\textunderscore\allowbreak{}busy\textunderscore\allowbreak{}shipped} & H-MARL Expert & stock & none &  & $-$136.43 & 49.60 \\
\texttt{hm\textunderscore\allowbreak{}s\textunderscore\allowbreak{}none} & H-MARL Expert & stock & none &  & $-$136.43 & 49.60 \\
\texttt{hm\textunderscore\allowbreak{}s\textunderscore\allowbreak{}none\textunderscore\allowbreak{}p9000} & H-MARL Expert & stock & seed 9000 &  & $-$136.36 & 48.56 \\
\texttt{hm\textunderscore\allowbreak{}s\textunderscore\allowbreak{}none\textunderscore\allowbreak{}p9001} & H-MARL Expert & stock & seed 9001 &  & $-$136.76 & 52.55 \\
\texttt{hm\textunderscore\allowbreak{}s\textunderscore\allowbreak{}none\textunderscore\allowbreak{}p9002} & H-MARL Expert & stock & seed 9002 &  & $-$134.60 & 49.54 \\
\texttt{hm\textunderscore\allowbreak{}s\textunderscore\allowbreak{}none\textunderscore\allowbreak{}p9003} & H-MARL Expert & stock & seed 9003 &  & $-$130.04 & 50.55 \\
\texttt{hm\textunderscore\allowbreak{}s\textunderscore\allowbreak{}none\textunderscore\allowbreak{}p9004} & H-MARL Expert & stock & seed 9004 &  & $-$137.63 & 54.58 \\
\texttt{hm\textunderscore\allowbreak{}so\textunderscore\allowbreak{}O1\textunderscore\allowbreak{}g32} & H-MARL Expert & stock & O1, gap 32 & fired & $-$160.69 & 38.29 \\
\texttt{hm\textunderscore\allowbreak{}so\textunderscore\allowbreak{}O1\textunderscore\allowbreak{}g32\textunderscore\allowbreak{}wb} & H-MARL Expert & stock & O1, gap 32 & held & $-$133.09 & 34.03 \\
\texttt{hm\textunderscore\allowbreak{}w\textunderscore\allowbreak{}none} & H-MARL Expert & Wrecker & none &  & $-$842.76 & 312.94 \\
\texttt{hm\textunderscore\allowbreak{}w\textunderscore\allowbreak{}none\textunderscore\allowbreak{}p9000} & H-MARL Expert & Wrecker & seed 9000 &  & $-$915.97 & 354.28 \\
\texttt{hm\textunderscore\allowbreak{}w\textunderscore\allowbreak{}none\textunderscore\allowbreak{}p9001} & H-MARL Expert & Wrecker & seed 9001 &  & $-$898.46 & 354.60 \\
\texttt{hm\textunderscore\allowbreak{}w\textunderscore\allowbreak{}none\textunderscore\allowbreak{}p9002} & H-MARL Expert & Wrecker & seed 9002 &  & $-$896.31 & 304.57 \\
\texttt{hm\textunderscore\allowbreak{}w\textunderscore\allowbreak{}none\textunderscore\allowbreak{}p9003} & H-MARL Expert & Wrecker & seed 9003 &  & $-$890.41 & 302.32 \\
\texttt{hm\textunderscore\allowbreak{}w\textunderscore\allowbreak{}none\textunderscore\allowbreak{}p9004} & H-MARL Expert & Wrecker & seed 9004 &  & $-$854.60 & 254.22 \\
\texttt{hm\textunderscore\allowbreak{}wo\textunderscore\allowbreak{}O1\textunderscore\allowbreak{}g32} & H-MARL Expert & Wrecker & O1, gap 32 & fired & $-$744.22 & 330.34 \\
\texttt{hm\textunderscore\allowbreak{}wo\textunderscore\allowbreak{}O1\textunderscore\allowbreak{}g32\textunderscore\allowbreak{}qf\textunderscore\allowbreak{}wb} & H-MARL Expert & Wrecker & O1, gap 32 & queue first & $-$457.81 & 188.28 \\
\texttt{hm\textunderscore\allowbreak{}wo\textunderscore\allowbreak{}O1\textunderscore\allowbreak{}g32\textunderscore\allowbreak{}wb} & H-MARL Expert & Wrecker & O1, gap 32 & held & $-$427.96 & 183.69 \\
\texttt{hmarl\textunderscore\allowbreak{}metrics\textunderscore\allowbreak{}s0} & H-MARL Expert & stock & none, authors' metrics script, seed 0 &  & $-$124.11 & 45.80 \\
\texttt{hmarl\textunderscore\allowbreak{}s0} & H-MARL Expert & stock & none, authors' path seed 0 &  & $-$124.11 & 45.80 \\
\texttt{hmarl\textunderscore\allowbreak{}s1} & H-MARL Expert & stock & none, authors' path seed 1 &  & $-$138.31 & 40.97 \\
\texttt{hmarl\textunderscore\allowbreak{}s2} & H-MARL Expert & stock & none, authors' path seed 2 &  & $-$132.10 & 45.40 \\
\texttt{hmarl\textunderscore\allowbreak{}s3} & H-MARL Expert & stock & none, authors' path seed 3 &  & $-$130.86 & 47.45 \\
\texttt{k\textunderscore\allowbreak{}kft\textunderscore\allowbreak{}WreckerQ32\textunderscore\allowbreak{}s1000\textunderscore\allowbreak{}final\textunderscore\allowbreak{}WreckerQ32\textunderscore\allowbreak{}none} & KEEP cont. 1000 & Wrecker Q32 & none &  & $-$838.79 & 272.15 \\
\texttt{k\textunderscore\allowbreak{}kft\textunderscore\allowbreak{}WreckerQ32\textunderscore\allowbreak{}s1000\textunderscore\allowbreak{}final\textunderscore\allowbreak{}WreckerQ32\textunderscore\allowbreak{}none\textunderscore\allowbreak{}p9000} & KEEP cont. 1000 & Wrecker Q32 & seed 9000 &  & $-$794.66 & 261.47 \\
\texttt{k\textunderscore\allowbreak{}kft\textunderscore\allowbreak{}WreckerQ32\textunderscore\allowbreak{}s1000\textunderscore\allowbreak{}final\textunderscore\allowbreak{}WreckerQ32\textunderscore\allowbreak{}none\textunderscore\allowbreak{}p9001} & KEEP cont. 1000 & Wrecker Q32 & seed 9001 &  & $-$842.15 & 285.20 \\
\texttt{k\textunderscore\allowbreak{}kft\textunderscore\allowbreak{}WreckerQ32\textunderscore\allowbreak{}s1001\textunderscore\allowbreak{}final\textunderscore\allowbreak{}WreckerQ32\textunderscore\allowbreak{}none} & KEEP cont. 1001 & Wrecker Q32 & none &  & $-$680.19 & 211.22 \\
\texttt{k\textunderscore\allowbreak{}kft\textunderscore\allowbreak{}WreckerQ32\textunderscore\allowbreak{}s1001\textunderscore\allowbreak{}final\textunderscore\allowbreak{}WreckerQ32\textunderscore\allowbreak{}none\textunderscore\allowbreak{}p9000} & KEEP cont. 1001 & Wrecker Q32 & seed 9000 &  & $-$688.79 & 225.45 \\
\texttt{k\textunderscore\allowbreak{}kft\textunderscore\allowbreak{}WreckerQ32\textunderscore\allowbreak{}s1001\textunderscore\allowbreak{}final\textunderscore\allowbreak{}WreckerQ32\textunderscore\allowbreak{}none\textunderscore\allowbreak{}p9001} & KEEP cont. 1001 & Wrecker Q32 & seed 9001 &  & $-$740.75 & 249.37 \\
\texttt{k\textunderscore\allowbreak{}kft\textunderscore\allowbreak{}WreckerQ32\textunderscore\allowbreak{}s1002\textunderscore\allowbreak{}final\textunderscore\allowbreak{}WreckerQ32\textunderscore\allowbreak{}none} & KEEP cont. 1002 & Wrecker Q32 & none &  & $-$892.35 & 314.63 \\
\texttt{k\textunderscore\allowbreak{}kft\textunderscore\allowbreak{}WreckerQ32\textunderscore\allowbreak{}s1002\textunderscore\allowbreak{}final\textunderscore\allowbreak{}WreckerQ32\textunderscore\allowbreak{}none\textunderscore\allowbreak{}p9000} & KEEP cont. 1002 & Wrecker Q32 & seed 9000 &  & $-$934.08 & 290.62 \\
\texttt{k\textunderscore\allowbreak{}kft\textunderscore\allowbreak{}WreckerQ32\textunderscore\allowbreak{}s1002\textunderscore\allowbreak{}final\textunderscore\allowbreak{}WreckerQ32\textunderscore\allowbreak{}none\textunderscore\allowbreak{}p9001} & KEEP cont. 1002 & Wrecker Q32 & seed 9001 &  & $-$865.55 & 240.50 \\
\texttt{k\textunderscore\allowbreak{}kft\textunderscore\allowbreak{}Wrecker\textunderscore\allowbreak{}s1000\textunderscore\allowbreak{}final\textunderscore\allowbreak{}w\textunderscore\allowbreak{}none} & KEEP cont. 1000 & Wrecker & none &  & $-$1020.23 & 301.79 \\
\texttt{k\textunderscore\allowbreak{}kft\textunderscore\allowbreak{}Wrecker\textunderscore\allowbreak{}s1000\textunderscore\allowbreak{}final\textunderscore\allowbreak{}w\textunderscore\allowbreak{}none\textunderscore\allowbreak{}p9000} & KEEP cont. 1000 & Wrecker & seed 9000 &  & $-$1084.14 & 291.08 \\
\texttt{k\textunderscore\allowbreak{}kft\textunderscore\allowbreak{}Wrecker\textunderscore\allowbreak{}s1000\textunderscore\allowbreak{}final\textunderscore\allowbreak{}w\textunderscore\allowbreak{}none\textunderscore\allowbreak{}p9001} & KEEP cont. 1000 & Wrecker & seed 9001 &  & $-$1025.35 & 270.09 \\
\texttt{k\textunderscore\allowbreak{}kft\textunderscore\allowbreak{}Wrecker\textunderscore\allowbreak{}s1001\textunderscore\allowbreak{}final\textunderscore\allowbreak{}w\textunderscore\allowbreak{}none} & KEEP cont. 1001 & Wrecker & none &  & $-$948.46 & 334.98 \\
\texttt{k\textunderscore\allowbreak{}kft\textunderscore\allowbreak{}Wrecker\textunderscore\allowbreak{}s1001\textunderscore\allowbreak{}final\textunderscore\allowbreak{}w\textunderscore\allowbreak{}none\textunderscore\allowbreak{}p9000} & KEEP cont. 1001 & Wrecker & seed 9000 &  & $-$1036.34 & 362.56 \\
\texttt{k\textunderscore\allowbreak{}kft\textunderscore\allowbreak{}Wrecker\textunderscore\allowbreak{}s1001\textunderscore\allowbreak{}final\textunderscore\allowbreak{}w\textunderscore\allowbreak{}none\textunderscore\allowbreak{}p9001} & KEEP cont. 1001 & Wrecker & seed 9001 &  & $-$1034.39 & 351.91 \\
\texttt{k\textunderscore\allowbreak{}kft\textunderscore\allowbreak{}Wrecker\textunderscore\allowbreak{}s1002\textunderscore\allowbreak{}final\textunderscore\allowbreak{}w\textunderscore\allowbreak{}none} & KEEP cont. 1002 & Wrecker & none &  & $-$1086.71 & 368.11 \\
\texttt{k\textunderscore\allowbreak{}kft\textunderscore\allowbreak{}Wrecker\textunderscore\allowbreak{}s1002\textunderscore\allowbreak{}final\textunderscore\allowbreak{}w\textunderscore\allowbreak{}none\textunderscore\allowbreak{}p9000} & KEEP cont. 1002 & Wrecker & seed 9000 &  & $-$1140.53 & 366.93 \\
\texttt{k\textunderscore\allowbreak{}kft\textunderscore\allowbreak{}Wrecker\textunderscore\allowbreak{}s1002\textunderscore\allowbreak{}final\textunderscore\allowbreak{}w\textunderscore\allowbreak{}none\textunderscore\allowbreak{}p9001} & KEEP cont. 1002 & Wrecker & seed 9001 &  & $-$1043.22 & 325.63 \\
\texttt{keep\textunderscore\allowbreak{}null} & KEEP & stock & seed 9000 &  & $-$196.09 & 51.96 \\
\texttt{keep\textunderscore\allowbreak{}null\textunderscore\allowbreak{}9001} & KEEP & stock & seed 9001 &  & $-$198.44 & 83.33 \\
\texttt{keep\textunderscore\allowbreak{}null\textunderscore\allowbreak{}9002} & KEEP & stock & seed 9002 &  & $-$183.51 & 49.40 \\
\texttt{keep\textunderscore\allowbreak{}released\textunderscore\allowbreak{}busy\textunderscore\allowbreak{}none} & KEEP, busy check off & stock & none &  & $-$286.90 & 62.62 \\
\texttt{keep\textunderscore\allowbreak{}released\textunderscore\allowbreak{}busy\textunderscore\allowbreak{}none\textunderscore\allowbreak{}FSRedAgent\allowbreak{}Combined} & KEEP, busy check off & Combined & none &  & $-$267.09 & 75.76 \\
\texttt{keep\textunderscore\allowbreak{}released\textunderscore\allowbreak{}busy\textunderscore\allowbreak{}none\textunderscore\allowbreak{}FSRedAgent\allowbreak{}FailureTracker} & KEEP, busy check off & Failure Tracker & none &  & $-$299.18 & 68.01 \\
\texttt{keep\textunderscore\allowbreak{}released\textunderscore\allowbreak{}busy\textunderscore\allowbreak{}none\textunderscore\allowbreak{}FSRedAgent\allowbreak{}TargetA} & KEEP, busy check off & Target A & none &  & $-$185.97 & 83.58 \\
\texttt{keep\textunderscore\allowbreak{}released\textunderscore\allowbreak{}busy\textunderscore\allowbreak{}none\textunderscore\allowbreak{}FSRedAgent\allowbreak{}TargetAOperational} & KEEP, busy check off & Target A Operational & none &  & $-$196.56 & 95.50 \\
\texttt{keep\textunderscore\allowbreak{}released\textunderscore\allowbreak{}busy\textunderscore\allowbreak{}none\textunderscore\allowbreak{}FSRedAgent\allowbreak{}TargetARestricted} & KEEP, busy check off & Target A Restricted & none &  & $-$185.88 & 80.92 \\
\texttt{keep\textunderscore\allowbreak{}released\textunderscore\allowbreak{}busy\textunderscore\allowbreak{}none\textunderscore\allowbreak{}FSRedAgent\allowbreak{}TargetB} & KEEP, busy check off & Target B & none &  & $-$206.86 & 80.87 \\
\texttt{keep\textunderscore\allowbreak{}released\textunderscore\allowbreak{}busy\textunderscore\allowbreak{}none\textunderscore\allowbreak{}FSRedAgent\allowbreak{}TargetBOperational} & KEEP, busy check off & Target B Operational & none &  & $-$211.39 & 78.90 \\
\texttt{keep\textunderscore\allowbreak{}released\textunderscore\allowbreak{}busy\textunderscore\allowbreak{}none\textunderscore\allowbreak{}FSRedAgent\allowbreak{}TargetBRestricted} & KEEP, busy check off & Target B Restricted & none &  & $-$213.24 & 86.12 \\
\texttt{keep\textunderscore\allowbreak{}released\textunderscore\allowbreak{}busy\textunderscore\allowbreak{}none\textunderscore\allowbreak{}FSRedAgent\allowbreak{}Wrecker} & KEEP, busy check off & Wrecker & none &  & $-$1270.70 & 308.69 \\
\texttt{keep\textunderscore\allowbreak{}released\textunderscore\allowbreak{}busy\textunderscore\allowbreak{}shipped} & KEEP & stock & none &  & $-$187.32 & 61.24 \\
\texttt{keep\textunderscore\allowbreak{}released\textunderscore\allowbreak{}busy\textunderscore\allowbreak{}shipped\textunderscore\allowbreak{}FSRedAgent\allowbreak{}Combined} & KEEP & Combined & none &  & $-$165.87 & 84.76 \\
\texttt{keep\textunderscore\allowbreak{}released\textunderscore\allowbreak{}busy\textunderscore\allowbreak{}shipped\textunderscore\allowbreak{}FSRedAgent\allowbreak{}FailureTracker} & KEEP & Failure Tracker & none &  & $-$205.66 & 90.26 \\
\texttt{keep\textunderscore\allowbreak{}released\textunderscore\allowbreak{}busy\textunderscore\allowbreak{}shipped\textunderscore\allowbreak{}FSRedAgent\allowbreak{}TargetA} & KEEP & Target A & none &  & $-$104.18 & 55.60 \\
\texttt{keep\textunderscore\allowbreak{}released\textunderscore\allowbreak{}busy\textunderscore\allowbreak{}shipped\textunderscore\allowbreak{}FSRedAgent\allowbreak{}TargetAOperational} & KEEP & Target A Operational & none &  & $-$99.89 & 55.40 \\
\texttt{keep\textunderscore\allowbreak{}released\textunderscore\allowbreak{}busy\textunderscore\allowbreak{}shipped\textunderscore\allowbreak{}FSRedAgent\allowbreak{}TargetARestricted} & KEEP & Target A Restricted & none &  & $-$98.52 & 48.61 \\
\texttt{keep\textunderscore\allowbreak{}released\textunderscore\allowbreak{}busy\textunderscore\allowbreak{}shipped\textunderscore\allowbreak{}FSRedAgent\allowbreak{}TargetB} & KEEP & Target B & none &  & $-$116.25 & 54.19 \\
\texttt{keep\textunderscore\allowbreak{}released\textunderscore\allowbreak{}busy\textunderscore\allowbreak{}shipped\textunderscore\allowbreak{}FSRedAgent\allowbreak{}TargetBOperational} & KEEP & Target B Operational & none &  & $-$127.31 & 88.17 \\
\texttt{keep\textunderscore\allowbreak{}released\textunderscore\allowbreak{}busy\textunderscore\allowbreak{}shipped\textunderscore\allowbreak{}FSRedAgent\allowbreak{}TargetBRestricted} & KEEP & Target B Restricted & none &  & $-$117.93 & 63.29 \\
\texttt{keep\textunderscore\allowbreak{}released\textunderscore\allowbreak{}busy\textunderscore\allowbreak{}shipped\textunderscore\allowbreak{}FSRedAgent\allowbreak{}Wrecker} & KEEP & Wrecker & none &  & $-$1115.65 & 272.91 \\
\texttt{o1\textunderscore\allowbreak{}g32\textunderscore\allowbreak{}FSRedAgent\allowbreak{}ConvOnly} & KEEP & ConvOnly & O1, gap 32 & fired & $-$268.19 & 81.52 \\
\texttt{o1\textunderscore\allowbreak{}g32\textunderscore\allowbreak{}FSRedAgent\allowbreak{}ConvOnly\textunderscore\allowbreak{}wb} & KEEP & ConvOnly & O1, gap 32 & held & $-$216.50 & 51.68 \\
\texttt{o1\textunderscore\allowbreak{}g32\textunderscore\allowbreak{}FSRedAgent\allowbreak{}WreckerQ04} & KEEP & Wrecker Q04 & O1, gap 32 & fired & $-$250.60 & 75.73 \\
\texttt{o1\textunderscore\allowbreak{}g32\textunderscore\allowbreak{}FSRedAgent\allowbreak{}WreckerQ04\textunderscore\allowbreak{}wb} & KEEP & Wrecker Q04 & O1, gap 32 & held & $-$200.80 & 67.77 \\
\texttt{o1\textunderscore\allowbreak{}g32\textunderscore\allowbreak{}FSRedAgent\allowbreak{}WreckerQ16} & KEEP & Wrecker Q16 & O1, gap 32 & fired & $-$376.15 & 102.11 \\
\texttt{o1\textunderscore\allowbreak{}g32\textunderscore\allowbreak{}FSRedAgent\allowbreak{}WreckerQ16\textunderscore\allowbreak{}wb} & KEEP & Wrecker Q16 & O1, gap 32 & held & $-$280.75 & 75.08 \\
\texttt{o1\textunderscore\allowbreak{}g32\textunderscore\allowbreak{}FSRedAgent\allowbreak{}WreckerQ32} & KEEP & Wrecker Q32 & O1, gap 32 & fired & $-$510.80 & 128.44 \\
\texttt{o1\textunderscore\allowbreak{}g32\textunderscore\allowbreak{}FSRedAgent\allowbreak{}WreckerQ32\textunderscore\allowbreak{}wb} & KEEP & Wrecker Q32 & O1, gap 32 & held & $-$379.19 & 139.25 \\
\texttt{o1\textunderscore\allowbreak{}g32\textunderscore\allowbreak{}FSRedAgent\allowbreak{}WreckerQ48} & KEEP & Wrecker Q48 & O1, gap 32 & fired & $-$579.43 & 148.02 \\
\texttt{o1\textunderscore\allowbreak{}g32\textunderscore\allowbreak{}FSRedAgent\allowbreak{}WreckerQ48\textunderscore\allowbreak{}wb} & KEEP & Wrecker Q48 & O1, gap 32 & held & $-$418.20 & 112.48 \\
\texttt{o2f\textunderscore\allowbreak{}b77} & KEEP & stock & O2f, budget 77 & fired & $-$262.71 & 74.78 \\
\texttt{oracle\textunderscore\allowbreak{}O1\textunderscore\allowbreak{}n100} & KEEP & stock & O1 & fired & $-$520.29 & 93.43 \\
\texttt{oracle\textunderscore\allowbreak{}O1\textunderscore\allowbreak{}n100t} & KEEP & stock & O1 & fired & $-$520.29 & 93.43 \\
\texttt{oracle\textunderscore\allowbreak{}O1\textunderscore\allowbreak{}n100x} & KEEP & stock & O1 & fired & $-$520.29 & 93.43 \\
\texttt{rd\textunderscore\allowbreak{}R2\textunderscore\allowbreak{}w\textunderscore\allowbreak{}O1\textunderscore\allowbreak{}g32} & rule defender R2 & Wrecker & O1, gap 32 & fired & $-$949.63 & 411.79 \\
\texttt{rd\textunderscore\allowbreak{}R2\textunderscore\allowbreak{}w\textunderscore\allowbreak{}O1\textunderscore\allowbreak{}g32\textunderscore\allowbreak{}wb} & rule defender R2 & Wrecker & O1, gap 32 & held & $-$372.64 & 77.86 \\
\texttt{rd\textunderscore\allowbreak{}R2d\textunderscore\allowbreak{}w\textunderscore\allowbreak{}O1\textunderscore\allowbreak{}g32\textunderscore\allowbreak{}wb} & rule defender R2d & Wrecker & O1, gap 32 & held & $-$380.56 & 81.66 \\
\texttt{rd\textunderscore\allowbreak{}R4tn\textunderscore\allowbreak{}w\textunderscore\allowbreak{}O1\textunderscore\allowbreak{}g32} & rule defender R4tn & Wrecker & O1, gap 32 & fired & $-$1193.23 & 446.56 \\
\texttt{rd\textunderscore\allowbreak{}R4tn\textunderscore\allowbreak{}w\textunderscore\allowbreak{}O1\textunderscore\allowbreak{}g32\textunderscore\allowbreak{}wb} & rule defender R4tn & Wrecker & O1, gap 32 & held & $-$413.37 & 115.04 \\
\texttt{rd\textunderscore\allowbreak{}RT\textunderscore\allowbreak{}w\textunderscore\allowbreak{}O1\textunderscore\allowbreak{}g32\textunderscore\allowbreak{}wb} & rule defender RT & Wrecker & O1, gap 32 & held & $-$276.91 & 34.54 \\
\texttt{rd\textunderscore\allowbreak{}RT\textunderscore\allowbreak{}w\textunderscore\allowbreak{}none\textunderscore\allowbreak{}wb} & rule defender RT & Wrecker & none & held & $-$274.88 & 34.58 \\
\texttt{red\textunderscore\allowbreak{}FSRedAgent\allowbreak{}Combined} & KEEP & Combined & none &  & $-$165.87 & 84.76 \\
\texttt{red\textunderscore\allowbreak{}FSRedAgent\allowbreak{}ConvOnly} & KEEP & ConvOnly & none &  & $-$316.62 & 150.48 \\
\texttt{red\textunderscore\allowbreak{}FSRedAgent\allowbreak{}TargetA} & KEEP & Target A & none &  & $-$104.18 & 55.60 \\
\texttt{red\textunderscore\allowbreak{}FSRedAgent\allowbreak{}TargetAOperational} & KEEP & Target A Operational & none &  & $-$99.89 & 55.40 \\
\texttt{red\textunderscore\allowbreak{}FSRedAgent\allowbreak{}TargetARestricted} & KEEP & Target A Restricted & none &  & $-$98.52 & 48.61 \\
\texttt{red\textunderscore\allowbreak{}FSRedAgent\allowbreak{}TargetB} & KEEP & Target B & none &  & $-$116.25 & 54.19 \\
\texttt{red\textunderscore\allowbreak{}FSRedAgent\allowbreak{}TargetBOperational} & KEEP & Target B Operational & none &  & $-$127.31 & 88.17 \\
\texttt{red\textunderscore\allowbreak{}FSRedAgent\allowbreak{}TargetBRestricted} & KEEP & Target B Restricted & none &  & $-$117.93 & 63.29 \\
\texttt{red\textunderscore\allowbreak{}FSRedAgent\allowbreak{}Wrecker} & KEEP & Wrecker & none &  & $-$1115.65 & 272.91 \\
\texttt{red\textunderscore\allowbreak{}FSRedAgent\allowbreak{}WreckerQ04} & KEEP & Wrecker Q04 & none &  & $-$325.11 & 132.96 \\
\texttt{red\textunderscore\allowbreak{}FSRedAgent\allowbreak{}WreckerQ16} & KEEP & Wrecker Q16 & none &  & $-$539.17 & 199.32 \\
\texttt{red\textunderscore\allowbreak{}FSRedAgent\allowbreak{}WreckerQ32} & KEEP & Wrecker Q32 & none &  & $-$773.08 & 242.81 \\
\texttt{red\textunderscore\allowbreak{}FSRedAgent\allowbreak{}WreckerQ48} & KEEP & Wrecker Q48 & none &  & $-$928.16 & 327.82 \\
\texttt{red\textunderscore\allowbreak{}FiniteStateRedAgent} & KEEP & stock & none &  & $-$187.32 & 61.24 \\
\texttt{so\textunderscore\allowbreak{}O1\textunderscore\allowbreak{}g32} & KEEP & stock & O1, gap 32 & fired & $-$200.16 & 54.72 \\
\texttt{so\textunderscore\allowbreak{}O1\textunderscore\allowbreak{}g32\textunderscore\allowbreak{}wb} & KEEP & stock & O1, gap 32 & held & $-$174.83 & 45.19 \\
\texttt{so\textunderscore\allowbreak{}O1\textunderscore\allowbreak{}n100\textunderscore\allowbreak{}wb} & KEEP & stock & O1 & held & $-$192.84 & 57.40 \\
\texttt{wo\textunderscore\allowbreak{}O1\textunderscore\allowbreak{}g32} & KEEP & Wrecker & O1, gap 32 & fired & $-$663.38 & 174.23 \\
\texttt{wo\textunderscore\allowbreak{}O1\textunderscore\allowbreak{}g32\textunderscore\allowbreak{}qf\textunderscore\allowbreak{}wb} & KEEP & Wrecker & O1, gap 32 & queue first & $-$491.90 & 130.46 \\
\texttt{wo\textunderscore\allowbreak{}O1\textunderscore\allowbreak{}g32\textunderscore\allowbreak{}wb} & KEEP & Wrecker & O1, gap 32 & held & $-$490.72 & 131.19 \\
\texttt{wo\textunderscore\allowbreak{}O2\textunderscore\allowbreak{}b77} & KEEP & Wrecker & O2, budget 77 & fired & $-$874.27 & 280.66 \\
\texttt{wo\textunderscore\allowbreak{}O2\textunderscore\allowbreak{}b77x} & KEEP & Wrecker & O2, budget 77 & fired & $-$874.27 & 280.66 \\
\texttt{wo\textunderscore\allowbreak{}O2\textunderscore\allowbreak{}g0\textunderscore\allowbreak{}b77\textunderscore\allowbreak{}wb} & KEEP & Wrecker & O2, gap 0, budget 77 & held & $-$327.54 & 44.92 \\
\texttt{wo\textunderscore\allowbreak{}O2\textunderscore\allowbreak{}g32} & KEEP & Wrecker & O2, gap 32 & fired & $-$680.42 & 158.46 \\
\texttt{wo\textunderscore\allowbreak{}O2\textunderscore\allowbreak{}g32\textunderscore\allowbreak{}wb} & KEEP & Wrecker & O2, gap 32 & held & $-$490.40 & 134.57 \\
\texttt{wo\textunderscore\allowbreak{}O2f\textunderscore\allowbreak{}b77} & KEEP & Wrecker & O2f, budget 77 & fired & $-$1080.73 & 261.52 \\
\texttt{wrecker\textunderscore\allowbreak{}null\textunderscore\allowbreak{}9000} & KEEP & Wrecker & seed 9000 &  & $-$1196.35 & 336.69 \\
\texttt{wrecker\textunderscore\allowbreak{}null\textunderscore\allowbreak{}9001} & KEEP & Wrecker & seed 9001 &  & $-$1107.05 & 297.93 \\
\texttt{wrecker\textunderscore\allowbreak{}null\textunderscore\allowbreak{}9002} & KEEP & Wrecker & seed 9002 &  & $-$1124.73 & 316.14 \\
\texttt{wrecker\textunderscore\allowbreak{}null\textunderscore\allowbreak{}9003} & KEEP & Wrecker & seed 9003 &  & $-$1159.41 & 347.01 \\
\texttt{wrecker\textunderscore\allowbreak{}null\textunderscore\allowbreak{}9004} & KEEP & Wrecker & seed 9004 &  & $-$1157.75 & 361.52 \\
\end{longtable}
\endgroup

\section{Proposed changes and their reproductions}\label{app:issues}
Each of the six defects of Table~\ref{tab:defects} has a one-command reproduction in the artifact, the Restore charge among them (\texttt{tools\_billing.py}). Nine changes follow for the two repositories concerned, seven for the challenge repository, which holds both the simulator and the template (a signal on the drop, the busy flag in the documented vector observation, an action cost charged only on accepted directives, the Restore charge in the README's reward tables, \texttt{get\_reward\_breakdown}, policy seeding, the per-step record), and two for KEEP's repository (the parallel path and the shared training seed), each with its paragraph in this paper and its reproduction command in the artifact, for the maintainers or anyone else to take up. A search of the three repositories' issue trackers in October 2026 found the drop's mechanism reported and closed as intended behavior in April 2024 \citep{cage4issue32}, with the challenge lead stating that additional actions are ignored while a blue agent is executing one, and no report of its cost, of the charge for a dropped directive or of the other five defects. The simulator's tracker also holds eleven open issues, four of them adjacent to ours and unanswered, that \texttt{set\_seed} does not reach random-number generators copied at construction (\#47), that the evaluation script sums the reward the submission's wrapper returns rather than the simulator's, so a wrapper can rewrite its own score (\#52), that collecting state for the reward calculators dominates step time (\#45), and that an invalid action is charged at the cost of the action requested (\#50 \citep{cage4issue50}), and two closed ones of the same kind, Restore costs untracked for simulator-managed blue agents, in a report that names Restore the only blue action with a cost (\#40 \citep{cage4issue40}, where the maintainers wrote that they would probably leave the code unchanged, since a change would make agents impossible to compare after the challenge, and would update the documentation) and \texttt{end\_episode} never called by the evaluation script (\#39, which the maintainers say is not needed). The two defenders' repositories have no issues.

\end{document}